%% file: tmi.tex
\documentclass[journal,twoside,web]{ieeecolor}
\usepackage{tmi}
\usepackage{cite}
\usepackage{url}
\usepackage{graphicx} 
\usepackage{amsmath,amssymb} 
\usepackage{algorithm} 
\usepackage{algorithmic}
\usepackage{textcomp} 
\usepackage{graphics}
\usepackage{threeparttable}
\usepackage{graphicx}
\usepackage{color}
\usepackage[normalem]{ulem}
\usepackage{multirow}
\usepackage{float}
\usepackage{overpic}
\usepackage{rotating}
\usepackage[table]{xcolor}
\usepackage[pagebackref=false,breaklinks=true,colorlinks, bookmarks=false]{hyperref}

\usepackage{tikz}
\usepackage{pgfplots}
\pgfdeclarelayer{background}
\pgfdeclarelayer{foreground}
\pgfsetlayers{background,main,foreground}

\usepackage{marginnote} 

\usepackage{pgfplots}
\pgfplotsset{compat=1.7}

\usepackage{silence}
\def\ie{\emph{i.e.}}
\def\eg{\emph{e.g.}}

\definecolor{bblue}{rgb}{0,150,230}
\definecolor{mygray}{gray}{.92}

\graphicspath{{./img/}{../../figure/}}
 
 \def\BibTeX{{\rm B\kern-.05em{\sc i\kern-.025em b}\kern-.08em
    T\kern-.1667em\lower.7ex\hbox{E}\kern-.125emX}}

\begin{document}

\title{Multi-scale Transformer Network with Edge-aware Pre-training for Cross-Modality MR Image Synthesis}

\author{
Yonghao~Li, 
Tao Zhou,~\IEEEmembership{Senior Member,~IEEE},
Kelei He,
Yi Zhou,
Dinggang Shen,~\IEEEmembership{Fellow,~IEEE}
\thanks{Y. Li is with the School of Biomedical Engineering, ShanghaiTech University, Shanghai, China. (e-mail: liyh2022@shanghaitech.edu.cn)}
\thanks{T. Zhou is with PCA Lab, and the School of Computer Science and Engineering, Nanjing University of Science and Technology, Nanjing 210094, China.
(e-mail: taozhou.dreams@gmail.com)
}
\thanks{K. He is with the Medical School, Nanjing University, Nanjing 210023, China, and also with the National Institute of Healthcare Data Science at Nanjing University, Nanjing 210023, China. (e-mail: hkl@nju.edu.cn)}
\thanks{Y. Zhou is with the School of Computer Science and Engineering, Southeast University, Nanjing 211189, China. (e-mail: yizhou.szcn@gmail.com)}
\thanks{D. Shen is with the School of Biomedical Engineering, ShanghaiTech University, Shanghai, China, Shanghai United Imaging Intelligence Co., Ltd., Shanghai, China, and also Shanghai Clinical Research and Trial Center, Shanghai, China. (e-mail: Dinggang.Shen@gmail.com)}
\thanks{
Corresponding authors: \textit{Tao Zhou}, \textit{Dinggang Shen}.}
}
 
\maketitle
 
\begin{abstract}

Cross-modality magnetic resonance (MR) image synthesis can be used to generate missing modalities from given ones. Existing (supervised learning) methods often require a large number of paired multi-modal data to train an effective synthesis model. However, it is often challenging to obtain sufficient paired data for supervised training. In reality, we often have a small number of paired data while a large number of unpaired data. To take advantage of both paired and unpaired data, in this paper, we propose a Multi-scale Transformer Network (MT-Net) with edge-aware pre-training for cross-modality MR image synthesis. Specifically, an Edge-preserving Masked AutoEncoder (Edge-MAE) is first pre-trained in a self-supervised manner to simultaneously perform 1) image imputation for randomly masked patches in each image and 2) whole edge map estimation, which effectively learns both contextual and structural information. Besides, a novel patch-wise loss is proposed to enhance the performance of Edge-MAE by treating different masked patches differently according to the difficulties of their respective imputations. Based on this proposed pre-training, in the subsequent fine-tuning stage, a Dual-scale Selective Fusion (DSF) module is designed (in our MT-Net) to synthesize missing-modality images by integrating multi-scale features extracted from the encoder of the pre-trained Edge-MAE. Furthermore, this pre-trained encoder is also employed to extract high-level features from the synthesized image and corresponding ground-truth image, which are required to be similar (consistent) in the training. Experimental results show that our MT-Net achieves comparable performance to the competing methods even using $70\%$ of all available paired data. Our code will be released at \href{https://github.com/lyhkevin/MT-Net}{https://github.com/lyhkevin/MT-Net}.

\end{abstract}


\section{Introduction}\label{sec:introduction}
\IEEEPARstart{M}{edical} imaging provides a visual approach to show the anatomy or function of the body part, which brings great convenience to medical research and clinical diagnosis~\cite{zhou2019latent,fan2020inf,ouyang2020dual}. In particular, magnetic resonance imaging (MRI) produces anatomical images utilizing magnetic fields and radio waves in a non-invasive manner. Various scanning parameters can be set to produce different modalities, such as T1-weighted (T1), T2-weighted (T2), T1-weighted dynamic contrast-enhanced (T1c), and T2-fluid-attenuated inversion recovery (FLAIR). Several existing works have shown that the use of multi-contrast MR imaging is beneficial for downstream tasks such as segmentation~\cite{yang2023flexible,srinivas2020segmentation,Jia2012IterativeMM, Ren2018Interleaved3F,Zhou2020HighResolutionEN}, morphological classification~\cite{liu2020enhancing,tong2017multi}, and disease prediction~\cite{Shi2012AlteredSC,Fan2007MultivariateEO,Liu2014HierarchicalFO}, as each modality exhibits a unique contrast of the same scanned tissue, which provides complementary information for clinical diagnosis. 

Several challenges, including high costs, limited scanning time, and image corruption, pose significant challenges to obtaining complete multi-modalities for each patient. The absence of certain modalities, as well as inconsistent modalities among medical institutions, can negatively affect the process of clinical diagnosis. As a potential solution, cross-modality medical image synthesis has received increasing attention in research. Medical image synthesis aims to predict missing-modality (or called target-modality) images from given source-modality ones. Currently, the majority of cross-modality medical image synthesis approaches~\cite{nie2017medical,nie2018medical,ge2019unpaired,zhou2020hi,yu2019ea,luo2021edge,dalmaz2021resvit,chen2021targan,zhang2021ptnet,liu2022one} are rooted in deep learning, which utilizes both source- and target-modality images to train artificial neural networks. Recently, generative adversarial networks (GANs)~\cite{goodfellow2014generative} have become the preferred method for cross-modality synthesis~\cite{nie2017medical,nie2018medical,ge2019unpaired,yu2019ea,luo2021edge,zhou2020hi,dalmaz2021resvit,chen2021targan}, where two neural networks compete with each other to produce more realistic predictions. In spite of their potential to enhance the resolution and realism of synthesized images, GANs are known to suffer from issues such as non-convergence, mode collapse, and vanishing gradients~\cite{Arjovsky2017WassersteinG}. 

Convolutional neural networks (CNNs) continue to dominate in terms of building blocks of the existing medical image synthesis methods\cite{nie2017medical,nie2018medical,zhou2020hi}. However, limited by the local receptive field in convolution operation, CNNs struggle with modeling long-range dependencies and contextual information~\cite{Shamshad2022TransformersIM}. In recent years, the attention-based Transformer model~\cite{vaswani2017attention,dosovitskiy2020image} has emerged as an alternative building block in computer vision. The design of Transformers is rooted in the self-attention mechanism~\cite{vaswani2017attention}, which effectively captures interdependence between elements of the input sequence for a better feature representation ability~\cite{Shamshad2022TransformersIM}. Due to this capability, a growing number of studies~\cite{dalmaz2021resvit,zhang2021ptnet,liu2022one} have focused on applying Transformers to medical image synthesis. 

With regard to optimization objectives, the majority of current methods aim to minimize the dissimilarity between synthesized and real images at pixel and voxel levels, while neglecting structural information contained in the images. More importantly, existing (supervised learning-based) synthesis methods heavily rely on fully paired modalities for end-to-end training. However, it is very difficult to collect a sufficient number of paired training samples. In reality, we often have a small number of paired data and a large number of unpaired data. In this case, a commonly adopted approach is to discard the samples of a patient with one or more missing modalities, followed by supervised learning using the remaining pair-wise aligned samples. However, this strategy ignores quite a bit of contextual information contained in the discarded unpaired samples, which further reduces the availability of scarce medical data. 

To address the aforementioned issues, we propose a novel Multi-scale Transformer Network (MT-Net), which takes a two-stage optimization: 1) edge-aware pre-training, and 2) multi-scale fine-tuning. Inspired by Masked Autoencoders~\cite{he2022masked}, an Edge-preserving Masked AutoEncoder (Edge-MAE) is presented, which is pre-trained using both paired and unpaired multimodal MR images in a self-supervised learning manner. During fine-tuning, our MT-Net effectively integrates multi-scale features extracted by the pre-trained encoder of Edge-MAE for cross-modality synthesis. Experimental results on two multi-modal MR datasets reveal that our method achieves comparable performance with other state-of-the-art synthesis methods when only using $70\%$ paired data. 

The main contributions are summarized as follows:

\begin{itemize}

\item We propose a novel cross-modality MR image synthesis framework with edge-aware pre-training, which can leverage self-supervised pre-training to handle the challenge of limited paired data.
 
\item An Edge-MAE is presented to simultaneously preserve intensity information as well as edge information by performing 1) image imputation for randomly masked patches and 2) whole edge map estimation. Edge-MAE is capable of efficiently encoding multi-modality MR images by utilizing a solitary pre-training process with both paired and unpaired data.
 
\item We design a novel patch-wise loss to augment the performance of the proposed Edge-MAE, by differentiating between input patches based on the difficulties of their respective imputation.

\item Our MT-Net is proposed for fine-tuning by integrating multi-scale features extracted by the pre-trained encoder of Edge-MAE. Besides, a Dual-scale Selective Fusion (DSF) module is proposed to adaptively aggregate multi-scale features for the synthesis of target-modality images. Moreover, we combine the $\ell_1$ loss with feature consistency loss to enhance the synthesized details and edges.

\end{itemize}

\section{Related works}\label{sec:Related works} 

\begin{figure*}[t!]
	\centering
    \footnotesize
	\begin{overpic}[width=1\textwidth]{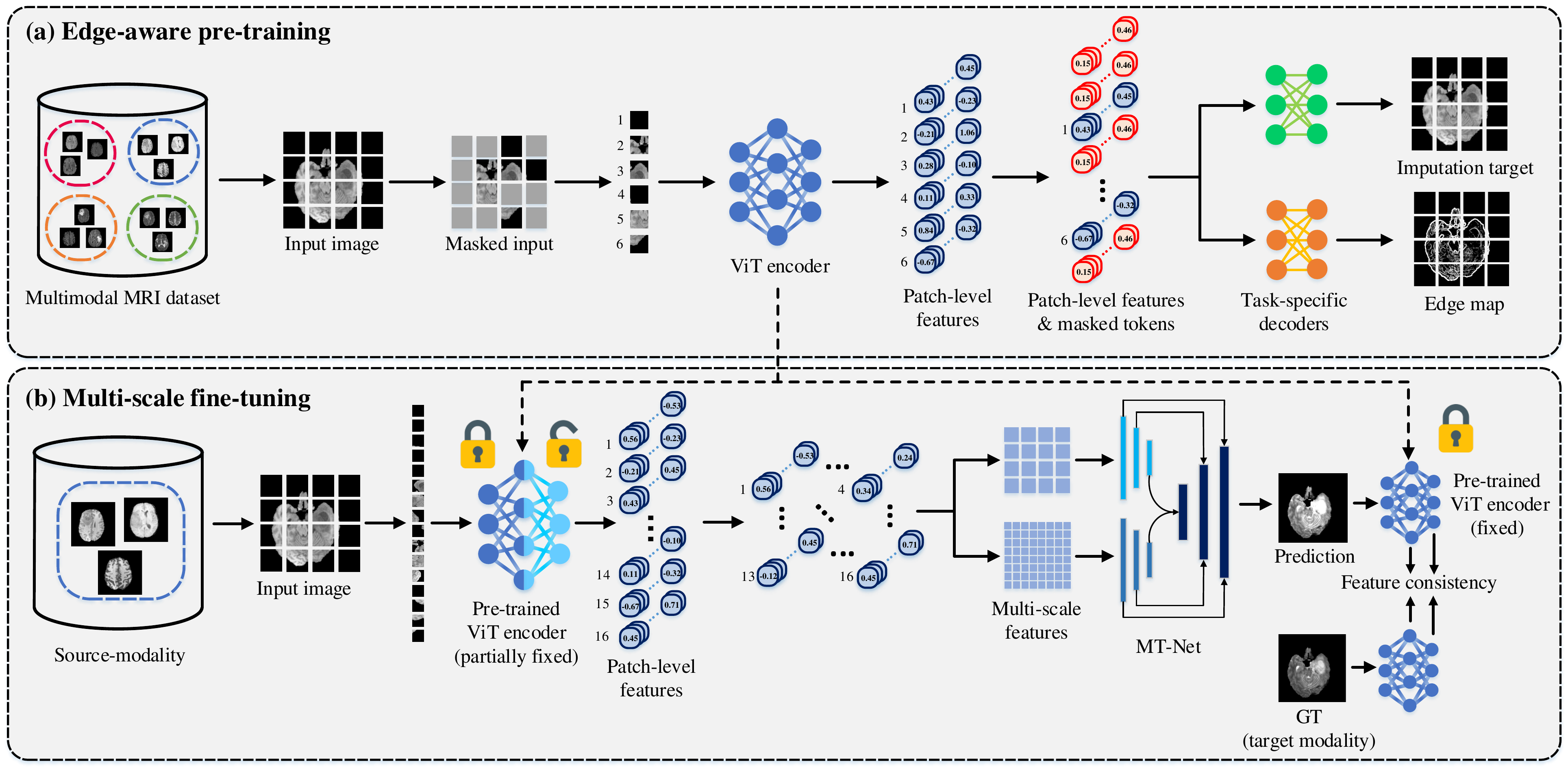}
    \end{overpic}\vspace{-0.25cm}
	\caption{\footnotesize An overview of the proposed framework for MR image synthesis, which consists of two key steps: edge-aware pre-training and multi-scale fine-tuning. (a) We conduct self-supervised pre-training using all available multi-modal data, rather than discarding samples with missing modalities. Our Edge-MAE consists of a Transformer-based encoder, and two task-specific decoders for 1) image imputation for randomly masked patches and 2) whole edge map estimation. (b) The proposed MT-Net is fine-tuned to synthesize the missing-modality images from the source-modality images. The first six layers of the pre-trained Edge-MAE encoder remain frozen during fine-tuning. See Sec.~\ref{Ourmethod} for details.}
    \label{fig:Framework}
\end{figure*}
 
\subsection{Medical Image Synthesis}

Cross-modality medical image synthesis aims to learn a mapping from the given source-modality images to the missing-modality images. A large portion of existing methods~\cite{nie2018medical, zhou2020hi, Qu2020Synthesized7M, yu2019ea, dalmaz2021resvit} are variants of conditional GANs (cGANs), such as Pix2Pix~\cite{isola2017image} and CycleGAN~\cite{zhu2017unpaired}. Cross-modality synthesis tasks include CT to MRI~\cite{nie2018medical}, CT to PET~\cite{BenCohen2018CrossModalitySF,Santini2020UnpairedPI}, ultrasound images to MRI~\cite{Jiao2020SelfSupervisedUT}, low-dose PET to high-dose ones~\cite{wang20183d}, and multi-contrast MR image synthesis~\cite{zhou2020hi, Qu2020Synthesized7M, yu2019ea,dalmaz2021resvit,zhang2021ptnet,liu2022one}. In addition, various applications can be divided into cross-modality synthesis with paired modalities~\cite{nie2018medical,zhang2021ptnet} or unpaired modalities~\cite{Jiao2020SelfSupervisedUT,Santini2020UnpairedPI}. Existing (supervised learning) synthesis approaches require the completely paired modalities to train a generator, by discarding the samples with missing modalities. The synthesis approaches with unpaired modalities~\cite{Jiao2020SelfSupervisedUT, zhu2017unpaired} often involve automatic training of image-to-image translation models without paired samples. Nevertheless, few studies consider the partially paired data scenario, where only part of the samples is paired. Additionally, most existing works minimize the pixel/voxel-wise distance between the synthesized missing-modality image and the ground truth, ignoring the preservation of critical edge information. To address this issue, some studies~\cite{Jiao2020SelfSupervisedUT,yu2019ea} simultaneously maintain voxel-wise intensity similarity and edge constraint. Different from existing MR image synthesis approaches, our method takes advantage of both paired and unpaired data for better feature representation. To be specific, a transformer-based encoder is pre-trained in a self-supervised manner to encode multi-modality MR images without discarding any unpaired samples. Besides, our model does not rely on adversarial training, making the training process more stable.

\subsection{Transformers in Medical Imaging }

Transformer was originally designed for natural language processing~\cite{vaswani2017attention}, which relies on self-attention mechanisms. For image classification tasks, \cite{dosovitskiy2020image} adopted a pure transformer model with the fewest possible modifications. Specifically, self-attention captures long-range dependencies by determining the correlation between the embeddings of all image patches.  Transformers have been successfully applied to the field of medical image analysis~\cite{chen2021transunet,cao2021swin,chen2021vit,perera2021pocformer}. The important applications of Transformers in medical image analysis include medical image segmentation~\cite{chen2021transunet, cao2021swin}, classification~\cite{perera2021pocformer}, registration~\cite{chen2021vit}, and cross-modality synthesis~\cite{zhang2021ptnet, dalmaz2021resvit}. To mitigate the high computational and memory shortcomings of self-attention, most current methods employ hybrid architecture with both CNN and Transformers~\cite{chen2021transunet}, or build a hierarchical representation of the input image by adding downsampling modules between transformer layers~\cite{zhang2021ptnet,cao2021swin}. 

\subsection{Self-supervised Learning}

Self-supervised learning paradigm is used to learn feature representations without requiring annotated datasets. Prior to fine-tuning for downstream tasks (\eg, image segmentation), the network is initialized by performing a pretext task, such as solving jigsaw puzzles~\cite{noroozi2016unsupervised}, masked pixel prediction~\cite{chen2020generative}, or image imputation for randomly masked image patches~\cite{he2022masked}. This process is referred to as pre-training. Self-supervised learning methods can be classified into different categories according to their network architectures and pretext tasks. For instance, contrastive learning methods~\cite{he2020momentum,chen2020simple,zbontar2021barlow} use encoder-only architecture to learn an embedding space that keeps similar samples close to each other, while keeping dissimilar samples far apart. Autoencoder-based methods~\cite{bao2021beit} map the input into the latent space with an encoder while reconstructing the input with a decoder. More recently, Masked Autoencoder (MAE)~\cite{he2022masked} is proposed, in which a transformer-based encoder learns the latent representation of a small subset of visible patches, while a lightweight decoder imputes the original input from mask tokens and latent representation. MAE achieves faster pre-training and reduced memory usage because of its high masking ratio and asymmetric architecture. Moreover, various studies~\cite{chen2019self,azizi2021big,Zhou2022SelfPW} have been developed to apply the self-supervised learning paradigm to medical image analysis.

\section{Proposed Method}
\label{Ourmethod}

\subsection{Overview}
As shown in Fig.~\ref{fig:Framework}, the proposed cross-modality framework consists of two key steps: edge-aware pre-training with the proposed Edge-MAE (Sec. \ref{subsection_B}), and multi-scale fine-tuning with our MT-Net (Sec. \ref{subsection_C}). Notably, the edge-aware pre-training is conducted using multi-modal MR images in a self-supervised learning manner. We will provide a detailed explanation of each key component below. 

\subsection{Edge-aware Pre-training}\label{subsection_B}

\subsubsection{Edge-preserving Masked AutoEncoder}

The architecture of the proposed Edge-MAE is illustrated in Fig.~\ref{fig:Framework}, which consists of a shared Transformer-based encoder and two task-specific decoders. Following ViT~\cite{dosovitskiy2020image}, the input image is first divided into a series of non-overlapping patches. Subsequently, we randomly mask a large portion (\eg, 70\%) of patches, and the remaining unmasked patches are projected into $D$-dimension embeddings. The masked patches are then discarded, while patch embeddings of unmasked patches are encoded by a standard ViT encoder. Following the encoding process, the learnable mask tokens are introduced to represent each patch of the masked image. After that, two task-specific decoders receive full sets of tokens, including latent representations of unmasked patches, as well as the learnable mask tokens. Notably, positional embeddings are added to the tokens in order to retain patch-wise positional information.

The original MAE~\cite{he2022masked} utilizes a transformer-based decoder to impute patches in the masked position. It is crucial for the encoder to effectively capture the relationships between unmasked patches, as the imputation performance is highly dependent on the quality of the representation from the encoder. The encoder, therefore, is capable of learning contextual information from the input image. %
As the original MAE minimizes only pixel-wise intensity difference, it fails to preserve structural information, such as edges, leading to ambiguous edges in the imputed images. However, edge information is crucial for depicting the boundaries between tissues~\cite{yu2019ea,luo2021edge}. For instance, without integrating edge information, the contours of synthesized lesions may appear fuzzy, which negatively impacts downstream tasks, such as morphology classification and lesion segmentation. Thus, another task-specific decoder is employed to enhance the edge-preserving ability of our framework, which estimates the corresponding edge maps of the input images. Specifically, the Sobel edge detector is applied to obtain the ground truth edge maps of source-modality images. Therefore, our Edge-MAE is pre-trained in a multi-task learning manner, which simultaneously preserves both contextual and structural information.

\begin{figure}[!t]
	\centering
	\footnotesize
	\begin{overpic}[width=1\columnwidth]{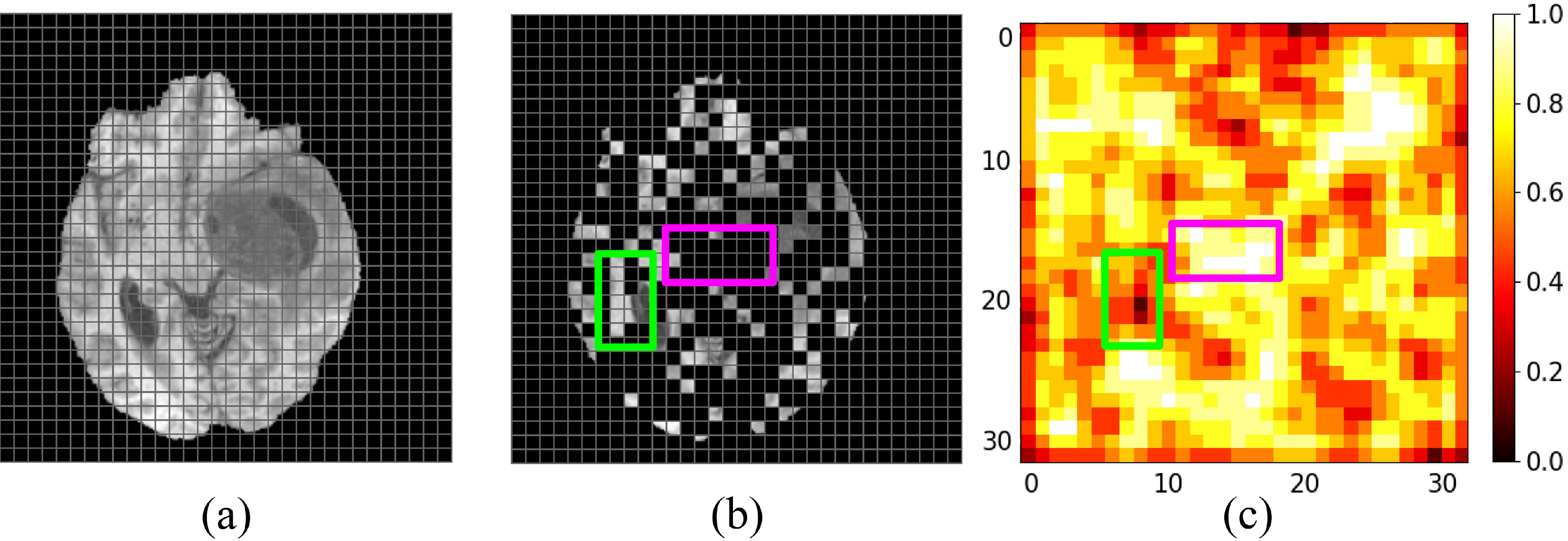}
	\end{overpic}\vspace{-0.25cm}
	\caption{\footnotesize Illustration of patch divisions and weights: (a) a $256\times{256}$ image is partitioned into a series of non-overlapping patches of size $8\times{8}$; (b) masked image; and (c) weight map. The green box contains only a few masked patches, whereas the purple box indicates a heavily masked area. Due to the lack of contextual information, imputing patches from heavily masked areas can be more challenging.
 }
	\label{fig:patchloss}\vspace{-0.25cm}
\end{figure}

\subsubsection{Patch-wise Loss}

The Mean Square Error (MSE) is a commonly used metric for quantifying the difference between the predicted image and the ground truth. Nevertheless, it treats all patches equally, regardless of varying levels of patch-wise imputation difficulties. This can be attributed to two reasons: a) semantic information varies from patch to patch (patches in the foreground, \eg, tumor patches, contain much more semantic information than those in the background); b) the employment of a random masking strategy also contributes to the diverse levels of image imputation difficulties. As shown in Fig.~\ref{fig:patchloss}, a $256\times{256}$ T1 modality image is partitioned into a series of non-overlapping $8\times{8}$ patches, with $70\%$ of patches randomly masked. Fig.~\ref{fig:patchloss} (b) illustrates that the majority of patches in the green box are visible, whereas the majority of patches in the purple box are masked. Consequently, the level of imputation difficulty of these masked patches varies. For instance, patches from heavily masked areas are particularly challenging to impute, due to the lack of contextual information from surrounding patches. Hence, patches from heavily masked areas are referred to as \textit{hard patches}, while patches from partially masked areas are referred to as \textit{easy patches}. Our goal is to prioritize the imputation of easy patches in the early stage of pre-training, while allocating greater attention to those hard patches during subsequent training epochs. Specifically, each input patch is first assigned a weight $\alpha$, for representing the level of difficulty for imputation. To obtain $\alpha$ for each patch, a binary mask is generated based on the random masking strategy~\cite{he2022masked}. For instance, the binary mask for the input image in Fig.~\ref{fig:patchloss} (b) has a size of $32\times{32}$, corresponding to the total number of patches, where a value of $1$ represents a masked patch, and $0$ stands for an unmasked patch. After that, average pooling is performed on the binary mask to obtain the patch weight $\alpha  \in {R^{H/P \times W/P}}$, where $H$ and $W$ denote respectively the height and the width of an input image, and $P$ denotes the patch size. Fig.~\ref{fig:patchloss} (c) shows that hard patches are assigned with a larger $\alpha$, while easy patches are assigned with a smaller $\alpha$. Furthermore, all pixels in the same patch share the same weight, and $\alpha  \in [0,{\rm{ 1]}}$ for all patches. 

During pre-training, the masked patches are imputed following the principle of ``easy to hard". During the initial stage of pre-training, we prioritize the imputation of \textit{easy patches} from partially masked areas. Following this, we shift our attention to imputing these \textit{hard patches} from heavily masked areas. A novel patch-wise loss based on the weighted $\ell_1$-norm is proposed, in which masked patches are treated differently according to their imputation difficulties. Subsequently, the loss function of Edge-MAE is formulated as follows:
\begin{equation}\label{equ:stage1}
\begin{split}
\mathcal{L}_{stage1} =& {\lambda _{imp}}[{\left\| {(2 - \alpha )(y - {D_{imp}}(E(x)))} \right\|_1}]+\\
 & {\lambda _{edge}}[{\left\| {(2 - \alpha )(S(y) - {D_{edge}}(E(x)))} \right\|_1}],
 \end{split}
\end{equation}
\begin{equation}\label{equ:stage2}
\begin{split}
\mathcal{L}_{stage2} = & {\lambda _{imp}}[{\left\| {(1 + \alpha )(y - {D_{imp}}(E(x)))} \right\|_1}] +\\
 & {\lambda _{edge}}[{\left\| {(1 + \alpha )(S(y) - {D_{edge}}(E(x)))} \right\|_1}],
 \end{split}
\end{equation}
where $x$ represents an input image, and $y$ represents the ground truth. ${\lambda}_{imp}$ and ${\lambda}_{edge}$ are hyper-parameters used to balance different tasks. $E$, ${D}_{imp}$, and ${D}_{edge}$ represent the encoder, the task-specific decoder for image imputation, and the decoder for edge map estimation, respectively. $S$ represents the Sobel edge detector, and $S(y)$ denotes the ground truth edge map of the input image. By optimization of $\mathcal{L}_{stage1}$ in Eq.~\eqref{equ:stage1}, in which the coefficient $2-\alpha$ assigns an easy patch with a larger weight $\alpha$, we prioritize the imputation of easy patches. Then, we shift our focus to hard patches, by optimizing $\mathcal{L}_{stage2}$ in Eq.~\eqref{equ:stage2}, in which the coefficient $1+\alpha$ assigns a hard patch with a larger weight. The proposed patch-wise loss results in expedited convergence, and facilitates the pre-training process. 

\begin{figure}[!t!]
	\centering
	\footnotesize
	\begin{overpic}[width=1\columnwidth]{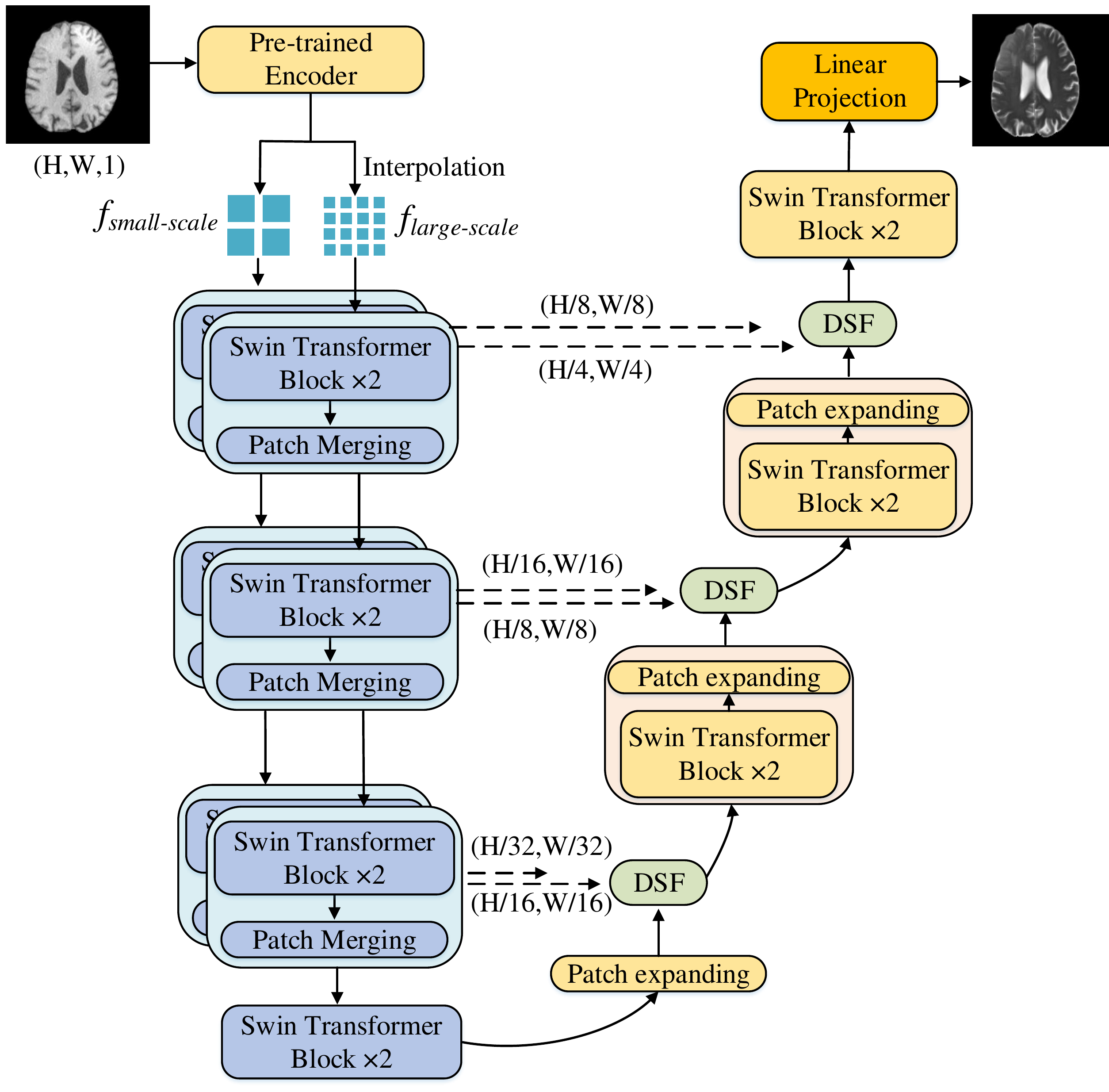}
	\end{overpic}\vspace{-0.05cm}
	\caption{\footnotesize Illustration of the proposed multi-scale transformer network. 
 }\vspace{-0.35cm}
	\label{fig:mtunet}
\end{figure}

\begin{figure*}[t!]
	\centering
    \footnotesize
	\begin{overpic}[width=0.95\textwidth]{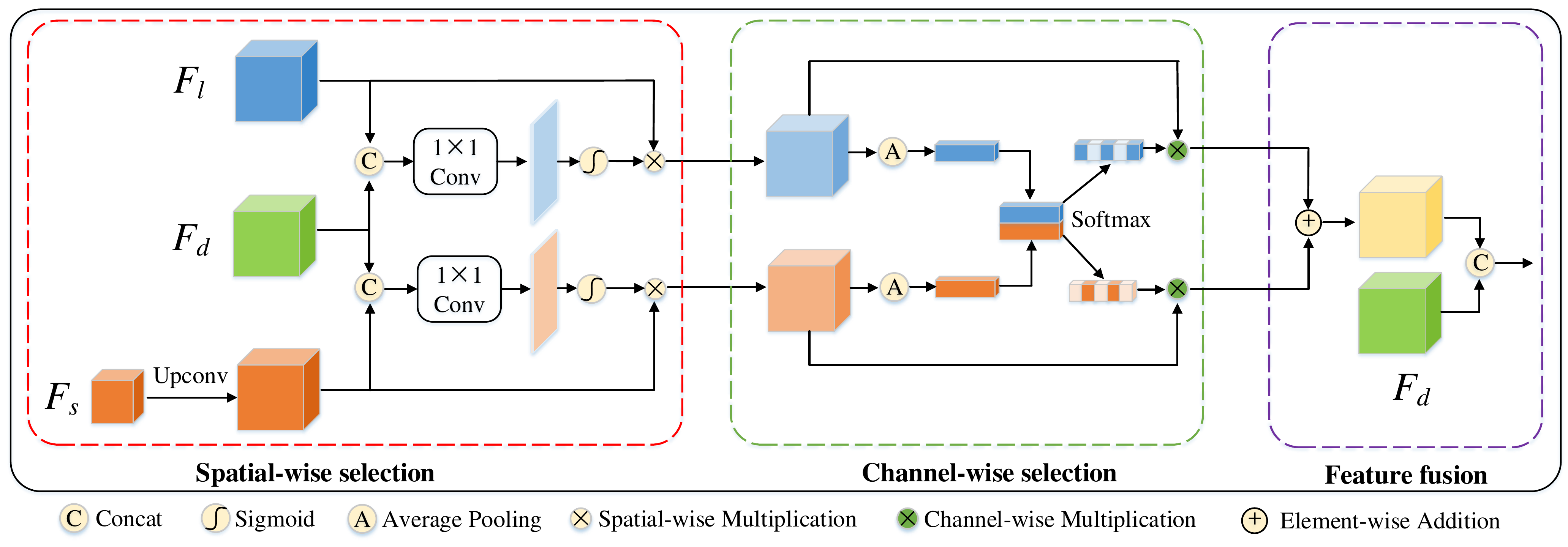}
    \end{overpic}\vspace{-0.15cm}
	\caption{\footnotesize Illustration of the proposed Dual-scale Selective Fusion (DSF) module, which consists of three parts: spatial-wise selection, channel-wise selection, and feature fusion.}\vspace{-0.35cm}
    \label{fig:dsf}
\end{figure*}

\subsection{Multi-scale Fine-tuning}\label{subsection_C}

Multi-scale fine-tuning is then performed to adapt our pre-trained framework for the downstream task of cross-modality synthesis. The incorporation of multi-scale features has been demonstrated to be beneficial in medical image analysis~\cite{Zhang2022MultiscaleFP,Fang2019UnifiedMF,Li2019MultiInstanceMC}. Thus, a Multi-scale Transformer Network (MT-Net) is proposed for fine-tuning, depicted in Fig.~\ref{fig:mtunet}, which employs an encoder-decoder architecture. The construction of multi-scale features is first performed with the single-scale output of the pre-trained Edge-MAE encoder. Specifically, we first conduct a bilinear interpolation to the output feature map ${f}_{small-scale}$, which yields a higher resolution feature ${f}_{large-scale}$. Next, ${f}_{small-scale}$ and ${f}_{large-scale}$ are fed into two independent encoder branches of our MT-Net, which comprises multiple downsampling stages. Within each stage, two consecutive Swin Transformer~\cite{cao2021swin} layers and a patch merging module lower the resolution of the feature maps, while simultaneously doubling the feature dimension, leading to a hierarchical feature representation. In comparison to the small-scale branch, the large-scale branch includes an additional downsampling stage. The decoder employs patch expanding operations~\cite{cao2021swin} and Swin Transformer layers to increase feature resolution, while skip connections~\cite{Ronneberger2015UNetCN} allow the encoders and the decoder to communicate, thereby preserving spatial information. 

\subsubsection{Dual-scale Selective Fusion Module} 
To adaptively fuse multi-scale features extracted by the dual-branch encoders, a Dual-scale Selective Fusion (DSF) module is proposed to effectively integrate multi-scale features, and then propagate the fused features into the decoder through skip connections. Inspired by the existing attention-based feature selection modules\cite{Sagar2021DMSANetDM}, the proposed DSF module consists of three key components, \ie, spatial-wise selection, channel-wise selection, and feature fusion. Specifically, as illustrated in Fig.~\ref{fig:dsf}, we denote the multi-scale features from dual-branch encoders as $F_{l}\in {R^{C \times (H \times W)}}$ and $F_{s} \in {R^{C \times (H/2 \times W/2)}}$, and the upsampled features from the decoder are denoted as $F_{d} \in {R^{C \times (H \times W)}}$. Due to the low resolution for $F_{s} \in {R^{C \times (H/2 \times W/2)}}$, we first conduct a transposed convolution layer on them to obtain the upsampled features $F_{s}^{up} \in {R^{C \times (H \times W)}}$. During the spatial-wise selection stage, we concatenate $F_d$ with $F_l$ and ${F_s^{up}}$, respectively, and then the concatenated features are fed into the point-wise convolution operations to generate two spatial attention maps ${M_A} \in {R^{H \times W}}$ and ${M_B} \in {R^{H \times W}}$. Then, the Sigmoid function is employed to scale the spatial attention maps into $[0,1]$. To suppress noisy features while emphasizing informative ones, we perform a spatial-wise multiplication between the spatial attention maps and the input features, thus we can obtain the filtered features $F_{l}^{att}$ and $F_s^{att}$. In the channel-wise selection stage, we apply a global average pooling to the filtered input features $F_{l}^{att}$ and $F_s^{att}$ to obtain channel context descriptors ${P_l} \in {R^{C \times 1}}$ and ${P_s} \in {R^{C \times 1}}$. Then, we combine them as ${F_p} = [{P_l},{P_s}] \in {R^{2C \times 1}}$, followed by a softmax operation along the channel-wise logits:

\begin{equation}\label{equ:channel}
a_c^i = \frac{{{e^{P_l^i}}}}{{{e^{P_l^i}} + {e^{P_s^i}}}},b_c^i = \frac{{{e^{P_s^i}}}}{{{e^{P_l^i}} + {e^{P_s^i}}}},
\end{equation}
where $a_c^i$ and $b_c^i$ denote the $i$-th element of the corresponding channel attention maps, respectively. Here, we have $a_c^i + b_c^i = 1$. $P_l^i$ and $P_s^i$ represent the $i$-th element in ${P_l}$ and ${P_s}$, respectively. We obtain informative features by fusing dual-branch features, which can be processed by
\begin{equation}\label{equ:fusion}
F = {a_c} \cdot F_{l}^{att} + {b_c} \cdot F_{s}^{att},
\end{equation}
where $F$ denotes the fused feature map. Finally, we concatenate $F$ and $F_{d}$ as the input of the next upsampling stage.

\subsubsection{Feature Consistency Module}
Currently, most cross-modality synthesis methods~\cite{nie2017medical,nie2018medical,ge2019unpaired,yu2019ea,luo2021edge,zhou2020hi,dalmaz2021resvit,chen2021targan} are based on GANs, which train both a generator and a discriminator simultaneously in a min-max game. Nevertheless, the instability of GANs frequently results in the mode collapse issue~\cite{Arjovsky2017WassersteinG}. In contrast to adversarial training, we employ the pre-trained Edge-MAE as a feature consistency module following ~\cite{johnson2016perceptual}, which leads to increased training stability and perceptually enhanced results. Note that our Edge-MAE can naturally act as a feature consistency module due to its capability of encoding multiple modalities and preserving edges. Therefore, edge-preserving loss~\cite{yu2019ea,luo2021edge} and adversarial loss~\cite{goodfellow2014generative} are no longer necessary during fine-tuning. Specifically, both the synthesized images $\hat y = G(E(x))$ and the corresponding missing-modality ground truth $y$ are fed into the pre-trained Edge-MAE encoder, where $x$, $E$, and $G$ denote the input image, the pre-trained encoder, and the proposed MT-Net, respectively. Letting ${F_j}(y)$ and ${F_j}(\hat y)$ be the outputs from the $j$-th transformer layer of the feature consistency module that extracts multi-level features from $y$ and $\hat y$, we use the feature consistent loss to measure the perceptual difference between the synthesized image and the ground truth, which can be defined by
\begin{equation}\label{equ:perceploss}
{\mathcal{L}^{Feature}}(\hat y,y) = \sum\nolimits_{j = 1}^l {{\mathbb{E}}[{{\left\| {{F_j}(\hat y) - {F_j}(y)} \right\|}_1}} ],
\end{equation}
where $\mathbb{E}$ denotes expectation, and $l$ denotes the number of transformer layers of the feature consistency module. Utilization of the feature consistency loss in our framework allows us to prioritize the similarity in content and style between images. In addition, the conventional per-pixel difference between the synthesized image and the ground truth can be measured by
\begin{equation}\label{equ:pixloss}
{\mathcal{L}^{Pix}}(\hat y  ,y) = {\mathbb{E}}[{\left\| {\hat y  - y} \right\|_1}].
\end{equation}

Then, $\mathcal{L}^{Pix}$ and $\mathcal{L}^{Feature}$ are linearly combined to form the overall objective function of the fine-tuning framework:
\begin{equation}\label{equ:totaloss}
\mathcal{L} = {\mathcal{L}^{Pix}} + {\mathcal{L}^{Feature}}.
\end{equation}

As a result, we manage to minimize pixel-wise intensity differences as well as enhance structural similarity.

\subsection{Detailed Architectures}

Our Edge-MAE has an encoder with 12 transformer layers, which divides the input image into non-overlapping patches of size $8\times{8}$. $70\%$ of image patches are randomly masked during pre-training, and each image patch has an embedding dimension of $128$. Each task-specific decoder consists of 8 transformer layers, with shared first three layers. A linear layer reduces the embedding dimension to $64$ before the decoding process. The patch expanding module~\cite{cao2021swin} of the proposed MT-Net consists of a linear layer that doubles the feature dimension, and a rearrange operation that expands the feature resolution. The final prediction is scaled into $[0,1]$ using a Sigmoid activation function. 

\section{Experiments AND Results}\label{sec:Experiments}

\subsection{Datasets}

We use the multi-modal brain tumor segmentation challenge 2020 (BraTS2020) dataset~\cite{menze2014multimodal} and the ischemic stroke lesion segmentation challenge 2015 (ISLES2015) dataset~\cite{maier2017isles} to validate the effectiveness of the proposed MT-Net. 

The BraTS2020 dataset consists of multi-parametric MRI (mpMRI) scans from 369 diffuse glioma patients with four modalities: T1-weighted (T1), contrast-enhanced (T1c), T2-weighted (T2), and FLAIR. We utilize 2D axial-plane slices of the volumes as inputs. Each 2D axial-plane slice ($240\times{240}$) is cropped to $200\times{200}$ from the image center, and further resized to $256\times{256}$. Then, the original intensity values are linearly scaled into $[0,1]$.  Besides, $295$ subjects are randomly selected as the training set and the remaining $74$ subjects are for testing. 

The ISLES2015 dataset consists of multi-spectral MR scans from $45$ subjects with four modalities: T1, T2, diffusion-weighted imaging (DWI), and FLAIR. For each 2D axial-plane slice ($230\times{230}$), we crop out an image of size $200\times{200}$ from the image center and resize it to $256\times{256}$. We also linearly scale the original intensity values into $[0,1]$. In this study, $28$ subjects are split as the training set and $17$ subjects for testing. 

During the pre-training, the Edge-MAE is trained with all four modalities in the training set. In the fine-tuning, a subset of subjects from the training set is used to assess the performance of MT-Net on different amounts of paired data.

\subsection{Comparison Methods and Evaluation Metrics}

\subsubsection{Comparison Methods}

We compare the proposed MT-Net with several image synthesis methods, including Pix2Pix~\cite{isola2017image}, PTNet~\cite{zhang2021ptnet}, ResViT~\cite{dalmaz2021resvit}, and TransUNet~\cite{chen2021transunet}. These methods can be summarized as follows.
1) \textbf{Pix2Pix}~\cite{isola2017image} is an architecture for conditional image-to-image translation, which uses a cGAN objective combined with a reconstruction loss.
2) \textbf{TransUNet}~\cite{chen2021transunet} is a U-shaped generator that extracts global contexts through transformer layers. An additional CNN-based discriminator enables us to train the TransUNet in an adversarial learning manner.
3) \textbf{PTNet}~\cite{zhang2021ptnet} is a transformer-based synthesis network with no convolution layers. 
4) \textbf{ResViT}~\cite{dalmaz2021resvit} employs transformers as the central bottleneck to preserve context extracted by CNN. %
Table \ref{tab:flops} shows the comparison of parameters and GFLOPs of different methods. Note that the pre-training process is considered when computing the number of parameters and GFLOPs of our framework. From Table \ref{tab:flops}, it can be observed that transformer-based methods, \eg, ResViT, require more computational complexity than CNN-based methods, \eg, Pix2Pix, primarily due to the inclusion of multi-head self-attention operations.

\begin{table}[!t]
	\centering
	\footnotesize
	\renewcommand{\arraystretch}{1.0}
	\setlength\tabcolsep{8pt}
	\caption{Comparison of parameters and GFLOPs. The GFLOPs of all methods are calculated with an input image size of 256$\times$256}. \vspace{-0.15cm}
 \label{tab:flops}
	\begin{tabular}{c|ccc}
\hline
\rowcolor{mygray}
Methods   & Pre-trained  & \multicolumn{1}{l}{Params~(M)} & \multicolumn{1}{l}{GFLOPs} \\ \hline
Pix2Pix   &   \scalebox{1.2}{$\times$}         & 41.2                          & 40.1                       \\
ResViT    & \scalebox{1.2}{$\times$}             & 123.4                         & 486.1                      \\
PTNet     & \scalebox{1.2}{$\times$}         & 27.69                          & 233.1                      \\
TransUNet & \scalebox{1.2}{$\times$}            & 96.07                         & 48.34                      \\
Ours      & \checkmark            & 138.2                         & 221.5                      \\ \hline
\end{tabular}
\end{table}

\subsubsection{Evaluation Metrics}

To quantitatively evaluate the synthesis performance of different models, we adopt three widely used evaluation metrics as follows: 1) Peak signal-to-noise ratio (PSNR). Given a ground truth missing-modality image $y(x)$ and the synthesized image $G(x)$, PSNR is defined as ${\rm{PSNR}} = 10{\log _{10}}\frac{{{{\max }^2}(y(x),G(x))}}{{\frac{1}{N}\sum {\left\| {y(x) - G(x)} \right\|_2^2} }}$, where ${\max ^2}(y(x),G(x))$ means the maximal intensity value of $y(x)$ and the generated image $G(x)$. A higher PSNR value indicates better performance of the synthesis model; 2)  Normalized mean squared error (NMSE). NMSE can be defined by ${\rm{NMSE}} = \frac{{\left\|y(x) - G(x) \right\|_2^2}}{{\left\| {y(x)} \right\|_2^2}}$, where a lower NMSE indicates better synthesis performance. 3) Structural similarity index measure (SSIM). SSIM can be defined by ${\rm{SSIM}} = \frac{{(2{\mu _{y(x)}}{\mu _{G(x)}} + {c_1})(2{\sigma _{y(x)G(x)}} + {c_2})}}{{(\mu _{y(x)}^2 + \mu _{G(x)}^2 + {c_1})(\sigma _{y(x)}^2 + \sigma _{G(x)}^2 + {c_2})}}$, where ${\mu _{y(x)}},{\mu _{G(x)}},{\sigma _{y(x)}}$, and ${\sigma _{G(x)}}$ are the mean and variance of $y(x)$ and $G(x)$, respectively. The positive constants $c_1$ and $c_2$ are used to avoid a null denominator. Note that higher SSIM indicates better synthesis performance. 

\subsection{Implementation Details}

The pre-training procedure costs over 200 epochs without data augmentation. The original learning rate is set to 0.0005 for the first 100 epochs and decreases to 0.00005 over the remaining epochs. Our Edge-MAE applies $\mathcal{L}_{stage1}$ and $\mathcal{L}_{stage2}$ over the first 100 epochs and the remaining epochs, respectively. Note that the patch-wise loss is only computed on masked patches. Besides, the hyper-parameters $\lambda_{rec}$ and $\lambda_{edge}$ are set to 5 and 1, respectively.

During the fine-tuning, the original learning rate is set to 0.0003 for the first 100 epochs and decreases to 0.00003 over the remaining 100 epochs. We adopt the partial fine-tuning strategy inspired by~\cite{he2022masked}, \ie, fine-tune the last six layers of the pre-trained encoder while freezing the others. We randomly adjust the contrast, intensity, and sharpness of the input image as data augmentation. The weight of the feature consistency module is fixed during the fine-tuning. Besides, the Adam solver is applied to minimize the objectives during the pre-training and fine-tuning stages.

The implementations of the comparison methods are all publicly available, and we directly run PTNet~\cite{zhang2021ptnet} and ResViT~\cite{dalmaz2021resvit} using their official PyTorch implementations. A coefficient $\lambda$ is required to balance the adversarial loss and the conventional per-pixel loss for both Pix2Pix~\cite{isola2017image} and TransUNet~\cite{chen2021transunet}. We set $\lambda$ to 100 for Pix2Pix and 50 for TransUNet, respectively. Both Pix2Pix and TransUNet are trained with a batch size of 30 for 200 epochs. The learning rate is set to 0.0002 in the first 100 epochs and decreases to 0.00002 in the remaining 100 epochs.

\subsection{Synthesis Results}

\begin{figure*}[t!]
	\centering
    \footnotesize
	\begin{overpic}[width=0.9\textwidth]{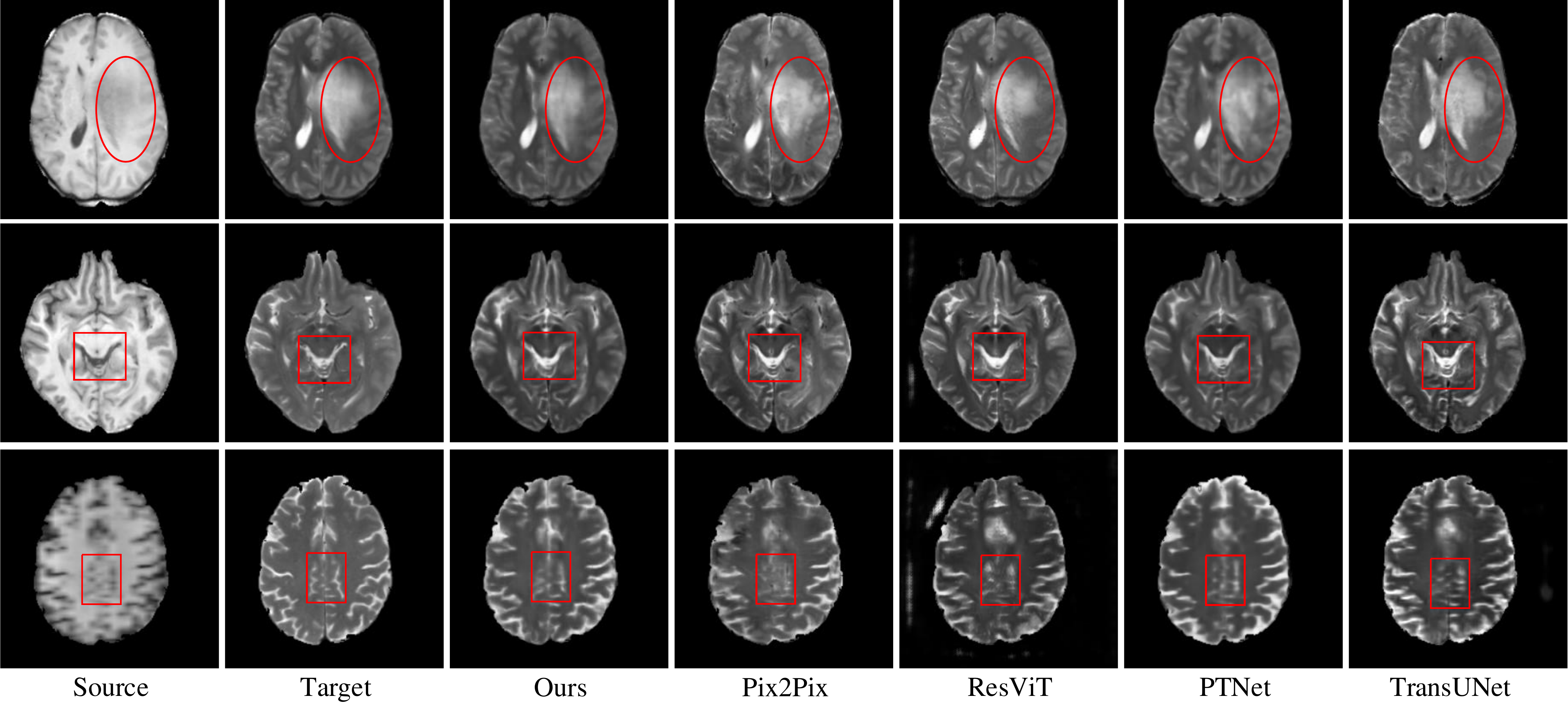}
    \end{overpic}\vspace{-0.25cm}
	\caption{\footnotesize Qualitative comparison between the proposed MT-Net and other state-of-the-art methods (T1 to T2 on the BraTS2020 dataset). MT-Net is fine-tuned using $70\%$ of the training data.}
    \label{fig:t1t2}
\end{figure*}

\begin{figure*}[t!]
	\centering
    \footnotesize
	\begin{overpic}[width=0.9\textwidth]{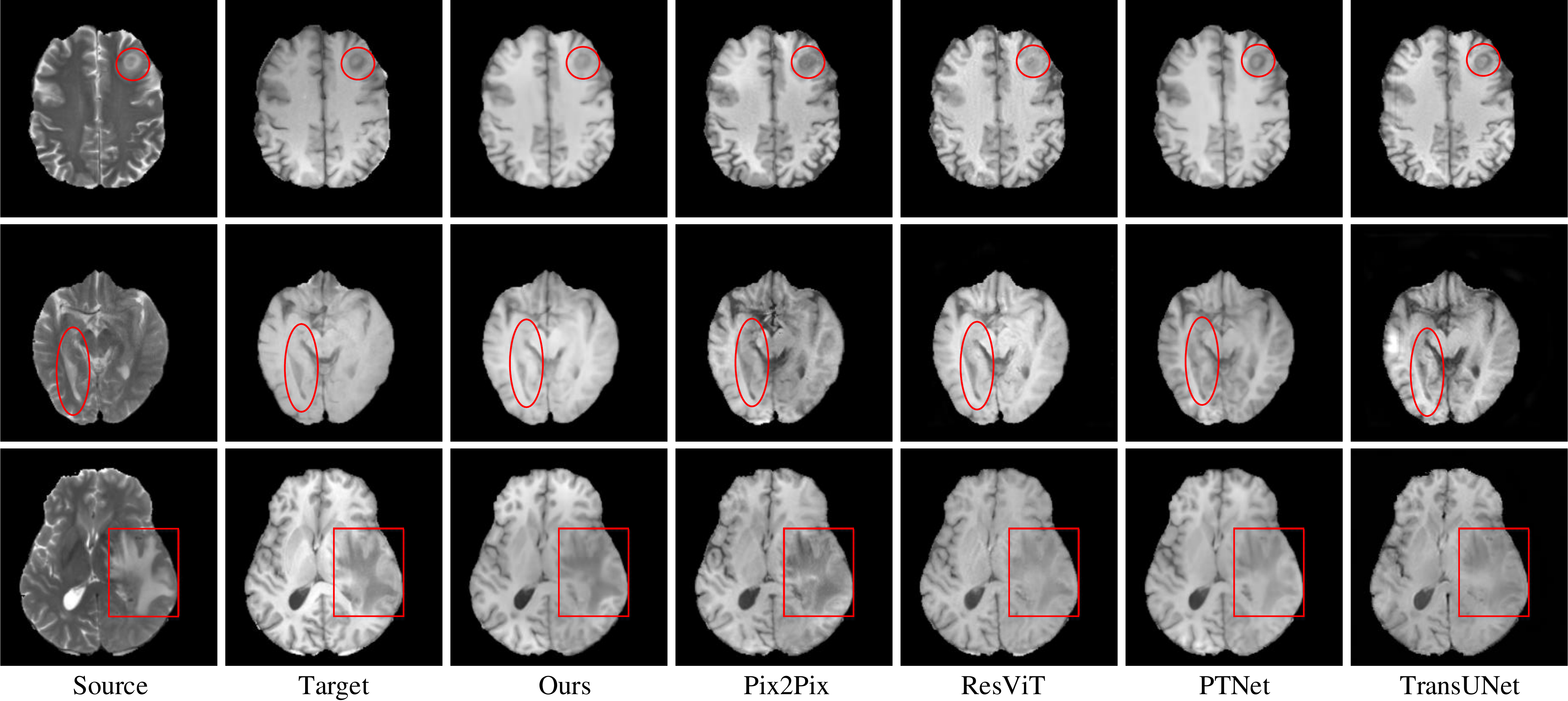}
    \end{overpic}\vspace{-0.25cm}
	\caption{\footnotesize Qualitative comparison between the proposed MT-Net and other state-of-the-art methods (T2 to T1 on the BraTS2020 dataset). Our MT-Net is fine-tuned using $70\%$ of the training data.}
    \label{fig:t2t1}
\end{figure*}

\subsubsection{Results on BraTS2020} We conduct three cross-modality synthesis tasks on the BraTS2020 dataset, \ie, synthesizing T2 from T1 (T1$\rightarrow$T2), T1 from T2 (T2$\rightarrow$T1), and T1c from T1 (T1$\rightarrow$T1c). All comparison models are trained on the entire training set (\ie, $295$ subjects), while our MT-Net is fine-tuned using only $70\%$ of the paired data, \ie, the first $206$ subjects. %
As measured by PSNR and NMSE, our MT-Net synthesizes T2 modality images from T1 modality images with higher quality than other models, which is shown in Table~\ref{tab:t1t2}. %
Furthermore, MT-Net maintains the characteristics of tumor regions, as shown in the $1^{st}$ row of Fig.~\ref{fig:t1t2}. Our MT-Net synthesizes tumor regions effectively and handles blurry input images well, demonstrating its strong generative ability without adversarial training. %
The performances of different methods are also evaluated on the T1 synthesis task, where our MT-Net obtains the best performance in terms of PSNR, as presented in Table~\ref{tab:t2t1}. Some representative synthesized images are shown in Fig.~\ref{fig:t2t1}, which demonstrates that our methods can obtain similar visual results with less number of paired data. %
This is also reflected by its superior performance on PSNR and NMSE compared to other models, as shown in Table~\ref{tab:t1t1c}. Moreover, the combination of cGANs and transformer-based models may exacerbate the problem of unstable training. The GAN-based TransUNet, for example, may experience mode collapse due to its sensitivity to hyper-parameters. By contrast, the training of our framework is much more stable, as our MT-Net does not rely on adversarial training.

It should be noted that the above implementations of the comparison methods do not require pre-training. To further demonstrate the superiority of our framework, we pre-train all comparison methods with an image reconstruction task. Specifically, the entire training set, which is composed of four different modalities (T1, T2, T1c, and FLAIR), is utilized to pre-train each comparison method as an autoencoder that reconstructs the input modality. Subsequently, the comparison methods are fine-tuned to synthesize T2 modality from T1 modality using $70\%$ of paired data from the training set. Our framework still exhibits the best performance in all metrics, as illustrated in Table~\ref{tab:t1t2_pretrain}. Compared to the results in Table~\ref{tab:t1t2}, the performance of the comparison methods degrades as less number of paired data are utilized for fine-tuning.

\begin{table}[!t]
	\centering
	\footnotesize
	\renewcommand{\arraystretch}{1.0}
	\setlength\tabcolsep{7pt}
	\caption{quantitative evaluation results of the synthesized t2 images using images on the brats2020 dataset.}
 \label{tab:t1t2}
	\begin{tabular}{l|ccc}
		\hline
		\rowcolor{mygray}
		Model (Paired data)&PSNR~$\uparrow$&NMSE~$\downarrow$&SSIM~$\uparrow$\\\hline
		Pix2Pix (100\%)&22.394±2.777&0.116±0.091&0.888±0.038\\
		PTNet (100\%)&22.794±3.061&0.114±0.129&\bf{0.904±0.041}\\
		ResViT (100\%)&22.214±3.150&0.120±0.125&0.878±0.042\\
		TransUNet (100\%)&22.251±3.149&0.119±0.097& 0.897±0.031\\
		\textbf{Ours (70\%)}&\bf{23.028±3.183}&\bf{0.103±0.070}& 0.903±0.039\\
		\hline
	\end{tabular}
\end{table}

\begin{table}[!t]
 	\centering
	\footnotesize
	\renewcommand{\arraystretch}{1.0}
	\setlength\tabcolsep{7pt}
	\caption{quantitative evaluation results of the synthesized t1 images using t2 images on the brats2020 dataset. 
	}\label{tab:t2t1}
	\begin{tabular}{l|ccc}
		\hline
		\rowcolor{mygray}
		  Model (Paired data)   &   PSNR~$\uparrow$   &   NMSE~$\downarrow$   &     SSIM~$\uparrow$  \\ \hline
		Pix2Pix (100\%)        &   21.652±2.130        &    0.063±0.079      &    0.898±0.028    \\
		PTNet (100\%) & 22.191±3.098    &  \bf{0.058±0.099}  &   \bf{0.915±0.029}   \\
		ResViT (100\%) &  21.994±2.619    &    0.070±0.174  &  0.888±0.033   \\
		TransUNet (100\%)   & 21.639±2.150     &     0.061±0.069   & 0.898±0.039  \\
		\textbf{Ours (70\%)}   & \bf{22.193±2.291} &	0.059±0.102 & 0.906±0.023\\  
		\hline
	\end{tabular}
\end{table}

\begin{table}[!t]
 	\centering
	\footnotesize
	\renewcommand{\arraystretch}{1.0}
	\setlength\tabcolsep{7pt}
	\caption{quantitative evaluation results of the synthesized t1c images using t1 images on the brats2020 dataset.
	}\label{tab:t1t1c}
	\begin{tabular}{l|ccc}
		\hline
		\rowcolor{mygray}
		  Model (Paired data)   &   PSNR~$\uparrow$   &   NMSE~$\downarrow$   &     SSIM~$\uparrow$  \\ \hline
		Pix2Pix (100\%)        &   21.874±2.343      &  0.115±0.062   &    0.872±0.034   \\
		PTNet (100\%) & 23.354±2.623   &  0.100±0.076  &   \bf{0.914±0.038}  \\
		ResViT (100\%) &  23.155±2.139  &  0.103±0.078 &  0.887±0.042   \\
		TransUNet (100\%)   & 22.247±2.530     &     0.131±0.058   & 0.882±0.048  \\
		\textbf{Ours (70\%)}   & \bf{23.415±2.258} &	\bf{0.096±0.081} & 0.913±0.034\\ 
		\hline
	\end{tabular}
\end{table}

\begin{table}[!t]
	\centering
	\footnotesize
	\renewcommand{\arraystretch}{1.0}
	\setlength\tabcolsep{7pt}
	\caption{quantitative evaluation results of the synthesized T2 images using t1 images on the brats2020 dataset. All methods are pre-trained using the same training set.}
 \label{tab:t1t2_pretrain}
	\begin{tabular}{l|ccc}
		\hline
		\rowcolor{mygray}
		Model (Paired data)&PSNR~$\uparrow$&NMSE~$\downarrow$&SSIM~$\uparrow$\\\hline
		Pix2Pix (70\%)&22.074±2.585&0.120±0.078&0.881±0.037\\
		PTNet (70\%)&22.242±3.114&0.119±0.066&0.903±0.041\\
		ResViT (70\%)&21.224±2.356&0.124±0.124&0.871±0.029\\
		TransUNet (70\%)&21.643±2.448&0.126±0.153& 0.883±0.048\\
		\textbf{Ours (70\%)}&\bf{23.028±3.183}&\bf{0.103±0.070}&\bf{0.903±0.039}\\
		\hline
	\end{tabular}
\end{table}

\begin{table}[!t]
 	\centering
	\footnotesize
	\renewcommand{\arraystretch}{1.0}
	\setlength\tabcolsep{7pt}
	\caption{quantitative evaluation results of the synthesized flair images using t1 images on the isles2015 dataset. 
	}\label{tab:t1f}
	\begin{tabular}{l|ccc}
		\hline
		\rowcolor{mygray}
		  Model (Paired data)   &   PSNR~$\uparrow$   &   NMSE~$\downarrow$   &     SSIM~$\uparrow$  \\ \hline
		Pix2Pix (100\%)      & 21.571±2.234     &  0.152±0.017  & 0.865±0.035   \\
		PTNet (100\%) & 23.233±3.899 & 0.137±0.023 &  0.889±0.027\\
		ResViT (100\%) & 22.579±3.342  & 0.144±0.024 & 0.871±0.034  \\
		TransUNet (100\%)   & 23.523±2.123    &   0.138±0.016  & \bf{0.893±0.043} \\
		\textbf{Ours (70\%)}   & \bf{23.890±2.348} &	\bf{0.133±0.018} & 0.891±0.032\\
		\hline
	\end{tabular}
\end{table}

\begin{table}[!t]
 	\centering
	\footnotesize
	\renewcommand{\arraystretch}{1.0}
	\setlength\tabcolsep{7pt}
	\caption{quantitative evaluation results of the synthesized t1 images using flair images on the isles2015 dataset. 
	}\label{tab:ft1}
	\begin{tabular}{l|ccc}
		\hline
		\rowcolor{mygray}
		  Model (Paired data)   &   PSNR~$\uparrow$   &   NMSE~$\downarrow$   &     SSIM~$\uparrow$  \\ \hline
		Pix2Pix (100\%)        &   23.572±1.719      &    0.159±0.021     &  0.879±0.038   \\
		PTNet (100\%) & 23.667±1.534   &  0.149±0.023 &   0.874±0.045 \\
		ResViT (100\%) & 24.392±1.286 &  0.133±0.022 &  0.892±0.025 \\
		TransUNet (100\%)   & 21.232±1.973    &   0.162±0.017  & 0.851±0.032 \\
		\textbf{Ours (70\%)}   & \bf{25.536±1.354} &	\bf{0.129±0.012} & \bf{0.898±0.043}\\
		\hline
	\end{tabular}
\end{table}

\subsubsection{Results on ISLES2015} In addition, we also evaluate the performance for synthesizing FLAIR modality from T1 images and synthesizing T1 modality using FLAIR images on the ISLES2015 dataset. We pre-train the Edge-MAE using the entire training set ($28$ subjects), while fine-tuning our MT-Net with $70\%$ of paired training data (\ie, the first $19$ subjects). Table~\ref{tab:t1f} and Table~\ref{tab:ft1} show the quantitative evaluation results. Our MT-Net surpasses other methods in terms of PSNR and NMSE in Table~\ref{tab:t1f} and all three metrics in Table~\ref{tab:ft1}. The results further validate the effectiveness of our model.

\subsection{Model Study}

\subsubsection{Performance on Pretext Tasks} 
\begin{figure}[!t!]
	\centering
	\footnotesize
	\begin{overpic}[width=\columnwidth]{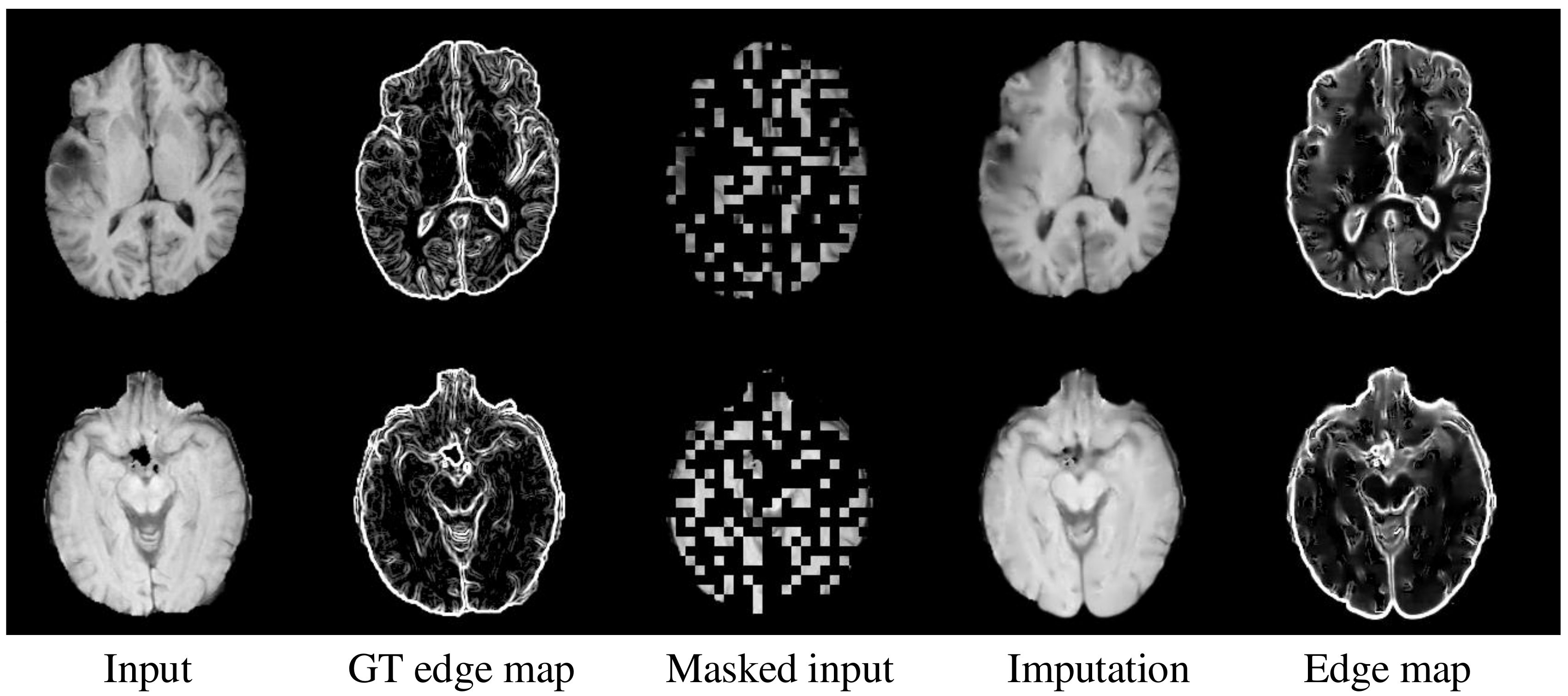}
	\end{overpic}\vspace{-0.15cm}
	\caption{\footnotesize Image imputation and edge map estimation results on the BraTS2020 testing dataset.
 }
	\label{fig:mae}
\end{figure}

A quantitative analysis of pretext tasks, \ie, image imputation and whole edge map estimation, is first performed after edge-aware pre-training. Our Edge-MAE effectively preserves edge and contextual information, as illustrated in Fig.~\ref{fig:mae}. Despite a large proportion of input image patches being masked (\ie, $70\%$), the imputed images and edge maps closely resemble the ground truth. Besides, the proposed patch-wise loss facilitates the imputation of hard patches from heavily masked areas.

\subsubsection{Results on Edge Maps} 
\begin{figure}[!t]
	\centering
	\footnotesize
	\begin{overpic}[width=1\columnwidth]{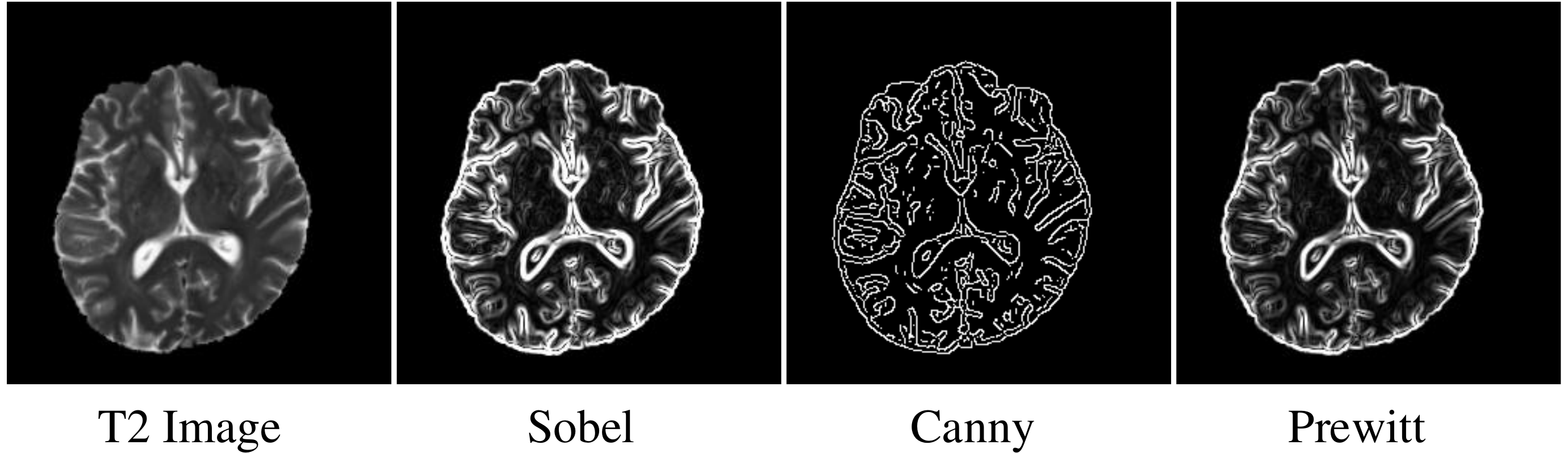}
	\end{overpic}\vspace{-0.15cm}
	\caption{\footnotesize A T2 image and the corresponding edge maps extracted by different edge detection methods}.\vspace{-0.35cm}
	\label{fig:edge}
\end{figure}

To evaluate the edge-preserving performance of our framework and other comparison methods, we extract three types of edge maps, \ie, Sobel, Prewitt, and Canny~\cite{canny1986computational}, from both the missing-modality images and the synthesized images. The edge maps from the synthesized images and the ground truths are compared with NMSE in Table~\ref{tab:brats_edge}. The results indicate that the extracted Sobel edge maps can more accurately resemble the ground truth, while the NMSE of Canny edge maps is much lower. Fig.~\ref{fig:edge} shows that the Canny edge detector yields a binary edge map with significantly thinner edges compared to other detection methods, due to its multi-stage design. In this way, it is more difficult to synthesize images from the Canny edge map.

\begin{table}[!t]
 	\centering
	\footnotesize
	\renewcommand{\arraystretch}{1.0}
	\setlength\tabcolsep{7pt}
	\caption{quantitative evaluation results of the synthesized t2 edge maps on the brats2020 dataset}.\vspace{-0.10cm}
	\label{tab:brats_edge}
	\begin{tabular}{c|ccc}
		\hline
		\rowcolor{mygray}
		  Task    &   Sobel   &   Prewitt   &     Canny  \\ \hline
		T1 $\rightarrow$ T2  & 22.312±1.948 &	20.090±1.798 & 11.107±0.998 \\ 
        T2 $\rightarrow$ T1  & 22.012±1.356 &	20.234±1.649 & 11.023±0.569 \\ 
        T1 $\rightarrow$ T1c  & 23.345±1.456 &	20.554±1.761 & 11.830±0.988 \\ 
		\hline
	\end{tabular}
\end{table}

\subsubsection{Results on Different Ratios of Training Data} 
\begin{figure}[!t]
	\centering
	\footnotesize
	\begin{overpic}[width=1\columnwidth]{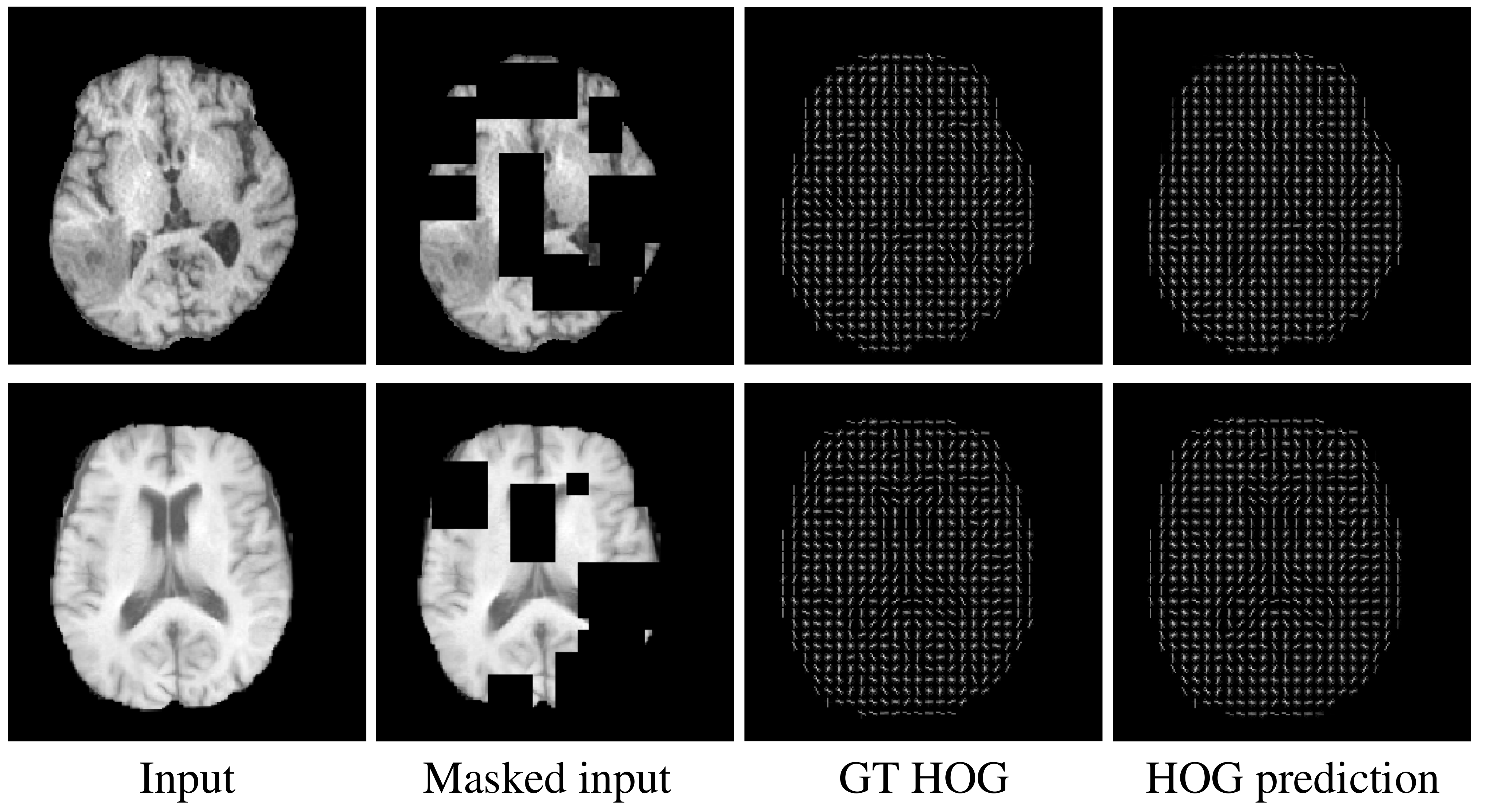}
	\end{overpic}\vspace{-0.25cm}
	\caption{\footnotesize Feature prediction results of MaskFeat~\cite{Wei2021MaskedFP} on the BraTS2020 dataset. MaskFeat adopts a block-wise masking strategy, and utilizes HOG features as the prediction target (by setting the mask ratio to $40\%$)}.\vspace{-0.15cm}
	\label{fig:hog}
\end{figure}

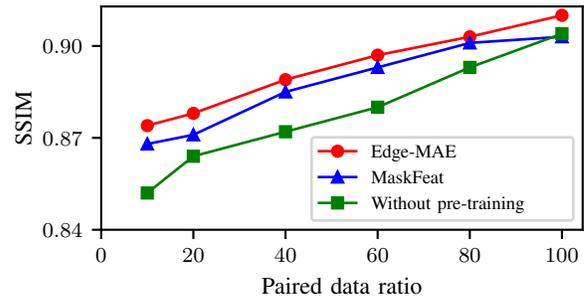
\begin{figure}[!t]
	\centering
	\footnotesize
	\begin{center}
        \input{./img/paired.pgf}
    \end{center}\vspace{-0.25cm}
	\caption{\footnotesize Quantitative comparison of our framework using different training strategies by varying amount of paired data (using T1 to synthesize T2 on the BraTS2020 dataset).} 
	\label{fig:paired}\vspace{-0.15cm}
\end{figure}

\begin{figure}[thp!]
	\centering
	\footnotesize
	\begin{overpic}[width=1\columnwidth]{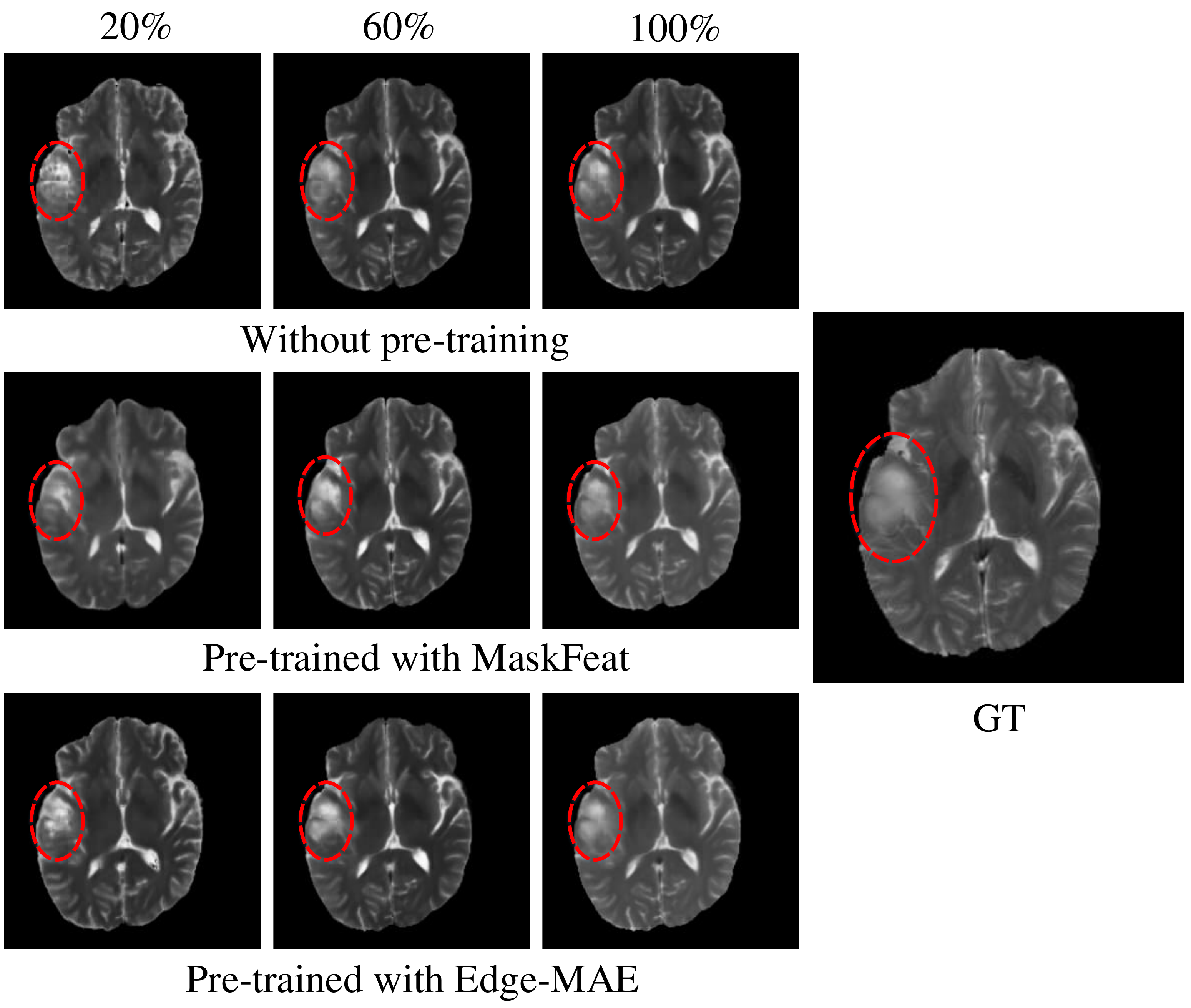}
	\end{overpic}\vspace{-0.15cm}
	\caption{\footnotesize Comparison of synthesized images obtained by our model and other frameworks under different ratios of paired data (using T1 images to synthesize T2 images on the BraTS2020 dataset).
}\vspace{-0.15cm}
	\label{fig:ratio_visualize}
\end{figure}

We further investigate the performance of our framework with varying the number of training samples. Specifically, our framework with the pre-trained Edge-MAE encoder is fine-tuned using paired training data from the first $10\%$, $20\%$, $40\%$, $60\%$, $80\%$, and $100\%$ subjects, denoted as ``Edge-MAE". Furthermore, to evaluate the impact of pre-training on synthesis performance, we apply a different self-supervised learning scheme, \ie, Masked Feature Prediction (MaskFeat)~\cite{Wei2021MaskedFP}, to pre-train the ViT encoder of our framework. MeatFeat first performs random block-wise masking~\cite{bao2021beit} to the input image, and predicts the Histogram-of-Oriented-Gradients (HOG) features that correspond to the masked region. As depicted in Fig.~\ref{fig:hog}, MaskFeat is capable of predicting the normalized HOG targets of the masked regions. Similarly, the framework is then fine-tuned using varying amounts of paired training data, denoted as ``MaskFeat". Additionally, we randomly initialize our framework during the fine-tuning to examine the effect of pre-training, denoted as ``without pre-training". Fig.~\ref{fig:paired} demonstrates that pre-training significantly enhances the performance of our framework, particularly when paired data is scarce, $\eg$, 10\%. In addition, our pre-trained framework with Edge-MAE outperforms the pre-trained version of MaskFeat, providing evidence for the superiority of the proposed Edge-MAE. Besides, Fig.~\ref{fig:ratio_visualize} displays the synthesized T2 images obtained by our framework trained using different strategies. Several approaches have comparable synthesis performance with a large number of paired data. However, the framework without pre-training cannot preserve the tumor structure with only $20\%$ of paired data. Our Edge-MAE pre-trained framework, on the other hand, can successfully reproduce the tumor’s characteristics.

\subsubsection{Segmentation Evaluation on BraTS2020}
\begin{figure}[thp!]
	\centering
	\footnotesize
	\begin{overpic}[width=1\columnwidth]{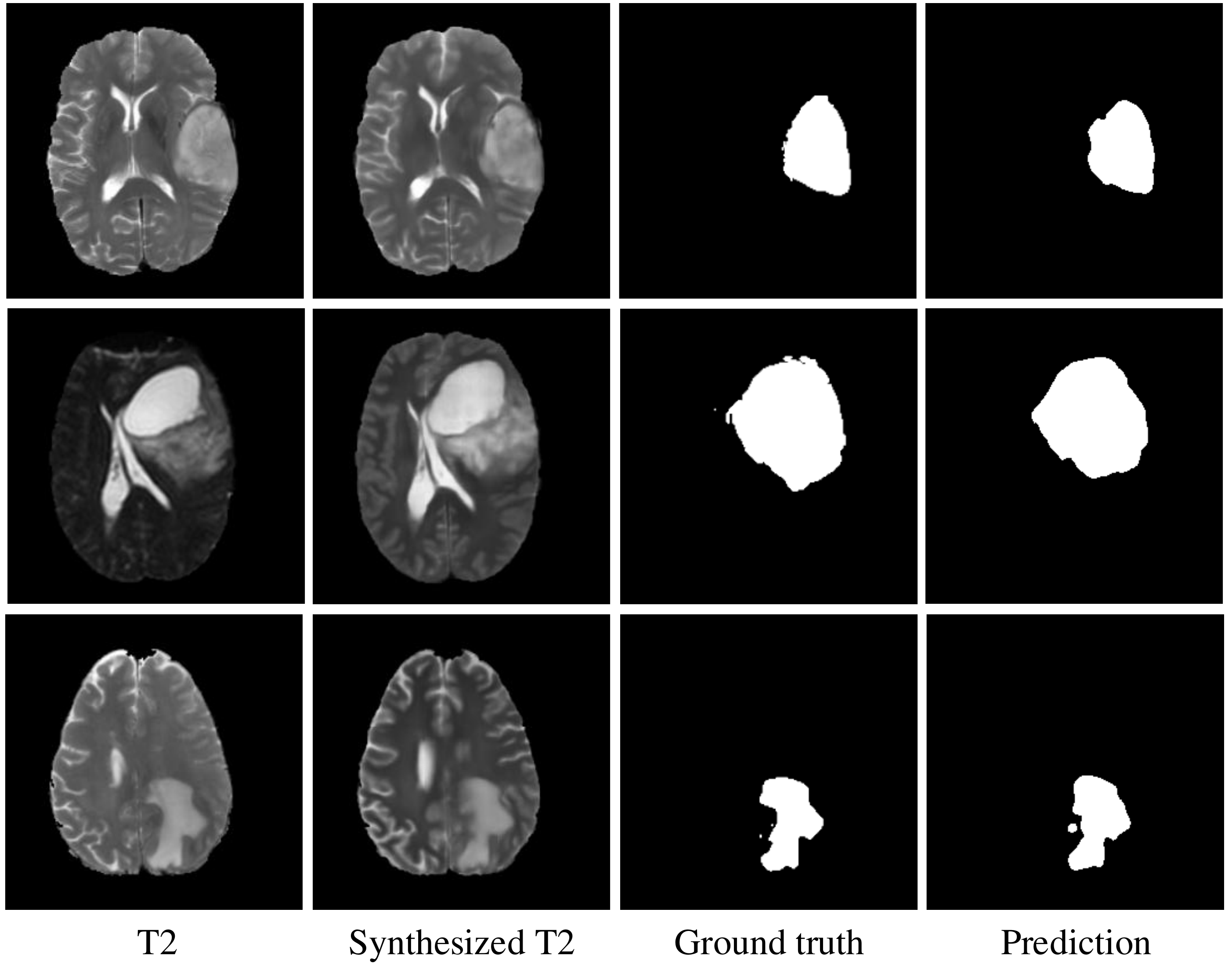}
	\end{overpic}\vspace{-0.15cm}
	\caption{\footnotesize Results of brain tumor segmentation on synthesized T2 images by the proposed framework on the BraTS2020 dataset}.\vspace{-0.25cm}
	\label{fig:seg}
\end{figure}

Segmenting brain tumors using MR images is an essential step in the diagnosis and treatment planning of patients, which can also be used to evaluate the quality of synthesized tumor regions. Following previous works~\cite{luo2021edge,Yu2020SampleAdaptiveGL}, a U-Net~\cite{ronneberger2015u} is trained and evaluated under two schemes on the BraTS2020 dataset. The first scheme is denoted as ``Syn\&Syn". In particular, for each comparison method, we synthesize T2 images with T1 images from the training and the testing sets. After that, the U-Net is trained with the synthesized training set and evaluated with the synthesized testing set. The second scheme is denoted as ``Real\&Syn", in which we train the segmentation model using real T2 images from the training set, and evaluate the model with the synthesized T2 images from the test set. The Dice scores of tumor segmentation in Table~\ref{tab:seg} show that our proposed framework yields superior segmentation performance compared to other methods. In addition, some examples of segmentation results on the synthesized T2 images are shown in Fig.~\ref{fig:seg}, which further demonstrate that the proposed framework can preserve the pathological information in the synthesized tumor regions. As a result, the tumor regions can be accurately segmented by the adopted U-Net model.

\begin{table}[!t]
 	\centering
	\footnotesize
	\renewcommand{\arraystretch}{1.0}
	\setlength\tabcolsep{9pt}
	\caption{Brain tumor segmentation results on the brats2020 dataset}.\label{tab:seg}\vspace{-0.15cm}
	\begin{tabular}{l|cc}
		\hline
		\rowcolor{mygray}
		  Model  &   Syn\&Syn   &   Real\&Syn  \\ \hline
		Pix2Pix   &   0.591±0.184    & 0.606±0.196  \\
		PTNet  & 0.590±0.192   &  0.635±0.180  \\
		ResViT  &  0.624±0.197 & 0.639±0.195 \\
		Ours    & \textbf{0.627±0.187} & \textbf{0.641±0.199}  \\ 
		\hline
	\end{tabular}
\end{table}

\subsection{Ablation Study}

We conduct ablation studies to validate the effectiveness of the primary components and training methodologies.

\begin{table*}[!t]
    \centering
    \renewcommand{\arraystretch}{1.2}
    \setlength\tabcolsep{6pt}
    \caption{quantitative evaluation results by comparing our full model with its ablated versions on the brats2020 dataset.}\label{tab:ablation}
\begin{tabular}{l|ccc|ccc}
\hline
\multicolumn{1}{c|}{}                                & \multicolumn{3}{c|}{T1~$\rightarrow$~T2}               & \multicolumn{3}{c}{T2~$\rightarrow$~T1}                 \\ \cline{2-7} 
Models (Ratio of paired data)     & PSNR~$\uparrow$         & NMSE~$\downarrow$        & SSIM~$\uparrow$        & PSNR~$\uparrow$         & NMSE~$\downarrow$         & SSIM~$\uparrow$        \\ \hline
w/o pre-training (100\%)                             & 23.231±3.108 & 0.102±0.077 & 0.904±0.045 & 22.216±2.779 & 0.060±0.098 & 0.907±0.027 \\ 
w/o pre-training (70\%)                              & 21.923±2.963 & 0.120±0.060 & 0.884±0.038 & 21.534±2.356 & 0.069±0.084  & 0.889±0.026 \\ \hline
With data augmentation (70\%)                        & 23.025±3.136 & 0.103±0.078 & 0.903±0.034 & 22.203±2.329 & 0.059±0.105  & 0.903±0.032 \\
\multicolumn{1}{c|}{With Canny edge detector (70\%)} & 22.032±3.152 & 0.108±0.064 & 0.894±0.031 & 21.890±3.073 & 0.062±0.100  & 0.902±0.026 \\
w/o patch-wise loss (70\%)                           & 22.398±2.811 & 0.115±0.079 & 0.896±0.034 & 21.623±2.232 & 0.065±0.112  & 0.903±0.023 \\
w/o edge map estimation                           & 22.965±2.937 & 0.104±0.067 & 0.903±0.041 & 22.182±2.590 & 0.061±0.095  & 0.906±0.021 \\ \hline
Single-scale (70\%)                                  & 21.790±2.846 & 0.110±0.073 & 0.882±0.036 & 21.541±2.432 & 0.065±0.104  & 0.895±0.024 \\
Downsampled-dual-scale (70\%)                                    & 22.782±2.939 & 0.105±0.075 & 0.900±0.041 & 21.985±2.250 & 0.062±0.112  & 0.899±0.028 \\
Triple-scale (70\%)                                  & 23.123±3.347 & 0.103±0.062 & 0.902±0.037 & 22.194±2.385 & 0.059±0.120  & 0.906±0.022 \\ \hline
Reversed DSF (70\%)                                  & 22.913±3.435 & 0.108±0.072 & 0.902±0.040 & 22.073±2.686 & 0.061±0.110  & 0.905±0.025 \\
Concat-based fusion (70\%)                           & 22.815±3.576 & 0.111±0.051 & 0.902±0.042 & 22.108±2.897 & 0.062±0.093  & 0.904±0.026 \\
Addition-based fusion (70\%)                         & 22.975±3.570 & 0.106±0.081 & 0.902±0.041 & 22.018±2.933 & 0.061±0.100  & 0.902±0.026 \\ \hline
w/o feature consistency (70\%)                       & 22.095±3.019 & 0.105±0.067 & 0.894±0.043 & 22.182±2.745 & 0.060±0.099  & 0.903±0.027 \\
With GAN (70\%)                                      & 20.800±2.885 & 0.132±0.060 & 0.860±0.035 & 20.355±2.339 & 0.078±0.121  & 0.881±0.032 \\ \hline
\textbf{Ours} (70\%)                                          & 23.028±3.183 & 0.103±0.070 & 0.903±0.039 & 22.193±2.291 & 0.059±0.102  & 0.906±0.023 \\ \hline
\end{tabular}
\end{table*}

\textbf{Effects of Masking Ratios}. The Edge-MAE is pre-trained using a masking ratio ranging from $40\%$ to $90\%$. Fig.~\ref{fig:ratio} indicates that the highest SSIM value is achieved using a masking ratio of $70\%$.  Our MT-Net is found to function effectively with masking ratios (\ie, $40\%$ to $80\%$), but a lower masking ratio often leads to longer training time. It should be noted that an excessively high masking ratio of $90\%$ could cause unstable training and thus hinder the convergence of the Edge-MAE.

\begin{figure}[!t]
	\centering
	\footnotesize
	\begin{center}
        \input{./img/ratio_mask.pgf}
    \end{center}\vspace{-0.25cm}
	\caption{\footnotesize Effects of different masking ratios.}
	\label{fig:ratio}\vspace{-0.25cm}
\end{figure}
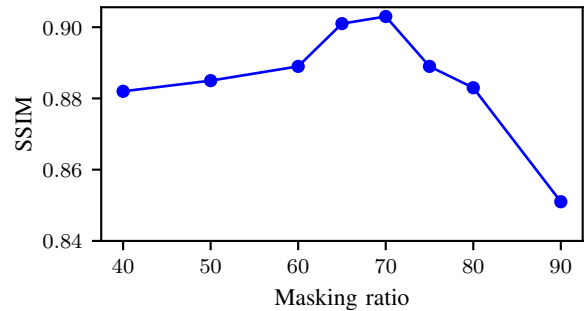

\begin{figure}[!t]
	\centering
	\footnotesize
	\begin{center}
        \input{./img/freeze.pgf}
    \end{center}\vspace{-0.25cm}
	\caption{\footnotesize Effects of partial fine-tuning on synthesizing T2 images from T1 images on the BraTS2020 dataset.  }
	\label{fig:freeze}
 \end{figure}
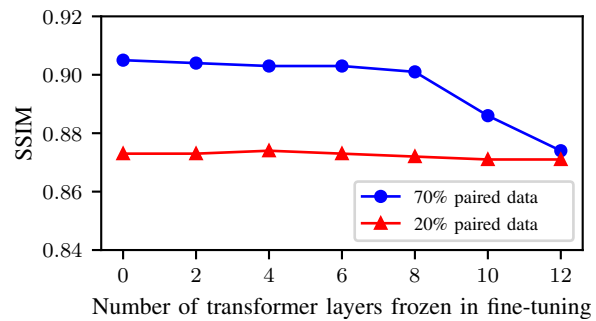

\textbf{Effects of Partial Fine-tuning}. To investigate the significance of partial fine-tuning, we freeze partial transformer layers of the pre-trained encoder during fine-tuning. As depicted in Fig.~\ref{fig:freeze}, the SSIM reaches convergence when 6 out of 12 transformer layers are gradually unfrozen during the fine-tuning. Fine-tuning the entire framework yields minor performance enhancement, but comes with the trade-off of longer training periods. We also observe that partial fine-tuning is beneficial in scenarios with limited paired training data, as it helps prevent overfitting and retains valuable pre-trained knowledge. It is worth noting that the number of frozen layers during fine-tuning is strongly influenced by the size of the dataset as well as the downstream task. For some scenarios with limited paired data, partial fine-tuning is a superior approach, as it accelerates training without compromising performance.

\textbf{Effectiveness of Edge-MAE}. In order to evaluate the significance of different modules in the Edge-MAE, we conduct the following ablative experiments. (1) We exclude the edge-aware pre-training, and utilize randomly-initialized weights for the patch encoder during the fine-tuning, denoted as ``w/o pre-training". (2) The effectiveness of the patch-wise loss is assessed by replacing it with a conventional $\ell_1$ loss, namely ``w/o patch-wise loss". (3) We discard the task-specific decoder that generates the edge map to evaluate the multi-task learning strategy, denoted as ``w/o edge map estimation". (4) To investigate the necessity of data augmentation, geometric transformations (\ie, random rotation and flipping) and color space transformations (\ie, adjustment of brightness and contrast), are applied during pre-training, denoted as ``With data augmentation". (5) The choice of the edge detectors is investigated, with the target edge maps generated by the Canny~\cite{canny1986computational} edge detector, denoted as ``With Canny edge detector". Table~\ref{tab:ablation} shows comparison results for the two synthesis tasks, \ie, using T1 images to synthesize T2 images, and using T2 images to synthesize T1 images. From the results, it can be observed that edge-aware pre-training substantially enhances the synthesis quality. Besides, the results also demonstrate that our framework fine-tuned on $70\%$ of paired data achieves comparable performance to the ablated version trained on $100\%$ paired data. Besides, the results of ``w/o patch-wise loss" and ``w/o edge map estimation" indicate that our patch-wise loss and the multi-task learning strategy further boost the synthesis performance. The results of ``With data augmentation" indicate that Edge-MAE performs effectively without data augmentation, which is consistent with the findings of~\cite{he2022masked}. That is because random masking itself is a kind of strong data augmentation, which greatly increases the diversity of the input. Furthermore, we adopt the Canny edge detector as the prediction target, as depicted in ``With Canny edge detector", demonstrating a performance degradation. Compared to the Sobel operator, the Canny edge detector involves additional steps, \ie, noise reduction, non-maximum suppression, and thresholding. However, these steps prevent the derivative from being obtained through back-propagation, which poses a challenge for our Edge-MAE in generating the corresponding edge maps.

\begin{figure}[!t]
	\centering
	\footnotesize
	\begin{overpic}[width=1\columnwidth]{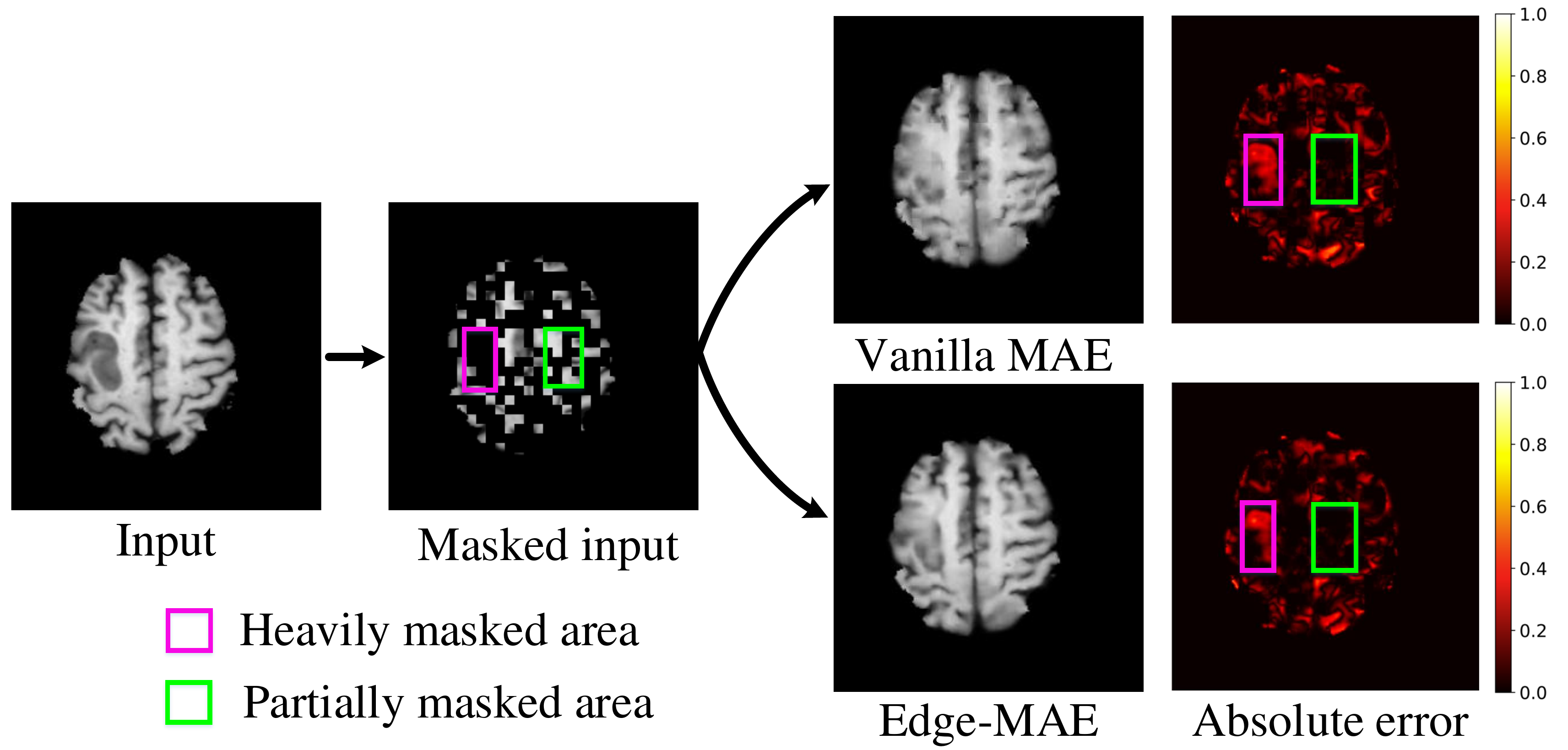}
	\end{overpic}\vspace{-0.15cm}
	\caption{\footnotesize Qualitative comparison between the proposed Edge-MAE and vanilla MAE. }
	\label{fig:reconstruction}\vspace{-0.25cm}
\end{figure}

Moreover, we compare the image imputation performance of our Edge-MAE with vanilla MAE~\cite{he2022masked}. The Edge-MAE and vanilla MAE are trained using an equivalent mask ratio of 70\%, and received the same masked inputs during inference. Fig.~\ref{fig:reconstruction} demonstrates a significant contrast between the imputation errors in the heavily masked areas and in the partially masked areas. This reveals that the imputation difficulty of different patches varies, with those from heavily masked areas (indicated by the purple box) posing a greater challenge. The better imputation performance in Fig.~\ref{fig:reconstruction} further illustrates the effectiveness of the proposed patch-wise loss.

\textbf{Effectiveness of Multi-scale Fine-tuning}. The effectiveness of multi-scale fine-tuning is evaluated by modifying the proposed MT-Net. (1) In place of a dual-encoder architecture, the MT-Net is replaced with a single-scale SwinUNet~\cite{cao2021swin}, which is fed with the features of the pre-trained Edge-MAE encoder, denoted as ``Single-scale". (2) The MT-Net takes in the output features of the pre-trained encoder with its downsampled version, denoted as ``Downsampled-dual-scale". (3) A triple-scale feature pyramid is constructed, which includes the final output features with its downsampled and upsampled versions. Accordingly, we modify the MT-Net into a triple-encoder architecture, denoted as ``Triple-scale". The results in Table~\ref{tab:ablation} demonstrate that multi-scale fine-tuning benefits the synthesis performance. It is critical to include an upsampled version of the feature map to preserve fine-grained details. In addition, compared with ``triple-scale", it can be observed that our dual-scale framework achieves comparable results with a simpler architecture.

\begin{figure}[thp!]
	\centering
	\footnotesize
	\begin{overpic}[width=.85\columnwidth]{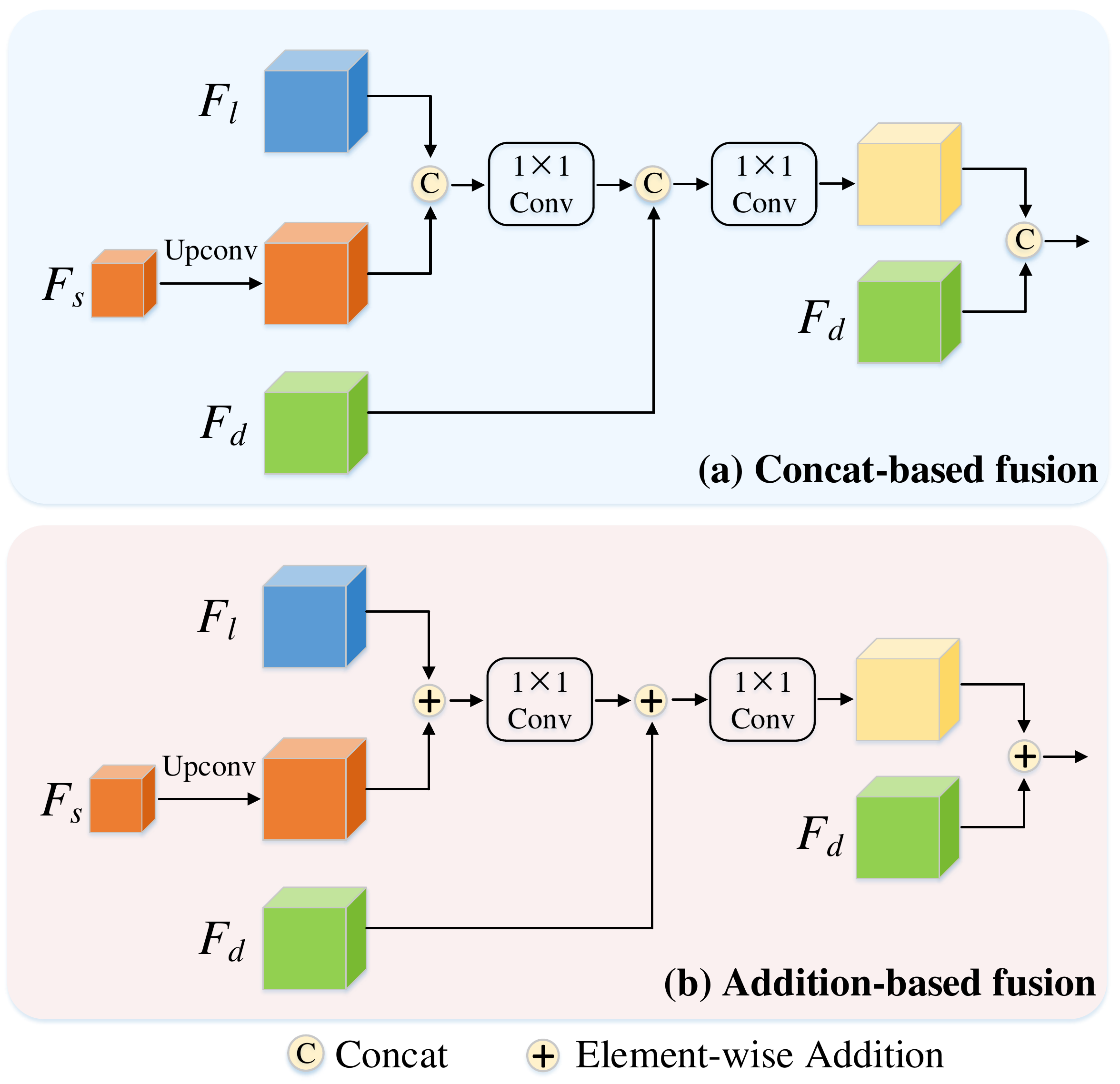}
	\end{overpic}\vspace{-0.05cm}
	\caption{\footnotesize Illustration of two fusion strategies: (a) concat-based fusion, and (b) addition-based fusion}.
	\label{fig:fusion}
\end{figure}

\textbf{Effectiveness of DSF Module}. The contribution of the proposed DSF module is explored through the derivation of three baselines. (1) The order of spatial-wise and channel-wise selection is reversed, denoted as ``Reversed DSF". (2) Inspired by~\cite{luo2021edge}, we designed two different feature fusion strategies as illustrated in Fig.~\ref{fig:fusion}, denoted as ``Concat-based fusion" and ``Addition-based fusion". It can be observed from Table~\ref{tab:ablation} that our DSF module outperforms the ``Reversed DSF", ``Concat-based fusion", and ``Addition-based fusion" methods in terms of synthesis performance. Our DSF module's superior performance stems from the selective fusion strategy.

\textbf{Importance of Feature Consistency Module}. We remove the feature consistency module, denoted as ``w/o feature consistency". 
Moreover, we investigate whether the feature consistency module is a better option for our MT-Net than adversarial training. Specifically, we substitute the feature consistency module with the PatchGAN~\cite{isola2017image} as a discriminator, and fine-tune our framework like a cGAN, denoted as ``With GAN". 
Our observations from Table~\ref{tab:ablation} indicate that incorporating adversarial training into our framework leads to performance degradation. 
The explanation for this is that stable training of GAN requires the comparable ability to represent features in both the discriminator and the generator. Initializing the generator with pre-trained weights from Edge-MAE, however, may cause an imbalance between the discriminator and generator. In this scenario, it would be more appropriate to use a feature consistency module as an alternative.

\section{Conclusion}

Our MT-Net with edge-aware pre-training can effectively utilize both paired and unpaired data to yield comparable performance with state-of-the-art methods, even with less paired training data, and no adversarial training. To our knowledge, this is the first work to improve cross-modality MR synthesis performance with self-supervised pre-training. It should also be noted that our MT-Net is a general framework that can be easily applied to other medical image analysis tasks, including semantic segmentation and classification, with limited labeled data. On the other hand, it is worth noting that the limitation of our MT-Net is that it is designed to learn a mapping from a single source-modality to multiple missing-modality images, resulting in the failure to merge complementary information from multiple source-modality images of a given patient. In future work, we will enable our framework to accept a varying number of source-modality images for image synthesis.

{
\bibliographystyle{IEEEtran}
\bibliography{tmi}
}
 
\end{document}

%% file: img/paired.pgf
\begingroup%
\makeatletter%
\begin{pgfpicture}%
\pgfpathrectangle{\pgfpointorigin}{\pgfqpoint{3.500000in}{1.720000in}}%
\pgfusepath{use as bounding box, clip}%
\begin{pgfscope}%
\pgfsetbuttcap%
\pgfsetmiterjoin%
\definecolor{currentfill}{rgb}{1.000000,1.000000,1.000000}%
\pgfsetfillcolor{currentfill}%
\pgfsetlinewidth{0.000000pt}%
\definecolor{currentstroke}{rgb}{1.000000,1.000000,1.000000}%
\pgfsetstrokecolor{currentstroke}%
\pgfsetdash{}{0pt}%
\pgfpathmoveto{\pgfqpoint{0.000000in}{0.000000in}}%
\pgfpathlineto{\pgfqpoint{3.500000in}{0.000000in}}%
\pgfpathlineto{\pgfqpoint{3.500000in}{1.720000in}}%
\pgfpathlineto{\pgfqpoint{0.000000in}{1.720000in}}%
\pgfpathlineto{\pgfqpoint{0.000000in}{0.000000in}}%
\pgfpathclose%
\pgfusepath{fill}%
\end{pgfscope}%
\begin{pgfscope}%
\pgfsetbuttcap%
\pgfsetmiterjoin%
\definecolor{currentfill}{rgb}{1.000000,1.000000,1.000000}%
\pgfsetfillcolor{currentfill}%
\pgfsetlinewidth{0.000000pt}%
\definecolor{currentstroke}{rgb}{0.000000,0.000000,0.000000}%
\pgfsetstrokecolor{currentstroke}%
\pgfsetstrokeopacity{0.000000}%
\pgfsetdash{}{0pt}%
\pgfpathmoveto{\pgfqpoint{0.630000in}{0.344000in}}%
\pgfpathlineto{\pgfqpoint{3.150000in}{0.344000in}}%
\pgfpathlineto{\pgfqpoint{3.150000in}{1.513600in}}%
\pgfpathlineto{\pgfqpoint{0.630000in}{1.513600in}}%
\pgfpathlineto{\pgfqpoint{0.630000in}{0.344000in}}%
\pgfpathclose%
\pgfusepath{fill}%
\end{pgfscope}%
\begin{pgfscope}%
\pgfsetbuttcap%
\pgfsetroundjoin%
\definecolor{currentfill}{rgb}{0.000000,0.000000,0.000000}%
\pgfsetfillcolor{currentfill}%
\pgfsetlinewidth{0.803000pt}%
\definecolor{currentstroke}{rgb}{0.000000,0.000000,0.000000}%
\pgfsetstrokecolor{currentstroke}%
\pgfsetdash{}{0pt}%
\pgfsys@defobject{currentmarker}{\pgfqpoint{0.000000in}{-0.048611in}}{\pgfqpoint{0.000000in}{0.000000in}}{%
\pgfpathmoveto{\pgfqpoint{0.000000in}{0.000000in}}%
\pgfpathlineto{\pgfqpoint{0.000000in}{-0.048611in}}%
\pgfusepath{stroke,fill}%
}%
\begin{pgfscope}%
\pgfsys@transformshift{0.630000in}{0.344000in}%
\pgfsys@useobject{currentmarker}{}%
\end{pgfscope}%
\end{pgfscope}%
\begin{pgfscope}%
\definecolor{textcolor}{rgb}{0.000000,0.000000,0.000000}%
\pgfsetstrokecolor{textcolor}%
\pgfsetfillcolor{textcolor}%
\pgftext[x=0.630000in,y=0.246778in,,top]{\color{textcolor}\fontsize{8.000000}{9.600000}\selectfont \(\displaystyle {0}\)}%
\end{pgfscope}%
\begin{pgfscope}%
\pgfsetbuttcap%
\pgfsetroundjoin%
\definecolor{currentfill}{rgb}{0.000000,0.000000,0.000000}%
\pgfsetfillcolor{currentfill}%
\pgfsetlinewidth{0.803000pt}%
\definecolor{currentstroke}{rgb}{0.000000,0.000000,0.000000}%
\pgfsetstrokecolor{currentstroke}%
\pgfsetdash{}{0pt}%
\pgfsys@defobject{currentmarker}{\pgfqpoint{0.000000in}{-0.048611in}}{\pgfqpoint{0.000000in}{0.000000in}}{%
\pgfpathmoveto{\pgfqpoint{0.000000in}{0.000000in}}%
\pgfpathlineto{\pgfqpoint{0.000000in}{-0.048611in}}%
\pgfusepath{stroke,fill}%
}%
\begin{pgfscope}%
\pgfsys@transformshift{1.112297in}{0.344000in}%
\pgfsys@useobject{currentmarker}{}%
\end{pgfscope}%
\end{pgfscope}%
\begin{pgfscope}%
\definecolor{textcolor}{rgb}{0.000000,0.000000,0.000000}%
\pgfsetstrokecolor{textcolor}%
\pgfsetfillcolor{textcolor}%
\pgftext[x=1.112297in,y=0.246778in,,top]{\color{textcolor}\fontsize{8.000000}{9.600000}\selectfont \(\displaystyle {20}\)}%
\end{pgfscope}%
\begin{pgfscope}%
\pgfsetbuttcap%
\pgfsetroundjoin%
\definecolor{currentfill}{rgb}{0.000000,0.000000,0.000000}%
\pgfsetfillcolor{currentfill}%
\pgfsetlinewidth{0.803000pt}%
\definecolor{currentstroke}{rgb}{0.000000,0.000000,0.000000}%
\pgfsetstrokecolor{currentstroke}%
\pgfsetdash{}{0pt}%
\pgfsys@defobject{currentmarker}{\pgfqpoint{0.000000in}{-0.048611in}}{\pgfqpoint{0.000000in}{0.000000in}}{%
\pgfpathmoveto{\pgfqpoint{0.000000in}{0.000000in}}%
\pgfpathlineto{\pgfqpoint{0.000000in}{-0.048611in}}%
\pgfusepath{stroke,fill}%
}%
\begin{pgfscope}%
\pgfsys@transformshift{1.594593in}{0.344000in}%
\pgfsys@useobject{currentmarker}{}%
\end{pgfscope}%
\end{pgfscope}%
\begin{pgfscope}%
\definecolor{textcolor}{rgb}{0.000000,0.000000,0.000000}%
\pgfsetstrokecolor{textcolor}%
\pgfsetfillcolor{textcolor}%
\pgftext[x=1.594593in,y=0.246778in,,top]{\color{textcolor}\fontsize{8.000000}{9.600000}\selectfont \(\displaystyle {40}\)}%
\end{pgfscope}%
\begin{pgfscope}%
\pgfsetbuttcap%
\pgfsetroundjoin%
\definecolor{currentfill}{rgb}{0.000000,0.000000,0.000000}%
\pgfsetfillcolor{currentfill}%
\pgfsetlinewidth{0.803000pt}%
\definecolor{currentstroke}{rgb}{0.000000,0.000000,0.000000}%
\pgfsetstrokecolor{currentstroke}%
\pgfsetdash{}{0pt}%
\pgfsys@defobject{currentmarker}{\pgfqpoint{0.000000in}{-0.048611in}}{\pgfqpoint{0.000000in}{0.000000in}}{%
\pgfpathmoveto{\pgfqpoint{0.000000in}{0.000000in}}%
\pgfpathlineto{\pgfqpoint{0.000000in}{-0.048611in}}%
\pgfusepath{stroke,fill}%
}%
\begin{pgfscope}%
\pgfsys@transformshift{2.076890in}{0.344000in}%
\pgfsys@useobject{currentmarker}{}%
\end{pgfscope}%
\end{pgfscope}%
\begin{pgfscope}%
\definecolor{textcolor}{rgb}{0.000000,0.000000,0.000000}%
\pgfsetstrokecolor{textcolor}%
\pgfsetfillcolor{textcolor}%
\pgftext[x=2.076890in,y=0.246778in,,top]{\color{textcolor}\fontsize{8.000000}{9.600000}\selectfont \(\displaystyle {60}\)}%
\end{pgfscope}%
\begin{pgfscope}%
\pgfsetbuttcap%
\pgfsetroundjoin%
\definecolor{currentfill}{rgb}{0.000000,0.000000,0.000000}%
\pgfsetfillcolor{currentfill}%
\pgfsetlinewidth{0.803000pt}%
\definecolor{currentstroke}{rgb}{0.000000,0.000000,0.000000}%
\pgfsetstrokecolor{currentstroke}%
\pgfsetdash{}{0pt}%
\pgfsys@defobject{currentmarker}{\pgfqpoint{0.000000in}{-0.048611in}}{\pgfqpoint{0.000000in}{0.000000in}}{%
\pgfpathmoveto{\pgfqpoint{0.000000in}{0.000000in}}%
\pgfpathlineto{\pgfqpoint{0.000000in}{-0.048611in}}%
\pgfusepath{stroke,fill}%
}%
\begin{pgfscope}%
\pgfsys@transformshift{2.559187in}{0.344000in}%
\pgfsys@useobject{currentmarker}{}%
\end{pgfscope}%
\end{pgfscope}%
\begin{pgfscope}%
\definecolor{textcolor}{rgb}{0.000000,0.000000,0.000000}%
\pgfsetstrokecolor{textcolor}%
\pgfsetfillcolor{textcolor}%
\pgftext[x=2.559187in,y=0.246778in,,top]{\color{textcolor}\fontsize{8.000000}{9.600000}\selectfont \(\displaystyle {80}\)}%
\end{pgfscope}%
\begin{pgfscope}%
\pgfsetbuttcap%
\pgfsetroundjoin%
\definecolor{currentfill}{rgb}{0.000000,0.000000,0.000000}%
\pgfsetfillcolor{currentfill}%
\pgfsetlinewidth{0.803000pt}%
\definecolor{currentstroke}{rgb}{0.000000,0.000000,0.000000}%
\pgfsetstrokecolor{currentstroke}%
\pgfsetdash{}{0pt}%
\pgfsys@defobject{currentmarker}{\pgfqpoint{0.000000in}{-0.048611in}}{\pgfqpoint{0.000000in}{0.000000in}}{%
\pgfpathmoveto{\pgfqpoint{0.000000in}{0.000000in}}%
\pgfpathlineto{\pgfqpoint{0.000000in}{-0.048611in}}%
\pgfusepath{stroke,fill}%
}%
\begin{pgfscope}%
\pgfsys@transformshift{3.041483in}{0.344000in}%
\pgfsys@useobject{currentmarker}{}%
\end{pgfscope}%
\end{pgfscope}%
\begin{pgfscope}%
\definecolor{textcolor}{rgb}{0.000000,0.000000,0.000000}%
\pgfsetstrokecolor{textcolor}%
\pgfsetfillcolor{textcolor}%
\pgftext[x=3.041483in,y=0.246778in,,top]{\color{textcolor}\fontsize{8.000000}{9.600000}\selectfont \(\displaystyle {100}\)}%
\end{pgfscope}%
\begin{pgfscope}%
\definecolor{textcolor}{rgb}{0.000000,0.000000,0.000000}%
\pgfsetstrokecolor{textcolor}%
\pgfsetfillcolor{textcolor}%
\pgftext[x=1.890000in,y=0.092457in,,top]{\color{textcolor}\fontsize{9.000000}{10.800000}\selectfont Paired data ratio}%
\end{pgfscope}%
\begin{pgfscope}%
\pgfsetbuttcap%
\pgfsetroundjoin%
\definecolor{currentfill}{rgb}{0.000000,0.000000,0.000000}%
\pgfsetfillcolor{currentfill}%
\pgfsetlinewidth{0.803000pt}%
\definecolor{currentstroke}{rgb}{0.000000,0.000000,0.000000}%
\pgfsetstrokecolor{currentstroke}%
\pgfsetdash{}{0pt}%
\pgfsys@defobject{currentmarker}{\pgfqpoint{-0.048611in}{0.000000in}}{\pgfqpoint{-0.000000in}{0.000000in}}{%
\pgfpathmoveto{\pgfqpoint{-0.000000in}{0.000000in}}%
\pgfpathlineto{\pgfqpoint{-0.048611in}{0.000000in}}%
\pgfusepath{stroke,fill}%
}%
\begin{pgfscope}%
\pgfsys@transformshift{0.630000in}{0.344000in}%
\pgfsys@useobject{currentmarker}{}%
\end{pgfscope}%
\end{pgfscope}%
\begin{pgfscope}%
\definecolor{textcolor}{rgb}{0.000000,0.000000,0.000000}%
\pgfsetstrokecolor{textcolor}%
\pgfsetfillcolor{textcolor}%
\pgftext[x=0.322898in, y=0.305420in, left, base]{\color{textcolor}\fontsize{8.000000}{9.600000}\selectfont \(\displaystyle {0.84}\)}%
\end{pgfscope}%
\begin{pgfscope}%
\pgfsetbuttcap%
\pgfsetroundjoin%
\definecolor{currentfill}{rgb}{0.000000,0.000000,0.000000}%
\pgfsetfillcolor{currentfill}%
\pgfsetlinewidth{0.803000pt}%
\definecolor{currentstroke}{rgb}{0.000000,0.000000,0.000000}%
\pgfsetstrokecolor{currentstroke}%
\pgfsetdash{}{0pt}%
\pgfsys@defobject{currentmarker}{\pgfqpoint{-0.048611in}{0.000000in}}{\pgfqpoint{-0.000000in}{0.000000in}}{%
\pgfpathmoveto{\pgfqpoint{-0.000000in}{0.000000in}}%
\pgfpathlineto{\pgfqpoint{-0.048611in}{0.000000in}}%
\pgfusepath{stroke,fill}%
}%
\begin{pgfscope}%
\pgfsys@transformshift{0.630000in}{0.825317in}%
\pgfsys@useobject{currentmarker}{}%
\end{pgfscope}%
\end{pgfscope}%
\begin{pgfscope}%
\definecolor{textcolor}{rgb}{0.000000,0.000000,0.000000}%
\pgfsetstrokecolor{textcolor}%
\pgfsetfillcolor{textcolor}%
\pgftext[x=0.322898in, y=0.786737in, left, base]{\color{textcolor}\fontsize{8.000000}{9.600000}\selectfont \(\displaystyle {0.87}\)}%
\end{pgfscope}%
\begin{pgfscope}%
\pgfsetbuttcap%
\pgfsetroundjoin%
\definecolor{currentfill}{rgb}{0.000000,0.000000,0.000000}%
\pgfsetfillcolor{currentfill}%
\pgfsetlinewidth{0.803000pt}%
\definecolor{currentstroke}{rgb}{0.000000,0.000000,0.000000}%
\pgfsetstrokecolor{currentstroke}%
\pgfsetdash{}{0pt}%
\pgfsys@defobject{currentmarker}{\pgfqpoint{-0.048611in}{0.000000in}}{\pgfqpoint{-0.000000in}{0.000000in}}{%
\pgfpathmoveto{\pgfqpoint{-0.000000in}{0.000000in}}%
\pgfpathlineto{\pgfqpoint{-0.048611in}{0.000000in}}%
\pgfusepath{stroke,fill}%
}%
\begin{pgfscope}%
\pgfsys@transformshift{0.630000in}{1.306634in}%
\pgfsys@useobject{currentmarker}{}%
\end{pgfscope}%
\end{pgfscope}%
\begin{pgfscope}%
\definecolor{textcolor}{rgb}{0.000000,0.000000,0.000000}%
\pgfsetstrokecolor{textcolor}%
\pgfsetfillcolor{textcolor}%
\pgftext[x=0.322898in, y=1.268053in, left, base]{\color{textcolor}\fontsize{8.000000}{9.600000}\selectfont \(\displaystyle {0.90}\)}%
\end{pgfscope}%
\begin{pgfscope}%
\definecolor{textcolor}{rgb}{0.000000,0.000000,0.000000}%
\pgfsetstrokecolor{textcolor}%
\pgfsetfillcolor{textcolor}%
\pgftext[x=0.267343in,y=0.928800in,,bottom,rotate=90.000000]{\color{textcolor}\fontsize{9.000000}{10.800000}\selectfont SSIM}%
\end{pgfscope}%
\begin{pgfscope}%
\pgfpathrectangle{\pgfqpoint{0.630000in}{0.344000in}}{\pgfqpoint{2.520000in}{1.169600in}}%
\pgfusepath{clip}%
\pgfsetrectcap%
\pgfsetroundjoin%
\pgfsetlinewidth{1.003750pt}%
\definecolor{currentstroke}{rgb}{1.000000,0.000000,0.000000}%
\pgfsetstrokecolor{currentstroke}%
\pgfsetdash{}{0pt}%
\pgfpathmoveto{\pgfqpoint{0.871148in}{0.889492in}}%
\pgfpathlineto{\pgfqpoint{1.112297in}{0.953668in}}%
\pgfpathlineto{\pgfqpoint{1.594593in}{1.130151in}}%
\pgfpathlineto{\pgfqpoint{2.076890in}{1.258502in}}%
\pgfpathlineto{\pgfqpoint{2.559187in}{1.354765in}}%
\pgfpathlineto{\pgfqpoint{3.041483in}{1.467073in}}%
\pgfusepath{stroke}%
\end{pgfscope}%
\begin{pgfscope}%
\pgfpathrectangle{\pgfqpoint{0.630000in}{0.344000in}}{\pgfqpoint{2.520000in}{1.169600in}}%
\pgfusepath{clip}%
\pgfsetbuttcap%
\pgfsetroundjoin%
\definecolor{currentfill}{rgb}{1.000000,0.000000,0.000000}%
\pgfsetfillcolor{currentfill}%
\pgfsetlinewidth{1.003750pt}%
\definecolor{currentstroke}{rgb}{1.000000,0.000000,0.000000}%
\pgfsetstrokecolor{currentstroke}%
\pgfsetdash{}{0pt}%
\pgfsys@defobject{currentmarker}{\pgfqpoint{-0.027778in}{-0.027778in}}{\pgfqpoint{0.027778in}{0.027778in}}{%
\pgfpathmoveto{\pgfqpoint{0.000000in}{-0.027778in}}%
\pgfpathcurveto{\pgfqpoint{0.007367in}{-0.027778in}}{\pgfqpoint{0.014433in}{-0.024851in}}{\pgfqpoint{0.019642in}{-0.019642in}}%
\pgfpathcurveto{\pgfqpoint{0.024851in}{-0.014433in}}{\pgfqpoint{0.027778in}{-0.007367in}}{\pgfqpoint{0.027778in}{0.000000in}}%
\pgfpathcurveto{\pgfqpoint{0.027778in}{0.007367in}}{\pgfqpoint{0.024851in}{0.014433in}}{\pgfqpoint{0.019642in}{0.019642in}}%
\pgfpathcurveto{\pgfqpoint{0.014433in}{0.024851in}}{\pgfqpoint{0.007367in}{0.027778in}}{\pgfqpoint{0.000000in}{0.027778in}}%
\pgfpathcurveto{\pgfqpoint{-0.007367in}{0.027778in}}{\pgfqpoint{-0.014433in}{0.024851in}}{\pgfqpoint{-0.019642in}{0.019642in}}%
\pgfpathcurveto{\pgfqpoint{-0.024851in}{0.014433in}}{\pgfqpoint{-0.027778in}{0.007367in}}{\pgfqpoint{-0.027778in}{0.000000in}}%
\pgfpathcurveto{\pgfqpoint{-0.027778in}{-0.007367in}}{\pgfqpoint{-0.024851in}{-0.014433in}}{\pgfqpoint{-0.019642in}{-0.019642in}}%
\pgfpathcurveto{\pgfqpoint{-0.014433in}{-0.024851in}}{\pgfqpoint{-0.007367in}{-0.027778in}}{\pgfqpoint{0.000000in}{-0.027778in}}%
\pgfpathlineto{\pgfqpoint{0.000000in}{-0.027778in}}%
\pgfpathclose%
\pgfusepath{stroke,fill}%
}%
\begin{pgfscope}%
\pgfsys@transformshift{0.871148in}{0.889492in}%
\pgfsys@useobject{currentmarker}{}%
\end{pgfscope}%
\begin{pgfscope}%
\pgfsys@transformshift{1.112297in}{0.953668in}%
\pgfsys@useobject{currentmarker}{}%
\end{pgfscope}%
\begin{pgfscope}%
\pgfsys@transformshift{1.594593in}{1.130151in}%
\pgfsys@useobject{currentmarker}{}%
\end{pgfscope}%
\begin{pgfscope}%
\pgfsys@transformshift{2.076890in}{1.258502in}%
\pgfsys@useobject{currentmarker}{}%
\end{pgfscope}%
\begin{pgfscope}%
\pgfsys@transformshift{2.559187in}{1.354765in}%
\pgfsys@useobject{currentmarker}{}%
\end{pgfscope}%
\begin{pgfscope}%
\pgfsys@transformshift{3.041483in}{1.467073in}%
\pgfsys@useobject{currentmarker}{}%
\end{pgfscope}%
\end{pgfscope}%
\begin{pgfscope}%
\pgfpathrectangle{\pgfqpoint{0.630000in}{0.344000in}}{\pgfqpoint{2.520000in}{1.169600in}}%
\pgfusepath{clip}%
\pgfsetrectcap%
\pgfsetroundjoin%
\pgfsetlinewidth{1.003750pt}%
\definecolor{currentstroke}{rgb}{0.000000,0.000000,1.000000}%
\pgfsetstrokecolor{currentstroke}%
\pgfsetdash{}{0pt}%
\pgfpathmoveto{\pgfqpoint{0.871148in}{0.793229in}}%
\pgfpathlineto{\pgfqpoint{1.112297in}{0.841361in}}%
\pgfpathlineto{\pgfqpoint{1.594593in}{1.065975in}}%
\pgfpathlineto{\pgfqpoint{2.076890in}{1.194326in}}%
\pgfpathlineto{\pgfqpoint{2.559187in}{1.322678in}}%
\pgfpathlineto{\pgfqpoint{3.041483in}{1.354765in}}%
\pgfusepath{stroke}%
\end{pgfscope}%
\begin{pgfscope}%
\pgfpathrectangle{\pgfqpoint{0.630000in}{0.344000in}}{\pgfqpoint{2.520000in}{1.169600in}}%
\pgfusepath{clip}%
\pgfsetbuttcap%
\pgfsetmiterjoin%
\definecolor{currentfill}{rgb}{0.000000,0.000000,1.000000}%
\pgfsetfillcolor{currentfill}%
\pgfsetlinewidth{1.003750pt}%
\definecolor{currentstroke}{rgb}{0.000000,0.000000,1.000000}%
\pgfsetstrokecolor{currentstroke}%
\pgfsetdash{}{0pt}%
\pgfsys@defobject{currentmarker}{\pgfqpoint{-0.027778in}{-0.027778in}}{\pgfqpoint{0.027778in}{0.027778in}}{%
\pgfpathmoveto{\pgfqpoint{0.000000in}{0.027778in}}%
\pgfpathlineto{\pgfqpoint{-0.027778in}{-0.027778in}}%
\pgfpathlineto{\pgfqpoint{0.027778in}{-0.027778in}}%
\pgfpathlineto{\pgfqpoint{0.000000in}{0.027778in}}%
\pgfpathclose%
\pgfusepath{stroke,fill}%
}%
\begin{pgfscope}%
\pgfsys@transformshift{0.871148in}{0.793229in}%
\pgfsys@useobject{currentmarker}{}%
\end{pgfscope}%
\begin{pgfscope}%
\pgfsys@transformshift{1.112297in}{0.841361in}%
\pgfsys@useobject{currentmarker}{}%
\end{pgfscope}%
\begin{pgfscope}%
\pgfsys@transformshift{1.594593in}{1.065975in}%
\pgfsys@useobject{currentmarker}{}%
\end{pgfscope}%
\begin{pgfscope}%
\pgfsys@transformshift{2.076890in}{1.194326in}%
\pgfsys@useobject{currentmarker}{}%
\end{pgfscope}%
\begin{pgfscope}%
\pgfsys@transformshift{2.559187in}{1.322678in}%
\pgfsys@useobject{currentmarker}{}%
\end{pgfscope}%
\begin{pgfscope}%
\pgfsys@transformshift{3.041483in}{1.354765in}%
\pgfsys@useobject{currentmarker}{}%
\end{pgfscope}%
\end{pgfscope}%
\begin{pgfscope}%
\pgfpathrectangle{\pgfqpoint{0.630000in}{0.344000in}}{\pgfqpoint{2.520000in}{1.169600in}}%
\pgfusepath{clip}%
\pgfsetrectcap%
\pgfsetroundjoin%
\pgfsetlinewidth{1.003750pt}%
\definecolor{currentstroke}{rgb}{0.000000,0.501961,0.000000}%
\pgfsetstrokecolor{currentstroke}%
\pgfsetdash{}{0pt}%
\pgfpathmoveto{\pgfqpoint{0.871148in}{0.536527in}}%
\pgfpathlineto{\pgfqpoint{1.112297in}{0.729053in}}%
\pgfpathlineto{\pgfqpoint{1.594593in}{0.857405in}}%
\pgfpathlineto{\pgfqpoint{2.076890in}{0.985756in}}%
\pgfpathlineto{\pgfqpoint{2.559187in}{1.194326in}}%
\pgfpathlineto{\pgfqpoint{3.041483in}{1.370809in}}%
\pgfusepath{stroke}%
\end{pgfscope}%
\begin{pgfscope}%
\pgfpathrectangle{\pgfqpoint{0.630000in}{0.344000in}}{\pgfqpoint{2.520000in}{1.169600in}}%
\pgfusepath{clip}%
\pgfsetbuttcap%
\pgfsetmiterjoin%
\definecolor{currentfill}{rgb}{0.000000,0.501961,0.000000}%
\pgfsetfillcolor{currentfill}%
\pgfsetlinewidth{1.003750pt}%
\definecolor{currentstroke}{rgb}{0.000000,0.501961,0.000000}%
\pgfsetstrokecolor{currentstroke}%
\pgfsetdash{}{0pt}%
\pgfsys@defobject{currentmarker}{\pgfqpoint{-0.027778in}{-0.027778in}}{\pgfqpoint{0.027778in}{0.027778in}}{%
\pgfpathmoveto{\pgfqpoint{-0.027778in}{-0.027778in}}%
\pgfpathlineto{\pgfqpoint{0.027778in}{-0.027778in}}%
\pgfpathlineto{\pgfqpoint{0.027778in}{0.027778in}}%
\pgfpathlineto{\pgfqpoint{-0.027778in}{0.027778in}}%
\pgfpathlineto{\pgfqpoint{-0.027778in}{-0.027778in}}%
\pgfpathclose%
\pgfusepath{stroke,fill}%
}%
\begin{pgfscope}%
\pgfsys@transformshift{0.871148in}{0.536527in}%
\pgfsys@useobject{currentmarker}{}%
\end{pgfscope}%
\begin{pgfscope}%
\pgfsys@transformshift{1.112297in}{0.729053in}%
\pgfsys@useobject{currentmarker}{}%
\end{pgfscope}%
\begin{pgfscope}%
\pgfsys@transformshift{1.594593in}{0.857405in}%
\pgfsys@useobject{currentmarker}{}%
\end{pgfscope}%
\begin{pgfscope}%
\pgfsys@transformshift{2.076890in}{0.985756in}%
\pgfsys@useobject{currentmarker}{}%
\end{pgfscope}%
\begin{pgfscope}%
\pgfsys@transformshift{2.559187in}{1.194326in}%
\pgfsys@useobject{currentmarker}{}%
\end{pgfscope}%
\begin{pgfscope}%
\pgfsys@transformshift{3.041483in}{1.370809in}%
\pgfsys@useobject{currentmarker}{}%
\end{pgfscope}%
\end{pgfscope}%
\begin{pgfscope}%
\pgfsetrectcap%
\pgfsetmiterjoin%
\pgfsetlinewidth{0.803000pt}%
\definecolor{currentstroke}{rgb}{0.000000,0.000000,0.000000}%
\pgfsetstrokecolor{currentstroke}%
\pgfsetdash{}{0pt}%
\pgfpathmoveto{\pgfqpoint{0.630000in}{0.344000in}}%
\pgfpathlineto{\pgfqpoint{0.630000in}{1.513600in}}%
\pgfusepath{stroke}%
\end{pgfscope}%
\begin{pgfscope}%
\pgfsetrectcap%
\pgfsetmiterjoin%
\pgfsetlinewidth{0.803000pt}%
\definecolor{currentstroke}{rgb}{0.000000,0.000000,0.000000}%
\pgfsetstrokecolor{currentstroke}%
\pgfsetdash{}{0pt}%
\pgfpathmoveto{\pgfqpoint{3.150000in}{0.344000in}}%
\pgfpathlineto{\pgfqpoint{3.150000in}{1.513600in}}%
\pgfusepath{stroke}%
\end{pgfscope}%
\begin{pgfscope}%
\pgfsetrectcap%
\pgfsetmiterjoin%
\pgfsetlinewidth{0.803000pt}%
\definecolor{currentstroke}{rgb}{0.000000,0.000000,0.000000}%
\pgfsetstrokecolor{currentstroke}%
\pgfsetdash{}{0pt}%
\pgfpathmoveto{\pgfqpoint{0.630000in}{0.344000in}}%
\pgfpathlineto{\pgfqpoint{3.150000in}{0.344000in}}%
\pgfusepath{stroke}%
\end{pgfscope}%
\begin{pgfscope}%
\pgfsetrectcap%
\pgfsetmiterjoin%
\pgfsetlinewidth{0.803000pt}%
\definecolor{currentstroke}{rgb}{0.000000,0.000000,0.000000}%
\pgfsetstrokecolor{currentstroke}%
\pgfsetdash{}{0pt}%
\pgfpathmoveto{\pgfqpoint{0.630000in}{1.513600in}}%
\pgfpathlineto{\pgfqpoint{3.150000in}{1.513600in}}%
\pgfusepath{stroke}%
\end{pgfscope}%
\begin{pgfscope}%
\pgfsetbuttcap%
\pgfsetmiterjoin%
\definecolor{currentfill}{rgb}{1.000000,1.000000,1.000000}%
\pgfsetfillcolor{currentfill}%
\pgfsetfillopacity{0.800000}%
\pgfsetlinewidth{1.003750pt}%
\definecolor{currentstroke}{rgb}{0.800000,0.800000,0.800000}%
\pgfsetstrokecolor{currentstroke}%
\pgfsetstrokeopacity{0.800000}%
\pgfsetdash{}{0pt}%
\pgfpathmoveto{\pgfqpoint{1.750574in}{0.392611in}}%
\pgfpathlineto{\pgfqpoint{3.081944in}{0.392611in}}%
\pgfpathquadraticcurveto{\pgfqpoint{3.101389in}{0.392611in}}{\pgfqpoint{3.101389in}{0.412056in}}%
\pgfpathlineto{\pgfqpoint{3.101389in}{0.809046in}}%
\pgfpathquadraticcurveto{\pgfqpoint{3.101389in}{0.828490in}}{\pgfqpoint{3.081944in}{0.828490in}}%
\pgfpathlineto{\pgfqpoint{1.750574in}{0.828490in}}%
\pgfpathquadraticcurveto{\pgfqpoint{1.731129in}{0.828490in}}{\pgfqpoint{1.731129in}{0.809046in}}%
\pgfpathlineto{\pgfqpoint{1.731129in}{0.412056in}}%
\pgfpathquadraticcurveto{\pgfqpoint{1.731129in}{0.392611in}}{\pgfqpoint{1.750574in}{0.392611in}}%
\pgfpathlineto{\pgfqpoint{1.750574in}{0.392611in}}%
\pgfpathclose%
\pgfusepath{stroke,fill}%
\end{pgfscope}%
\begin{pgfscope}%
\pgfsetrectcap%
\pgfsetroundjoin%
\pgfsetlinewidth{1.003750pt}%
\definecolor{currentstroke}{rgb}{1.000000,0.000000,0.000000}%
\pgfsetstrokecolor{currentstroke}%
\pgfsetdash{}{0pt}%
\pgfpathmoveto{\pgfqpoint{1.770018in}{0.755574in}}%
\pgfpathlineto{\pgfqpoint{1.867240in}{0.755574in}}%
\pgfpathlineto{\pgfqpoint{1.964462in}{0.755574in}}%
\pgfusepath{stroke}%
\end{pgfscope}%
\begin{pgfscope}%
\pgfsetbuttcap%
\pgfsetroundjoin%
\definecolor{currentfill}{rgb}{1.000000,0.000000,0.000000}%
\pgfsetfillcolor{currentfill}%
\pgfsetlinewidth{1.003750pt}%
\definecolor{currentstroke}{rgb}{1.000000,0.000000,0.000000}%
\pgfsetstrokecolor{currentstroke}%
\pgfsetdash{}{0pt}%
\pgfsys@defobject{currentmarker}{\pgfqpoint{-0.027778in}{-0.027778in}}{\pgfqpoint{0.027778in}{0.027778in}}{%
\pgfpathmoveto{\pgfqpoint{0.000000in}{-0.027778in}}%
\pgfpathcurveto{\pgfqpoint{0.007367in}{-0.027778in}}{\pgfqpoint{0.014433in}{-0.024851in}}{\pgfqpoint{0.019642in}{-0.019642in}}%
\pgfpathcurveto{\pgfqpoint{0.024851in}{-0.014433in}}{\pgfqpoint{0.027778in}{-0.007367in}}{\pgfqpoint{0.027778in}{0.000000in}}%
\pgfpathcurveto{\pgfqpoint{0.027778in}{0.007367in}}{\pgfqpoint{0.024851in}{0.014433in}}{\pgfqpoint{0.019642in}{0.019642in}}%
\pgfpathcurveto{\pgfqpoint{0.014433in}{0.024851in}}{\pgfqpoint{0.007367in}{0.027778in}}{\pgfqpoint{0.000000in}{0.027778in}}%
\pgfpathcurveto{\pgfqpoint{-0.007367in}{0.027778in}}{\pgfqpoint{-0.014433in}{0.024851in}}{\pgfqpoint{-0.019642in}{0.019642in}}%
\pgfpathcurveto{\pgfqpoint{-0.024851in}{0.014433in}}{\pgfqpoint{-0.027778in}{0.007367in}}{\pgfqpoint{-0.027778in}{0.000000in}}%
\pgfpathcurveto{\pgfqpoint{-0.027778in}{-0.007367in}}{\pgfqpoint{-0.024851in}{-0.014433in}}{\pgfqpoint{-0.019642in}{-0.019642in}}%
\pgfpathcurveto{\pgfqpoint{-0.014433in}{-0.024851in}}{\pgfqpoint{-0.007367in}{-0.027778in}}{\pgfqpoint{0.000000in}{-0.027778in}}%
\pgfpathlineto{\pgfqpoint{0.000000in}{-0.027778in}}%
\pgfpathclose%
\pgfusepath{stroke,fill}%
}%
\begin{pgfscope}%
\pgfsys@transformshift{1.867240in}{0.755574in}%
\pgfsys@useobject{currentmarker}{}%
\end{pgfscope}%
\end{pgfscope}%
\begin{pgfscope}%
\definecolor{textcolor}{rgb}{0.000000,0.000000,0.000000}%
\pgfsetstrokecolor{textcolor}%
\pgfsetfillcolor{textcolor}%
\pgftext[x=2.042240in,y=0.721546in,left,base]{\color{textcolor}\fontsize{7.000000}{8.400000}\selectfont Edge-MAE}%
\end{pgfscope}%
\begin{pgfscope}%
\pgfsetrectcap%
\pgfsetroundjoin%
\pgfsetlinewidth{1.003750pt}%
\definecolor{currentstroke}{rgb}{0.000000,0.000000,1.000000}%
\pgfsetstrokecolor{currentstroke}%
\pgfsetdash{}{0pt}%
\pgfpathmoveto{\pgfqpoint{1.770018in}{0.620003in}}%
\pgfpathlineto{\pgfqpoint{1.867240in}{0.620003in}}%
\pgfpathlineto{\pgfqpoint{1.964462in}{0.620003in}}%
\pgfusepath{stroke}%
\end{pgfscope}%
\begin{pgfscope}%
\pgfsetbuttcap%
\pgfsetmiterjoin%
\definecolor{currentfill}{rgb}{0.000000,0.000000,1.000000}%
\pgfsetfillcolor{currentfill}%
\pgfsetlinewidth{1.003750pt}%
\definecolor{currentstroke}{rgb}{0.000000,0.000000,1.000000}%
\pgfsetstrokecolor{currentstroke}%
\pgfsetdash{}{0pt}%
\pgfsys@defobject{currentmarker}{\pgfqpoint{-0.027778in}{-0.027778in}}{\pgfqpoint{0.027778in}{0.027778in}}{%
\pgfpathmoveto{\pgfqpoint{0.000000in}{0.027778in}}%
\pgfpathlineto{\pgfqpoint{-0.027778in}{-0.027778in}}%
\pgfpathlineto{\pgfqpoint{0.027778in}{-0.027778in}}%
\pgfpathlineto{\pgfqpoint{0.000000in}{0.027778in}}%
\pgfpathclose%
\pgfusepath{stroke,fill}%
}%
\begin{pgfscope}%
\pgfsys@transformshift{1.867240in}{0.620003in}%
\pgfsys@useobject{currentmarker}{}%
\end{pgfscope}%
\end{pgfscope}%
\begin{pgfscope}%
\definecolor{textcolor}{rgb}{0.000000,0.000000,0.000000}%
\pgfsetstrokecolor{textcolor}%
\pgfsetfillcolor{textcolor}%
\pgftext[x=2.042240in,y=0.585975in,left,base]{\color{textcolor}\fontsize{7.000000}{8.400000}\selectfont MaskFeat}%
\end{pgfscope}%
\begin{pgfscope}%
\pgfsetrectcap%
\pgfsetroundjoin%
\pgfsetlinewidth{1.003750pt}%
\definecolor{currentstroke}{rgb}{0.000000,0.501961,0.000000}%
\pgfsetstrokecolor{currentstroke}%
\pgfsetdash{}{0pt}%
\pgfpathmoveto{\pgfqpoint{1.770018in}{0.484432in}}%
\pgfpathlineto{\pgfqpoint{1.867240in}{0.484432in}}%
\pgfpathlineto{\pgfqpoint{1.964462in}{0.484432in}}%
\pgfusepath{stroke}%
\end{pgfscope}%
\begin{pgfscope}%
\pgfsetbuttcap%
\pgfsetmiterjoin%
\definecolor{currentfill}{rgb}{0.000000,0.501961,0.000000}%
\pgfsetfillcolor{currentfill}%
\pgfsetlinewidth{1.003750pt}%
\definecolor{currentstroke}{rgb}{0.000000,0.501961,0.000000}%
\pgfsetstrokecolor{currentstroke}%
\pgfsetdash{}{0pt}%
\pgfsys@defobject{currentmarker}{\pgfqpoint{-0.027778in}{-0.027778in}}{\pgfqpoint{0.027778in}{0.027778in}}{%
\pgfpathmoveto{\pgfqpoint{-0.027778in}{-0.027778in}}%
\pgfpathlineto{\pgfqpoint{0.027778in}{-0.027778in}}%
\pgfpathlineto{\pgfqpoint{0.027778in}{0.027778in}}%
\pgfpathlineto{\pgfqpoint{-0.027778in}{0.027778in}}%
\pgfpathlineto{\pgfqpoint{-0.027778in}{-0.027778in}}%
\pgfpathclose%
\pgfusepath{stroke,fill}%
}%
\begin{pgfscope}%
\pgfsys@transformshift{1.867240in}{0.484432in}%
\pgfsys@useobject{currentmarker}{}%
\end{pgfscope}%
\end{pgfscope}%
\begin{pgfscope}%
\definecolor{textcolor}{rgb}{0.000000,0.000000,0.000000}%
\pgfsetstrokecolor{textcolor}%
\pgfsetfillcolor{textcolor}%
\pgftext[x=2.042240in,y=0.450404in,left,base]{\color{textcolor}\fontsize{7.000000}{8.400000}\selectfont Without pre-training}%
\end{pgfscope}%
\end{pgfpicture}%
\makeatother%
\endgroup%

%% file: img/ratio_mask.pgf
\begingroup%
\makeatletter%
\begin{pgfpicture}%
\pgfpathrectangle{\pgfpointorigin}{\pgfqpoint{3.500000in}{1.800000in}}%
\pgfusepath{use as bounding box, clip}%
\begin{pgfscope}%
\pgfsetbuttcap%
\pgfsetmiterjoin%
\definecolor{currentfill}{rgb}{1.000000,1.000000,1.000000}%
\pgfsetfillcolor{currentfill}%
\pgfsetlinewidth{0.000000pt}%
\definecolor{currentstroke}{rgb}{1.000000,1.000000,1.000000}%
\pgfsetstrokecolor{currentstroke}%
\pgfsetdash{}{0pt}%
\pgfpathmoveto{\pgfqpoint{0.000000in}{0.000000in}}%
\pgfpathlineto{\pgfqpoint{3.500000in}{0.000000in}}%
\pgfpathlineto{\pgfqpoint{3.500000in}{1.800000in}}%
\pgfpathlineto{\pgfqpoint{0.000000in}{1.800000in}}%
\pgfpathlineto{\pgfqpoint{0.000000in}{0.000000in}}%
\pgfpathclose%
\pgfusepath{fill}%
\end{pgfscope}%
\begin{pgfscope}%
\pgfsetbuttcap%
\pgfsetmiterjoin%
\definecolor{currentfill}{rgb}{1.000000,1.000000,1.000000}%
\pgfsetfillcolor{currentfill}%
\pgfsetlinewidth{0.000000pt}%
\definecolor{currentstroke}{rgb}{0.000000,0.000000,0.000000}%
\pgfsetstrokecolor{currentstroke}%
\pgfsetstrokeopacity{0.000000}%
\pgfsetdash{}{0pt}%
\pgfpathmoveto{\pgfqpoint{0.630000in}{0.360000in}}%
\pgfpathlineto{\pgfqpoint{3.150000in}{0.360000in}}%
\pgfpathlineto{\pgfqpoint{3.150000in}{1.584000in}}%
\pgfpathlineto{\pgfqpoint{0.630000in}{1.584000in}}%
\pgfpathlineto{\pgfqpoint{0.630000in}{0.360000in}}%
\pgfpathclose%
\pgfusepath{fill}%
\end{pgfscope}%
\begin{pgfscope}%
\pgfsetbuttcap%
\pgfsetroundjoin%
\definecolor{currentfill}{rgb}{0.000000,0.000000,0.000000}%
\pgfsetfillcolor{currentfill}%
\pgfsetlinewidth{0.803000pt}%
\definecolor{currentstroke}{rgb}{0.000000,0.000000,0.000000}%
\pgfsetstrokecolor{currentstroke}%
\pgfsetdash{}{0pt}%
\pgfsys@defobject{currentmarker}{\pgfqpoint{0.000000in}{-0.048611in}}{\pgfqpoint{0.000000in}{0.000000in}}{%
\pgfpathmoveto{\pgfqpoint{0.000000in}{0.000000in}}%
\pgfpathlineto{\pgfqpoint{0.000000in}{-0.048611in}}%
\pgfusepath{stroke,fill}%
}%
\begin{pgfscope}%
\pgfsys@transformshift{0.744545in}{0.360000in}%
\pgfsys@useobject{currentmarker}{}%
\end{pgfscope}%
\end{pgfscope}%
\begin{pgfscope}%
\definecolor{textcolor}{rgb}{0.000000,0.000000,0.000000}%
\pgfsetstrokecolor{textcolor}%
\pgfsetfillcolor{textcolor}%
\pgftext[x=0.744545in,y=0.262778in,,top]{\color{textcolor}\rmfamily\fontsize{8.000000}{9.600000}\selectfont \(\displaystyle {40}\)}%
\end{pgfscope}%
\begin{pgfscope}%
\pgfsetbuttcap%
\pgfsetroundjoin%
\definecolor{currentfill}{rgb}{0.000000,0.000000,0.000000}%
\pgfsetfillcolor{currentfill}%
\pgfsetlinewidth{0.803000pt}%
\definecolor{currentstroke}{rgb}{0.000000,0.000000,0.000000}%
\pgfsetstrokecolor{currentstroke}%
\pgfsetdash{}{0pt}%
\pgfsys@defobject{currentmarker}{\pgfqpoint{0.000000in}{-0.048611in}}{\pgfqpoint{0.000000in}{0.000000in}}{%
\pgfpathmoveto{\pgfqpoint{0.000000in}{0.000000in}}%
\pgfpathlineto{\pgfqpoint{0.000000in}{-0.048611in}}%
\pgfusepath{stroke,fill}%
}%
\begin{pgfscope}%
\pgfsys@transformshift{1.202727in}{0.360000in}%
\pgfsys@useobject{currentmarker}{}%
\end{pgfscope}%
\end{pgfscope}%
\begin{pgfscope}%
\definecolor{textcolor}{rgb}{0.000000,0.000000,0.000000}%
\pgfsetstrokecolor{textcolor}%
\pgfsetfillcolor{textcolor}%
\pgftext[x=1.202727in,y=0.262778in,,top]{\color{textcolor}\rmfamily\fontsize{8.000000}{9.600000}\selectfont \(\displaystyle {50}\)}%
\end{pgfscope}%
\begin{pgfscope}%
\pgfsetbuttcap%
\pgfsetroundjoin%
\definecolor{currentfill}{rgb}{0.000000,0.000000,0.000000}%
\pgfsetfillcolor{currentfill}%
\pgfsetlinewidth{0.803000pt}%
\definecolor{currentstroke}{rgb}{0.000000,0.000000,0.000000}%
\pgfsetstrokecolor{currentstroke}%
\pgfsetdash{}{0pt}%
\pgfsys@defobject{currentmarker}{\pgfqpoint{0.000000in}{-0.048611in}}{\pgfqpoint{0.000000in}{0.000000in}}{%
\pgfpathmoveto{\pgfqpoint{0.000000in}{0.000000in}}%
\pgfpathlineto{\pgfqpoint{0.000000in}{-0.048611in}}%
\pgfusepath{stroke,fill}%
}%
\begin{pgfscope}%
\pgfsys@transformshift{1.660909in}{0.360000in}%
\pgfsys@useobject{currentmarker}{}%
\end{pgfscope}%
\end{pgfscope}%
\begin{pgfscope}%
\definecolor{textcolor}{rgb}{0.000000,0.000000,0.000000}%
\pgfsetstrokecolor{textcolor}%
\pgfsetfillcolor{textcolor}%
\pgftext[x=1.660909in,y=0.262778in,,top]{\color{textcolor}\rmfamily\fontsize{8.000000}{9.600000}\selectfont \(\displaystyle {60}\)}%
\end{pgfscope}%
\begin{pgfscope}%
\pgfsetbuttcap%
\pgfsetroundjoin%
\definecolor{currentfill}{rgb}{0.000000,0.000000,0.000000}%
\pgfsetfillcolor{currentfill}%
\pgfsetlinewidth{0.803000pt}%
\definecolor{currentstroke}{rgb}{0.000000,0.000000,0.000000}%
\pgfsetstrokecolor{currentstroke}%
\pgfsetdash{}{0pt}%
\pgfsys@defobject{currentmarker}{\pgfqpoint{0.000000in}{-0.048611in}}{\pgfqpoint{0.000000in}{0.000000in}}{%
\pgfpathmoveto{\pgfqpoint{0.000000in}{0.000000in}}%
\pgfpathlineto{\pgfqpoint{0.000000in}{-0.048611in}}%
\pgfusepath{stroke,fill}%
}%
\begin{pgfscope}%
\pgfsys@transformshift{2.119091in}{0.360000in}%
\pgfsys@useobject{currentmarker}{}%
\end{pgfscope}%
\end{pgfscope}%
\begin{pgfscope}%
\definecolor{textcolor}{rgb}{0.000000,0.000000,0.000000}%
\pgfsetstrokecolor{textcolor}%
\pgfsetfillcolor{textcolor}%
\pgftext[x=2.119091in,y=0.262778in,,top]{\color{textcolor}\rmfamily\fontsize{8.000000}{9.600000}\selectfont \(\displaystyle {70}\)}%
\end{pgfscope}%
\begin{pgfscope}%
\pgfsetbuttcap%
\pgfsetroundjoin%
\definecolor{currentfill}{rgb}{0.000000,0.000000,0.000000}%
\pgfsetfillcolor{currentfill}%
\pgfsetlinewidth{0.803000pt}%
\definecolor{currentstroke}{rgb}{0.000000,0.000000,0.000000}%
\pgfsetstrokecolor{currentstroke}%
\pgfsetdash{}{0pt}%
\pgfsys@defobject{currentmarker}{\pgfqpoint{0.000000in}{-0.048611in}}{\pgfqpoint{0.000000in}{0.000000in}}{%
\pgfpathmoveto{\pgfqpoint{0.000000in}{0.000000in}}%
\pgfpathlineto{\pgfqpoint{0.000000in}{-0.048611in}}%
\pgfusepath{stroke,fill}%
}%
\begin{pgfscope}%
\pgfsys@transformshift{2.577273in}{0.360000in}%
\pgfsys@useobject{currentmarker}{}%
\end{pgfscope}%
\end{pgfscope}%
\begin{pgfscope}%
\definecolor{textcolor}{rgb}{0.000000,0.000000,0.000000}%
\pgfsetstrokecolor{textcolor}%
\pgfsetfillcolor{textcolor}%
\pgftext[x=2.577273in,y=0.262778in,,top]{\color{textcolor}\rmfamily\fontsize{8.000000}{9.600000}\selectfont \(\displaystyle {80}\)}%
\end{pgfscope}%
\begin{pgfscope}%
\pgfsetbuttcap%
\pgfsetroundjoin%
\definecolor{currentfill}{rgb}{0.000000,0.000000,0.000000}%
\pgfsetfillcolor{currentfill}%
\pgfsetlinewidth{0.803000pt}%
\definecolor{currentstroke}{rgb}{0.000000,0.000000,0.000000}%
\pgfsetstrokecolor{currentstroke}%
\pgfsetdash{}{0pt}%
\pgfsys@defobject{currentmarker}{\pgfqpoint{0.000000in}{-0.048611in}}{\pgfqpoint{0.000000in}{0.000000in}}{%
\pgfpathmoveto{\pgfqpoint{0.000000in}{0.000000in}}%
\pgfpathlineto{\pgfqpoint{0.000000in}{-0.048611in}}%
\pgfusepath{stroke,fill}%
}%
\begin{pgfscope}%
\pgfsys@transformshift{3.035455in}{0.360000in}%
\pgfsys@useobject{currentmarker}{}%
\end{pgfscope}%
\end{pgfscope}%
\begin{pgfscope}%
\definecolor{textcolor}{rgb}{0.000000,0.000000,0.000000}%
\pgfsetstrokecolor{textcolor}%
\pgfsetfillcolor{textcolor}%
\pgftext[x=3.035455in,y=0.262778in,,top]{\color{textcolor}\rmfamily\fontsize{8.000000}{9.600000}\selectfont \(\displaystyle {90}\)}%
\end{pgfscope}%
\begin{pgfscope}%
\definecolor{textcolor}{rgb}{0.000000,0.000000,0.000000}%
\pgfsetstrokecolor{textcolor}%
\pgfsetfillcolor{textcolor}%
\pgftext[x=1.890000in,y=0.108457in,,top]{\color{textcolor}\rmfamily\fontsize{9.000000}{10.800000}\selectfont Masking ratio}%
\end{pgfscope}%
\begin{pgfscope}%
\pgfsetbuttcap%
\pgfsetroundjoin%
\definecolor{currentfill}{rgb}{0.000000,0.000000,0.000000}%
\pgfsetfillcolor{currentfill}%
\pgfsetlinewidth{0.803000pt}%
\definecolor{currentstroke}{rgb}{0.000000,0.000000,0.000000}%
\pgfsetstrokecolor{currentstroke}%
\pgfsetdash{}{0pt}%
\pgfsys@defobject{currentmarker}{\pgfqpoint{-0.048611in}{0.000000in}}{\pgfqpoint{-0.000000in}{0.000000in}}{%
\pgfpathmoveto{\pgfqpoint{-0.000000in}{0.000000in}}%
\pgfpathlineto{\pgfqpoint{-0.048611in}{0.000000in}}%
\pgfusepath{stroke,fill}%
}%
\begin{pgfscope}%
\pgfsys@transformshift{0.630000in}{0.360000in}%
\pgfsys@useobject{currentmarker}{}%
\end{pgfscope}%
\end{pgfscope}%
\begin{pgfscope}%
\definecolor{textcolor}{rgb}{0.000000,0.000000,0.000000}%
\pgfsetstrokecolor{textcolor}%
\pgfsetfillcolor{textcolor}%
\pgftext[x=0.322898in, y=0.321420in, left, base]{\color{textcolor}\rmfamily\fontsize{8.000000}{9.600000}\selectfont \(\displaystyle {0.84}\)}%
\end{pgfscope}%
\begin{pgfscope}%
\pgfsetbuttcap%
\pgfsetroundjoin%
\definecolor{currentfill}{rgb}{0.000000,0.000000,0.000000}%
\pgfsetfillcolor{currentfill}%
\pgfsetlinewidth{0.803000pt}%
\definecolor{currentstroke}{rgb}{0.000000,0.000000,0.000000}%
\pgfsetstrokecolor{currentstroke}%
\pgfsetdash{}{0pt}%
\pgfsys@defobject{currentmarker}{\pgfqpoint{-0.048611in}{0.000000in}}{\pgfqpoint{-0.000000in}{0.000000in}}{%
\pgfpathmoveto{\pgfqpoint{-0.000000in}{0.000000in}}%
\pgfpathlineto{\pgfqpoint{-0.048611in}{0.000000in}}%
\pgfusepath{stroke,fill}%
}%
\begin{pgfscope}%
\pgfsys@transformshift{0.630000in}{0.733171in}%
\pgfsys@useobject{currentmarker}{}%
\end{pgfscope}%
\end{pgfscope}%
\begin{pgfscope}%
\definecolor{textcolor}{rgb}{0.000000,0.000000,0.000000}%
\pgfsetstrokecolor{textcolor}%
\pgfsetfillcolor{textcolor}%
\pgftext[x=0.322898in, y=0.694590in, left, base]{\color{textcolor}\rmfamily\fontsize{8.000000}{9.600000}\selectfont \(\displaystyle {0.86}\)}%
\end{pgfscope}%
\begin{pgfscope}%
\pgfsetbuttcap%
\pgfsetroundjoin%
\definecolor{currentfill}{rgb}{0.000000,0.000000,0.000000}%
\pgfsetfillcolor{currentfill}%
\pgfsetlinewidth{0.803000pt}%
\definecolor{currentstroke}{rgb}{0.000000,0.000000,0.000000}%
\pgfsetstrokecolor{currentstroke}%
\pgfsetdash{}{0pt}%
\pgfsys@defobject{currentmarker}{\pgfqpoint{-0.048611in}{0.000000in}}{\pgfqpoint{-0.000000in}{0.000000in}}{%
\pgfpathmoveto{\pgfqpoint{-0.000000in}{0.000000in}}%
\pgfpathlineto{\pgfqpoint{-0.048611in}{0.000000in}}%
\pgfusepath{stroke,fill}%
}%
\begin{pgfscope}%
\pgfsys@transformshift{0.630000in}{1.106341in}%
\pgfsys@useobject{currentmarker}{}%
\end{pgfscope}%
\end{pgfscope}%
\begin{pgfscope}%
\definecolor{textcolor}{rgb}{0.000000,0.000000,0.000000}%
\pgfsetstrokecolor{textcolor}%
\pgfsetfillcolor{textcolor}%
\pgftext[x=0.322898in, y=1.067761in, left, base]{\color{textcolor}\rmfamily\fontsize{8.000000}{9.600000}\selectfont \(\displaystyle {0.88}\)}%
\end{pgfscope}%
\begin{pgfscope}%
\pgfsetbuttcap%
\pgfsetroundjoin%
\definecolor{currentfill}{rgb}{0.000000,0.000000,0.000000}%
\pgfsetfillcolor{currentfill}%
\pgfsetlinewidth{0.803000pt}%
\definecolor{currentstroke}{rgb}{0.000000,0.000000,0.000000}%
\pgfsetstrokecolor{currentstroke}%
\pgfsetdash{}{0pt}%
\pgfsys@defobject{currentmarker}{\pgfqpoint{-0.048611in}{0.000000in}}{\pgfqpoint{-0.000000in}{0.000000in}}{%
\pgfpathmoveto{\pgfqpoint{-0.000000in}{0.000000in}}%
\pgfpathlineto{\pgfqpoint{-0.048611in}{0.000000in}}%
\pgfusepath{stroke,fill}%
}%
\begin{pgfscope}%
\pgfsys@transformshift{0.630000in}{1.479512in}%
\pgfsys@useobject{currentmarker}{}%
\end{pgfscope}%
\end{pgfscope}%
\begin{pgfscope}%
\definecolor{textcolor}{rgb}{0.000000,0.000000,0.000000}%
\pgfsetstrokecolor{textcolor}%
\pgfsetfillcolor{textcolor}%
\pgftext[x=0.322898in, y=1.440932in, left, base]{\color{textcolor}\rmfamily\fontsize{8.000000}{9.600000}\selectfont \(\displaystyle {0.90}\)}%
\end{pgfscope}%
\begin{pgfscope}%
\definecolor{textcolor}{rgb}{0.000000,0.000000,0.000000}%
\pgfsetstrokecolor{textcolor}%
\pgfsetfillcolor{textcolor}%
\pgftext[x=0.267343in,y=0.972000in,,bottom,rotate=90.000000]{\color{textcolor}\rmfamily\fontsize{9.000000}{10.800000}\selectfont SSIM}%
\end{pgfscope}%
\begin{pgfscope}%
\pgfpathrectangle{\pgfqpoint{0.630000in}{0.360000in}}{\pgfqpoint{2.520000in}{1.224000in}}%
\pgfusepath{clip}%
\pgfsetrectcap%
\pgfsetroundjoin%
\pgfsetlinewidth{1.003750pt}%
\definecolor{currentstroke}{rgb}{0.000000,0.000000,1.000000}%
\pgfsetstrokecolor{currentstroke}%
\pgfsetdash{}{0pt}%
\pgfpathmoveto{\pgfqpoint{0.744545in}{1.143659in}}%
\pgfpathlineto{\pgfqpoint{1.202727in}{1.199634in}}%
\pgfpathlineto{\pgfqpoint{1.660909in}{1.274268in}}%
\pgfpathlineto{\pgfqpoint{1.890000in}{1.498171in}}%
\pgfpathlineto{\pgfqpoint{2.119091in}{1.535488in}}%
\pgfpathlineto{\pgfqpoint{2.348182in}{1.274268in}}%
\pgfpathlineto{\pgfqpoint{2.577273in}{1.162317in}}%
\pgfpathlineto{\pgfqpoint{3.035455in}{0.565244in}}%
\pgfusepath{stroke}%
\end{pgfscope}%
\begin{pgfscope}%
\pgfpathrectangle{\pgfqpoint{0.630000in}{0.360000in}}{\pgfqpoint{2.520000in}{1.224000in}}%
\pgfusepath{clip}%
\pgfsetbuttcap%
\pgfsetroundjoin%
\definecolor{currentfill}{rgb}{0.000000,0.000000,1.000000}%
\pgfsetfillcolor{currentfill}%
\pgfsetlinewidth{1.003750pt}%
\definecolor{currentstroke}{rgb}{0.000000,0.000000,1.000000}%
\pgfsetstrokecolor{currentstroke}%
\pgfsetdash{}{0pt}%
\pgfsys@defobject{currentmarker}{\pgfqpoint{-0.027778in}{-0.027778in}}{\pgfqpoint{0.027778in}{0.027778in}}{%
\pgfpathmoveto{\pgfqpoint{0.000000in}{-0.027778in}}%
\pgfpathcurveto{\pgfqpoint{0.007367in}{-0.027778in}}{\pgfqpoint{0.014433in}{-0.024851in}}{\pgfqpoint{0.019642in}{-0.019642in}}%
\pgfpathcurveto{\pgfqpoint{0.024851in}{-0.014433in}}{\pgfqpoint{0.027778in}{-0.007367in}}{\pgfqpoint{0.027778in}{0.000000in}}%
\pgfpathcurveto{\pgfqpoint{0.027778in}{0.007367in}}{\pgfqpoint{0.024851in}{0.014433in}}{\pgfqpoint{0.019642in}{0.019642in}}%
\pgfpathcurveto{\pgfqpoint{0.014433in}{0.024851in}}{\pgfqpoint{0.007367in}{0.027778in}}{\pgfqpoint{0.000000in}{0.027778in}}%
\pgfpathcurveto{\pgfqpoint{-0.007367in}{0.027778in}}{\pgfqpoint{-0.014433in}{0.024851in}}{\pgfqpoint{-0.019642in}{0.019642in}}%
\pgfpathcurveto{\pgfqpoint{-0.024851in}{0.014433in}}{\pgfqpoint{-0.027778in}{0.007367in}}{\pgfqpoint{-0.027778in}{0.000000in}}%
\pgfpathcurveto{\pgfqpoint{-0.027778in}{-0.007367in}}{\pgfqpoint{-0.024851in}{-0.014433in}}{\pgfqpoint{-0.019642in}{-0.019642in}}%
\pgfpathcurveto{\pgfqpoint{-0.014433in}{-0.024851in}}{\pgfqpoint{-0.007367in}{-0.027778in}}{\pgfqpoint{0.000000in}{-0.027778in}}%
\pgfpathlineto{\pgfqpoint{0.000000in}{-0.027778in}}%
\pgfpathclose%
\pgfusepath{stroke,fill}%
}%
\begin{pgfscope}%
\pgfsys@transformshift{0.744545in}{1.143659in}%
\pgfsys@useobject{currentmarker}{}%
\end{pgfscope}%
\begin{pgfscope}%
\pgfsys@transformshift{1.202727in}{1.199634in}%
\pgfsys@useobject{currentmarker}{}%
\end{pgfscope}%
\begin{pgfscope}%
\pgfsys@transformshift{1.660909in}{1.274268in}%
\pgfsys@useobject{currentmarker}{}%
\end{pgfscope}%
\begin{pgfscope}%
\pgfsys@transformshift{1.890000in}{1.498171in}%
\pgfsys@useobject{currentmarker}{}%
\end{pgfscope}%
\begin{pgfscope}%
\pgfsys@transformshift{2.119091in}{1.535488in}%
\pgfsys@useobject{currentmarker}{}%
\end{pgfscope}%
\begin{pgfscope}%
\pgfsys@transformshift{2.348182in}{1.274268in}%
\pgfsys@useobject{currentmarker}{}%
\end{pgfscope}%
\begin{pgfscope}%
\pgfsys@transformshift{2.577273in}{1.162317in}%
\pgfsys@useobject{currentmarker}{}%
\end{pgfscope}%
\begin{pgfscope}%
\pgfsys@transformshift{3.035455in}{0.565244in}%
\pgfsys@useobject{currentmarker}{}%
\end{pgfscope}%
\end{pgfscope}%
\begin{pgfscope}%
\pgfsetrectcap%
\pgfsetmiterjoin%
\pgfsetlinewidth{0.803000pt}%
\definecolor{currentstroke}{rgb}{0.000000,0.000000,0.000000}%
\pgfsetstrokecolor{currentstroke}%
\pgfsetdash{}{0pt}%
\pgfpathmoveto{\pgfqpoint{0.630000in}{0.360000in}}%
\pgfpathlineto{\pgfqpoint{0.630000in}{1.584000in}}%
\pgfusepath{stroke}%
\end{pgfscope}%
\begin{pgfscope}%
\pgfsetrectcap%
\pgfsetmiterjoin%
\pgfsetlinewidth{0.803000pt}%
\definecolor{currentstroke}{rgb}{0.000000,0.000000,0.000000}%
\pgfsetstrokecolor{currentstroke}%
\pgfsetdash{}{0pt}%
\pgfpathmoveto{\pgfqpoint{3.150000in}{0.360000in}}%
\pgfpathlineto{\pgfqpoint{3.150000in}{1.584000in}}%
\pgfusepath{stroke}%
\end{pgfscope}%
\begin{pgfscope}%
\pgfsetrectcap%
\pgfsetmiterjoin%
\pgfsetlinewidth{0.803000pt}%
\definecolor{currentstroke}{rgb}{0.000000,0.000000,0.000000}%
\pgfsetstrokecolor{currentstroke}%
\pgfsetdash{}{0pt}%
\pgfpathmoveto{\pgfqpoint{0.630000in}{0.360000in}}%
\pgfpathlineto{\pgfqpoint{3.150000in}{0.360000in}}%
\pgfusepath{stroke}%
\end{pgfscope}%
\begin{pgfscope}%
\pgfsetrectcap%
\pgfsetmiterjoin%
\pgfsetlinewidth{0.803000pt}%
\definecolor{currentstroke}{rgb}{0.000000,0.000000,0.000000}%
\pgfsetstrokecolor{currentstroke}%
\pgfsetdash{}{0pt}%
\pgfpathmoveto{\pgfqpoint{0.630000in}{1.584000in}}%
\pgfpathlineto{\pgfqpoint{3.150000in}{1.584000in}}%
\pgfusepath{stroke}%
\end{pgfscope}%
\end{pgfpicture}%
\makeatother%
\endgroup%

%% file: img/freeze.pgf
\begingroup%
\makeatletter%
\begin{pgfpicture}%
\pgfpathrectangle{\pgfpointorigin}{\pgfqpoint{3.500000in}{1.800000in}}%
\pgfusepath{use as bounding box, clip}%
\begin{pgfscope}%
\pgfsetbuttcap%
\pgfsetmiterjoin%
\definecolor{currentfill}{rgb}{1.000000,1.000000,1.000000}%
\pgfsetfillcolor{currentfill}%
\pgfsetlinewidth{0.000000pt}%
\definecolor{currentstroke}{rgb}{1.000000,1.000000,1.000000}%
\pgfsetstrokecolor{currentstroke}%
\pgfsetdash{}{0pt}%
\pgfpathmoveto{\pgfqpoint{0.000000in}{0.000000in}}%
\pgfpathlineto{\pgfqpoint{3.500000in}{0.000000in}}%
\pgfpathlineto{\pgfqpoint{3.500000in}{1.800000in}}%
\pgfpathlineto{\pgfqpoint{0.000000in}{1.800000in}}%
\pgfpathlineto{\pgfqpoint{0.000000in}{0.000000in}}%
\pgfpathclose%
\pgfusepath{fill}%
\end{pgfscope}%
\begin{pgfscope}%
\pgfsetbuttcap%
\pgfsetmiterjoin%
\definecolor{currentfill}{rgb}{1.000000,1.000000,1.000000}%
\pgfsetfillcolor{currentfill}%
\pgfsetlinewidth{0.000000pt}%
\definecolor{currentstroke}{rgb}{0.000000,0.000000,0.000000}%
\pgfsetstrokecolor{currentstroke}%
\pgfsetstrokeopacity{0.000000}%
\pgfsetdash{}{0pt}%
\pgfpathmoveto{\pgfqpoint{0.630000in}{0.360000in}}%
\pgfpathlineto{\pgfqpoint{3.150000in}{0.360000in}}%
\pgfpathlineto{\pgfqpoint{3.150000in}{1.584000in}}%
\pgfpathlineto{\pgfqpoint{0.630000in}{1.584000in}}%
\pgfpathlineto{\pgfqpoint{0.630000in}{0.360000in}}%
\pgfpathclose%
\pgfusepath{fill}%
\end{pgfscope}%
\begin{pgfscope}%
\pgfsetbuttcap%
\pgfsetroundjoin%
\definecolor{currentfill}{rgb}{0.000000,0.000000,0.000000}%
\pgfsetfillcolor{currentfill}%
\pgfsetlinewidth{0.803000pt}%
\definecolor{currentstroke}{rgb}{0.000000,0.000000,0.000000}%
\pgfsetstrokecolor{currentstroke}%
\pgfsetdash{}{0pt}%
\pgfsys@defobject{currentmarker}{\pgfqpoint{0.000000in}{-0.048611in}}{\pgfqpoint{0.000000in}{0.000000in}}{%
\pgfpathmoveto{\pgfqpoint{0.000000in}{0.000000in}}%
\pgfpathlineto{\pgfqpoint{0.000000in}{-0.048611in}}%
\pgfusepath{stroke,fill}%
}%
\begin{pgfscope}%
\pgfsys@transformshift{0.744545in}{0.360000in}%
\pgfsys@useobject{currentmarker}{}%
\end{pgfscope}%
\end{pgfscope}%
\begin{pgfscope}%
\definecolor{textcolor}{rgb}{0.000000,0.000000,0.000000}%
\pgfsetstrokecolor{textcolor}%
\pgfsetfillcolor{textcolor}%
\pgftext[x=0.744545in,y=0.262778in,,top]{\color{textcolor}\rmfamily\fontsize{8.000000}{9.600000}\selectfont \(\displaystyle {0}\)}%
\end{pgfscope}%
\begin{pgfscope}%
\pgfsetbuttcap%
\pgfsetroundjoin%
\definecolor{currentfill}{rgb}{0.000000,0.000000,0.000000}%
\pgfsetfillcolor{currentfill}%
\pgfsetlinewidth{0.803000pt}%
\definecolor{currentstroke}{rgb}{0.000000,0.000000,0.000000}%
\pgfsetstrokecolor{currentstroke}%
\pgfsetdash{}{0pt}%
\pgfsys@defobject{currentmarker}{\pgfqpoint{0.000000in}{-0.048611in}}{\pgfqpoint{0.000000in}{0.000000in}}{%
\pgfpathmoveto{\pgfqpoint{0.000000in}{0.000000in}}%
\pgfpathlineto{\pgfqpoint{0.000000in}{-0.048611in}}%
\pgfusepath{stroke,fill}%
}%
\begin{pgfscope}%
\pgfsys@transformshift{1.126364in}{0.360000in}%
\pgfsys@useobject{currentmarker}{}%
\end{pgfscope}%
\end{pgfscope}%
\begin{pgfscope}%
\definecolor{textcolor}{rgb}{0.000000,0.000000,0.000000}%
\pgfsetstrokecolor{textcolor}%
\pgfsetfillcolor{textcolor}%
\pgftext[x=1.126364in,y=0.262778in,,top]{\color{textcolor}\rmfamily\fontsize{8.000000}{9.600000}\selectfont \(\displaystyle {2}\)}%
\end{pgfscope}%
\begin{pgfscope}%
\pgfsetbuttcap%
\pgfsetroundjoin%
\definecolor{currentfill}{rgb}{0.000000,0.000000,0.000000}%
\pgfsetfillcolor{currentfill}%
\pgfsetlinewidth{0.803000pt}%
\definecolor{currentstroke}{rgb}{0.000000,0.000000,0.000000}%
\pgfsetstrokecolor{currentstroke}%
\pgfsetdash{}{0pt}%
\pgfsys@defobject{currentmarker}{\pgfqpoint{0.000000in}{-0.048611in}}{\pgfqpoint{0.000000in}{0.000000in}}{%
\pgfpathmoveto{\pgfqpoint{0.000000in}{0.000000in}}%
\pgfpathlineto{\pgfqpoint{0.000000in}{-0.048611in}}%
\pgfusepath{stroke,fill}%
}%
\begin{pgfscope}%
\pgfsys@transformshift{1.508182in}{0.360000in}%
\pgfsys@useobject{currentmarker}{}%
\end{pgfscope}%
\end{pgfscope}%
\begin{pgfscope}%
\definecolor{textcolor}{rgb}{0.000000,0.000000,0.000000}%
\pgfsetstrokecolor{textcolor}%
\pgfsetfillcolor{textcolor}%
\pgftext[x=1.508182in,y=0.262778in,,top]{\color{textcolor}\rmfamily\fontsize{8.000000}{9.600000}\selectfont \(\displaystyle {4}\)}%
\end{pgfscope}%
\begin{pgfscope}%
\pgfsetbuttcap%
\pgfsetroundjoin%
\definecolor{currentfill}{rgb}{0.000000,0.000000,0.000000}%
\pgfsetfillcolor{currentfill}%
\pgfsetlinewidth{0.803000pt}%
\definecolor{currentstroke}{rgb}{0.000000,0.000000,0.000000}%
\pgfsetstrokecolor{currentstroke}%
\pgfsetdash{}{0pt}%
\pgfsys@defobject{currentmarker}{\pgfqpoint{0.000000in}{-0.048611in}}{\pgfqpoint{0.000000in}{0.000000in}}{%
\pgfpathmoveto{\pgfqpoint{0.000000in}{0.000000in}}%
\pgfpathlineto{\pgfqpoint{0.000000in}{-0.048611in}}%
\pgfusepath{stroke,fill}%
}%
\begin{pgfscope}%
\pgfsys@transformshift{1.890000in}{0.360000in}%
\pgfsys@useobject{currentmarker}{}%
\end{pgfscope}%
\end{pgfscope}%
\begin{pgfscope}%
\definecolor{textcolor}{rgb}{0.000000,0.000000,0.000000}%
\pgfsetstrokecolor{textcolor}%
\pgfsetfillcolor{textcolor}%
\pgftext[x=1.890000in,y=0.262778in,,top]{\color{textcolor}\rmfamily\fontsize{8.000000}{9.600000}\selectfont \(\displaystyle {6}\)}%
\end{pgfscope}%
\begin{pgfscope}%
\pgfsetbuttcap%
\pgfsetroundjoin%
\definecolor{currentfill}{rgb}{0.000000,0.000000,0.000000}%
\pgfsetfillcolor{currentfill}%
\pgfsetlinewidth{0.803000pt}%
\definecolor{currentstroke}{rgb}{0.000000,0.000000,0.000000}%
\pgfsetstrokecolor{currentstroke}%
\pgfsetdash{}{0pt}%
\pgfsys@defobject{currentmarker}{\pgfqpoint{0.000000in}{-0.048611in}}{\pgfqpoint{0.000000in}{0.000000in}}{%
\pgfpathmoveto{\pgfqpoint{0.000000in}{0.000000in}}%
\pgfpathlineto{\pgfqpoint{0.000000in}{-0.048611in}}%
\pgfusepath{stroke,fill}%
}%
\begin{pgfscope}%
\pgfsys@transformshift{2.271818in}{0.360000in}%
\pgfsys@useobject{currentmarker}{}%
\end{pgfscope}%
\end{pgfscope}%
\begin{pgfscope}%
\definecolor{textcolor}{rgb}{0.000000,0.000000,0.000000}%
\pgfsetstrokecolor{textcolor}%
\pgfsetfillcolor{textcolor}%
\pgftext[x=2.271818in,y=0.262778in,,top]{\color{textcolor}\rmfamily\fontsize{8.000000}{9.600000}\selectfont \(\displaystyle {8}\)}%
\end{pgfscope}%
\begin{pgfscope}%
\pgfsetbuttcap%
\pgfsetroundjoin%
\definecolor{currentfill}{rgb}{0.000000,0.000000,0.000000}%
\pgfsetfillcolor{currentfill}%
\pgfsetlinewidth{0.803000pt}%
\definecolor{currentstroke}{rgb}{0.000000,0.000000,0.000000}%
\pgfsetstrokecolor{currentstroke}%
\pgfsetdash{}{0pt}%
\pgfsys@defobject{currentmarker}{\pgfqpoint{0.000000in}{-0.048611in}}{\pgfqpoint{0.000000in}{0.000000in}}{%
\pgfpathmoveto{\pgfqpoint{0.000000in}{0.000000in}}%
\pgfpathlineto{\pgfqpoint{0.000000in}{-0.048611in}}%
\pgfusepath{stroke,fill}%
}%
\begin{pgfscope}%
\pgfsys@transformshift{2.653636in}{0.360000in}%
\pgfsys@useobject{currentmarker}{}%
\end{pgfscope}%
\end{pgfscope}%
\begin{pgfscope}%
\definecolor{textcolor}{rgb}{0.000000,0.000000,0.000000}%
\pgfsetstrokecolor{textcolor}%
\pgfsetfillcolor{textcolor}%
\pgftext[x=2.653636in,y=0.262778in,,top]{\color{textcolor}\rmfamily\fontsize{8.000000}{9.600000}\selectfont \(\displaystyle {10}\)}%
\end{pgfscope}%
\begin{pgfscope}%
\pgfsetbuttcap%
\pgfsetroundjoin%
\definecolor{currentfill}{rgb}{0.000000,0.000000,0.000000}%
\pgfsetfillcolor{currentfill}%
\pgfsetlinewidth{0.803000pt}%
\definecolor{currentstroke}{rgb}{0.000000,0.000000,0.000000}%
\pgfsetstrokecolor{currentstroke}%
\pgfsetdash{}{0pt}%
\pgfsys@defobject{currentmarker}{\pgfqpoint{0.000000in}{-0.048611in}}{\pgfqpoint{0.000000in}{0.000000in}}{%
\pgfpathmoveto{\pgfqpoint{0.000000in}{0.000000in}}%
\pgfpathlineto{\pgfqpoint{0.000000in}{-0.048611in}}%
\pgfusepath{stroke,fill}%
}%
\begin{pgfscope}%
\pgfsys@transformshift{3.035455in}{0.360000in}%
\pgfsys@useobject{currentmarker}{}%
\end{pgfscope}%
\end{pgfscope}%
\begin{pgfscope}%
\definecolor{textcolor}{rgb}{0.000000,0.000000,0.000000}%
\pgfsetstrokecolor{textcolor}%
\pgfsetfillcolor{textcolor}%
\pgftext[x=3.035455in,y=0.262778in,,top]{\color{textcolor}\rmfamily\fontsize{8.000000}{9.600000}\selectfont \(\displaystyle {12}\)}%
\end{pgfscope}%
\begin{pgfscope}%
\definecolor{textcolor}{rgb}{0.000000,0.000000,0.000000}%
\pgfsetstrokecolor{textcolor}%
\pgfsetfillcolor{textcolor}%
\pgftext[x=1.890000in,y=0.108457in,,top]{\color{textcolor}\rmfamily\fontsize{9.000000}{10.800000}\selectfont Number of transformer layers frozen in fine-tuning}%
\end{pgfscope}%
\begin{pgfscope}%
\pgfsetbuttcap%
\pgfsetroundjoin%
\definecolor{currentfill}{rgb}{0.000000,0.000000,0.000000}%
\pgfsetfillcolor{currentfill}%
\pgfsetlinewidth{0.803000pt}%
\definecolor{currentstroke}{rgb}{0.000000,0.000000,0.000000}%
\pgfsetstrokecolor{currentstroke}%
\pgfsetdash{}{0pt}%
\pgfsys@defobject{currentmarker}{\pgfqpoint{-0.048611in}{0.000000in}}{\pgfqpoint{-0.000000in}{0.000000in}}{%
\pgfpathmoveto{\pgfqpoint{-0.000000in}{0.000000in}}%
\pgfpathlineto{\pgfqpoint{-0.048611in}{0.000000in}}%
\pgfusepath{stroke,fill}%
}%
\begin{pgfscope}%
\pgfsys@transformshift{0.630000in}{0.360000in}%
\pgfsys@useobject{currentmarker}{}%
\end{pgfscope}%
\end{pgfscope}%
\begin{pgfscope}%
\definecolor{textcolor}{rgb}{0.000000,0.000000,0.000000}%
\pgfsetstrokecolor{textcolor}%
\pgfsetfillcolor{textcolor}%
\pgftext[x=0.322898in, y=0.321420in, left, base]{\color{textcolor}\rmfamily\fontsize{8.000000}{9.600000}\selectfont \(\displaystyle {0.84}\)}%
\end{pgfscope}%
\begin{pgfscope}%
\pgfsetbuttcap%
\pgfsetroundjoin%
\definecolor{currentfill}{rgb}{0.000000,0.000000,0.000000}%
\pgfsetfillcolor{currentfill}%
\pgfsetlinewidth{0.803000pt}%
\definecolor{currentstroke}{rgb}{0.000000,0.000000,0.000000}%
\pgfsetstrokecolor{currentstroke}%
\pgfsetdash{}{0pt}%
\pgfsys@defobject{currentmarker}{\pgfqpoint{-0.048611in}{0.000000in}}{\pgfqpoint{-0.000000in}{0.000000in}}{%
\pgfpathmoveto{\pgfqpoint{-0.000000in}{0.000000in}}%
\pgfpathlineto{\pgfqpoint{-0.048611in}{0.000000in}}%
\pgfusepath{stroke,fill}%
}%
\begin{pgfscope}%
\pgfsys@transformshift{0.630000in}{0.666000in}%
\pgfsys@useobject{currentmarker}{}%
\end{pgfscope}%
\end{pgfscope}%
\begin{pgfscope}%
\definecolor{textcolor}{rgb}{0.000000,0.000000,0.000000}%
\pgfsetstrokecolor{textcolor}%
\pgfsetfillcolor{textcolor}%
\pgftext[x=0.322898in, y=0.627420in, left, base]{\color{textcolor}\rmfamily\fontsize{8.000000}{9.600000}\selectfont \(\displaystyle {0.86}\)}%
\end{pgfscope}%
\begin{pgfscope}%
\pgfsetbuttcap%
\pgfsetroundjoin%
\definecolor{currentfill}{rgb}{0.000000,0.000000,0.000000}%
\pgfsetfillcolor{currentfill}%
\pgfsetlinewidth{0.803000pt}%
\definecolor{currentstroke}{rgb}{0.000000,0.000000,0.000000}%
\pgfsetstrokecolor{currentstroke}%
\pgfsetdash{}{0pt}%
\pgfsys@defobject{currentmarker}{\pgfqpoint{-0.048611in}{0.000000in}}{\pgfqpoint{-0.000000in}{0.000000in}}{%
\pgfpathmoveto{\pgfqpoint{-0.000000in}{0.000000in}}%
\pgfpathlineto{\pgfqpoint{-0.048611in}{0.000000in}}%
\pgfusepath{stroke,fill}%
}%
\begin{pgfscope}%
\pgfsys@transformshift{0.630000in}{0.972000in}%
\pgfsys@useobject{currentmarker}{}%
\end{pgfscope}%
\end{pgfscope}%
\begin{pgfscope}%
\definecolor{textcolor}{rgb}{0.000000,0.000000,0.000000}%
\pgfsetstrokecolor{textcolor}%
\pgfsetfillcolor{textcolor}%
\pgftext[x=0.322898in, y=0.933420in, left, base]{\color{textcolor}\rmfamily\fontsize{8.000000}{9.600000}\selectfont \(\displaystyle {0.88}\)}%
\end{pgfscope}%
\begin{pgfscope}%
\pgfsetbuttcap%
\pgfsetroundjoin%
\definecolor{currentfill}{rgb}{0.000000,0.000000,0.000000}%
\pgfsetfillcolor{currentfill}%
\pgfsetlinewidth{0.803000pt}%
\definecolor{currentstroke}{rgb}{0.000000,0.000000,0.000000}%
\pgfsetstrokecolor{currentstroke}%
\pgfsetdash{}{0pt}%
\pgfsys@defobject{currentmarker}{\pgfqpoint{-0.048611in}{0.000000in}}{\pgfqpoint{-0.000000in}{0.000000in}}{%
\pgfpathmoveto{\pgfqpoint{-0.000000in}{0.000000in}}%
\pgfpathlineto{\pgfqpoint{-0.048611in}{0.000000in}}%
\pgfusepath{stroke,fill}%
}%
\begin{pgfscope}%
\pgfsys@transformshift{0.630000in}{1.278000in}%
\pgfsys@useobject{currentmarker}{}%
\end{pgfscope}%
\end{pgfscope}%
\begin{pgfscope}%
\definecolor{textcolor}{rgb}{0.000000,0.000000,0.000000}%
\pgfsetstrokecolor{textcolor}%
\pgfsetfillcolor{textcolor}%
\pgftext[x=0.322898in, y=1.239420in, left, base]{\color{textcolor}\rmfamily\fontsize{8.000000}{9.600000}\selectfont \(\displaystyle {0.90}\)}%
\end{pgfscope}%
\begin{pgfscope}%
\pgfsetbuttcap%
\pgfsetroundjoin%
\definecolor{currentfill}{rgb}{0.000000,0.000000,0.000000}%
\pgfsetfillcolor{currentfill}%
\pgfsetlinewidth{0.803000pt}%
\definecolor{currentstroke}{rgb}{0.000000,0.000000,0.000000}%
\pgfsetstrokecolor{currentstroke}%
\pgfsetdash{}{0pt}%
\pgfsys@defobject{currentmarker}{\pgfqpoint{-0.048611in}{0.000000in}}{\pgfqpoint{-0.000000in}{0.000000in}}{%
\pgfpathmoveto{\pgfqpoint{-0.000000in}{0.000000in}}%
\pgfpathlineto{\pgfqpoint{-0.048611in}{0.000000in}}%
\pgfusepath{stroke,fill}%
}%
\begin{pgfscope}%
\pgfsys@transformshift{0.630000in}{1.584000in}%
\pgfsys@useobject{currentmarker}{}%
\end{pgfscope}%
\end{pgfscope}%
\begin{pgfscope}%
\definecolor{textcolor}{rgb}{0.000000,0.000000,0.000000}%
\pgfsetstrokecolor{textcolor}%
\pgfsetfillcolor{textcolor}%
\pgftext[x=0.322898in, y=1.545420in, left, base]{\color{textcolor}\rmfamily\fontsize{8.000000}{9.600000}\selectfont \(\displaystyle {0.92}\)}%
\end{pgfscope}%
\begin{pgfscope}%
\definecolor{textcolor}{rgb}{0.000000,0.000000,0.000000}%
\pgfsetstrokecolor{textcolor}%
\pgfsetfillcolor{textcolor}%
\pgftext[x=0.267343in,y=0.972000in,,bottom,rotate=90.000000]{\color{textcolor}\rmfamily\fontsize{9.000000}{10.800000}\selectfont SSIM}%
\end{pgfscope}%
\begin{pgfscope}%
\pgfpathrectangle{\pgfqpoint{0.630000in}{0.360000in}}{\pgfqpoint{2.520000in}{1.224000in}}%
\pgfusepath{clip}%
\pgfsetrectcap%
\pgfsetroundjoin%
\pgfsetlinewidth{1.003750pt}%
\definecolor{currentstroke}{rgb}{0.000000,0.000000,1.000000}%
\pgfsetstrokecolor{currentstroke}%
\pgfsetdash{}{0pt}%
\pgfpathmoveto{\pgfqpoint{0.744545in}{1.354500in}}%
\pgfpathlineto{\pgfqpoint{1.126364in}{1.339200in}}%
\pgfpathlineto{\pgfqpoint{1.508182in}{1.323900in}}%
\pgfpathlineto{\pgfqpoint{1.890000in}{1.323900in}}%
\pgfpathlineto{\pgfqpoint{2.271818in}{1.293300in}}%
\pgfpathlineto{\pgfqpoint{2.653636in}{1.063800in}}%
\pgfpathlineto{\pgfqpoint{3.035455in}{0.880200in}}%
\pgfusepath{stroke}%
\end{pgfscope}%
\begin{pgfscope}%
\pgfpathrectangle{\pgfqpoint{0.630000in}{0.360000in}}{\pgfqpoint{2.520000in}{1.224000in}}%
\pgfusepath{clip}%
\pgfsetbuttcap%
\pgfsetroundjoin%
\definecolor{currentfill}{rgb}{0.000000,0.000000,1.000000}%
\pgfsetfillcolor{currentfill}%
\pgfsetlinewidth{1.003750pt}%
\definecolor{currentstroke}{rgb}{0.000000,0.000000,1.000000}%
\pgfsetstrokecolor{currentstroke}%
\pgfsetdash{}{0pt}%
\pgfsys@defobject{currentmarker}{\pgfqpoint{-0.027778in}{-0.027778in}}{\pgfqpoint{0.027778in}{0.027778in}}{%
\pgfpathmoveto{\pgfqpoint{0.000000in}{-0.027778in}}%
\pgfpathcurveto{\pgfqpoint{0.007367in}{-0.027778in}}{\pgfqpoint{0.014433in}{-0.024851in}}{\pgfqpoint{0.019642in}{-0.019642in}}%
\pgfpathcurveto{\pgfqpoint{0.024851in}{-0.014433in}}{\pgfqpoint{0.027778in}{-0.007367in}}{\pgfqpoint{0.027778in}{0.000000in}}%
\pgfpathcurveto{\pgfqpoint{0.027778in}{0.007367in}}{\pgfqpoint{0.024851in}{0.014433in}}{\pgfqpoint{0.019642in}{0.019642in}}%
\pgfpathcurveto{\pgfqpoint{0.014433in}{0.024851in}}{\pgfqpoint{0.007367in}{0.027778in}}{\pgfqpoint{0.000000in}{0.027778in}}%
\pgfpathcurveto{\pgfqpoint{-0.007367in}{0.027778in}}{\pgfqpoint{-0.014433in}{0.024851in}}{\pgfqpoint{-0.019642in}{0.019642in}}%
\pgfpathcurveto{\pgfqpoint{-0.024851in}{0.014433in}}{\pgfqpoint{-0.027778in}{0.007367in}}{\pgfqpoint{-0.027778in}{0.000000in}}%
\pgfpathcurveto{\pgfqpoint{-0.027778in}{-0.007367in}}{\pgfqpoint{-0.024851in}{-0.014433in}}{\pgfqpoint{-0.019642in}{-0.019642in}}%
\pgfpathcurveto{\pgfqpoint{-0.014433in}{-0.024851in}}{\pgfqpoint{-0.007367in}{-0.027778in}}{\pgfqpoint{0.000000in}{-0.027778in}}%
\pgfpathlineto{\pgfqpoint{0.000000in}{-0.027778in}}%
\pgfpathclose%
\pgfusepath{stroke,fill}%
}%
\begin{pgfscope}%
\pgfsys@transformshift{0.744545in}{1.354500in}%
\pgfsys@useobject{currentmarker}{}%
\end{pgfscope}%
\begin{pgfscope}%
\pgfsys@transformshift{1.126364in}{1.339200in}%
\pgfsys@useobject{currentmarker}{}%
\end{pgfscope}%
\begin{pgfscope}%
\pgfsys@transformshift{1.508182in}{1.323900in}%
\pgfsys@useobject{currentmarker}{}%
\end{pgfscope}%
\begin{pgfscope}%
\pgfsys@transformshift{1.890000in}{1.323900in}%
\pgfsys@useobject{currentmarker}{}%
\end{pgfscope}%
\begin{pgfscope}%
\pgfsys@transformshift{2.271818in}{1.293300in}%
\pgfsys@useobject{currentmarker}{}%
\end{pgfscope}%
\begin{pgfscope}%
\pgfsys@transformshift{2.653636in}{1.063800in}%
\pgfsys@useobject{currentmarker}{}%
\end{pgfscope}%
\begin{pgfscope}%
\pgfsys@transformshift{3.035455in}{0.880200in}%
\pgfsys@useobject{currentmarker}{}%
\end{pgfscope}%
\end{pgfscope}%
\begin{pgfscope}%
\pgfpathrectangle{\pgfqpoint{0.630000in}{0.360000in}}{\pgfqpoint{2.520000in}{1.224000in}}%
\pgfusepath{clip}%
\pgfsetrectcap%
\pgfsetroundjoin%
\pgfsetlinewidth{1.003750pt}%
\definecolor{currentstroke}{rgb}{1.000000,0.000000,0.000000}%
\pgfsetstrokecolor{currentstroke}%
\pgfsetdash{}{0pt}%
\pgfpathmoveto{\pgfqpoint{0.744545in}{0.864900in}}%
\pgfpathlineto{\pgfqpoint{1.126364in}{0.864900in}}%
\pgfpathlineto{\pgfqpoint{1.508182in}{0.880200in}}%
\pgfpathlineto{\pgfqpoint{1.890000in}{0.864900in}}%
\pgfpathlineto{\pgfqpoint{2.271818in}{0.849600in}}%
\pgfpathlineto{\pgfqpoint{2.653636in}{0.834300in}}%
\pgfpathlineto{\pgfqpoint{3.035455in}{0.834300in}}%
\pgfusepath{stroke}%
\end{pgfscope}%
\begin{pgfscope}%
\pgfpathrectangle{\pgfqpoint{0.630000in}{0.360000in}}{\pgfqpoint{2.520000in}{1.224000in}}%
\pgfusepath{clip}%
\pgfsetbuttcap%
\pgfsetmiterjoin%
\definecolor{currentfill}{rgb}{1.000000,0.000000,0.000000}%
\pgfsetfillcolor{currentfill}%
\pgfsetlinewidth{1.003750pt}%
\definecolor{currentstroke}{rgb}{1.000000,0.000000,0.000000}%
\pgfsetstrokecolor{currentstroke}%
\pgfsetdash{}{0pt}%
\pgfsys@defobject{currentmarker}{\pgfqpoint{-0.027778in}{-0.027778in}}{\pgfqpoint{0.027778in}{0.027778in}}{%
\pgfpathmoveto{\pgfqpoint{0.000000in}{0.027778in}}%
\pgfpathlineto{\pgfqpoint{-0.027778in}{-0.027778in}}%
\pgfpathlineto{\pgfqpoint{0.027778in}{-0.027778in}}%
\pgfpathlineto{\pgfqpoint{0.000000in}{0.027778in}}%
\pgfpathclose%
\pgfusepath{stroke,fill}%
}%
\begin{pgfscope}%
\pgfsys@transformshift{0.744545in}{0.864900in}%
\pgfsys@useobject{currentmarker}{}%
\end{pgfscope}%
\begin{pgfscope}%
\pgfsys@transformshift{1.126364in}{0.864900in}%
\pgfsys@useobject{currentmarker}{}%
\end{pgfscope}%
\begin{pgfscope}%
\pgfsys@transformshift{1.508182in}{0.880200in}%
\pgfsys@useobject{currentmarker}{}%
\end{pgfscope}%
\begin{pgfscope}%
\pgfsys@transformshift{1.890000in}{0.864900in}%
\pgfsys@useobject{currentmarker}{}%
\end{pgfscope}%
\begin{pgfscope}%
\pgfsys@transformshift{2.271818in}{0.849600in}%
\pgfsys@useobject{currentmarker}{}%
\end{pgfscope}%
\begin{pgfscope}%
\pgfsys@transformshift{2.653636in}{0.834300in}%
\pgfsys@useobject{currentmarker}{}%
\end{pgfscope}%
\begin{pgfscope}%
\pgfsys@transformshift{3.035455in}{0.834300in}%
\pgfsys@useobject{currentmarker}{}%
\end{pgfscope}%
\end{pgfscope}%
\begin{pgfscope}%
\pgfsetrectcap%
\pgfsetmiterjoin%
\pgfsetlinewidth{0.803000pt}%
\definecolor{currentstroke}{rgb}{0.000000,0.000000,0.000000}%
\pgfsetstrokecolor{currentstroke}%
\pgfsetdash{}{0pt}%
\pgfpathmoveto{\pgfqpoint{0.630000in}{0.360000in}}%
\pgfpathlineto{\pgfqpoint{0.630000in}{1.584000in}}%
\pgfusepath{stroke}%
\end{pgfscope}%
\begin{pgfscope}%
\pgfsetrectcap%
\pgfsetmiterjoin%
\pgfsetlinewidth{0.803000pt}%
\definecolor{currentstroke}{rgb}{0.000000,0.000000,0.000000}%
\pgfsetstrokecolor{currentstroke}%
\pgfsetdash{}{0pt}%
\pgfpathmoveto{\pgfqpoint{3.150000in}{0.360000in}}%
\pgfpathlineto{\pgfqpoint{3.150000in}{1.584000in}}%
\pgfusepath{stroke}%
\end{pgfscope}%
\begin{pgfscope}%
\pgfsetrectcap%
\pgfsetmiterjoin%
\pgfsetlinewidth{0.803000pt}%
\definecolor{currentstroke}{rgb}{0.000000,0.000000,0.000000}%
\pgfsetstrokecolor{currentstroke}%
\pgfsetdash{}{0pt}%
\pgfpathmoveto{\pgfqpoint{0.630000in}{0.360000in}}%
\pgfpathlineto{\pgfqpoint{3.150000in}{0.360000in}}%
\pgfusepath{stroke}%
\end{pgfscope}%
\begin{pgfscope}%
\pgfsetrectcap%
\pgfsetmiterjoin%
\pgfsetlinewidth{0.803000pt}%
\definecolor{currentstroke}{rgb}{0.000000,0.000000,0.000000}%
\pgfsetstrokecolor{currentstroke}%
\pgfsetdash{}{0pt}%
\pgfpathmoveto{\pgfqpoint{0.630000in}{1.584000in}}%
\pgfpathlineto{\pgfqpoint{3.150000in}{1.584000in}}%
\pgfusepath{stroke}%
\end{pgfscope}%
\begin{pgfscope}%
\pgfsetbuttcap%
\pgfsetmiterjoin%
\definecolor{currentfill}{rgb}{1.000000,1.000000,1.000000}%
\pgfsetfillcolor{currentfill}%
\pgfsetfillopacity{0.800000}%
\pgfsetlinewidth{1.003750pt}%
\definecolor{currentstroke}{rgb}{0.800000,0.800000,0.800000}%
\pgfsetstrokecolor{currentstroke}%
\pgfsetstrokeopacity{0.800000}%
\pgfsetdash{}{0pt}%
\pgfpathmoveto{\pgfqpoint{1.976269in}{0.408611in}}%
\pgfpathlineto{\pgfqpoint{3.081944in}{0.408611in}}%
\pgfpathquadraticcurveto{\pgfqpoint{3.101389in}{0.408611in}}{\pgfqpoint{3.101389in}{0.428056in}}%
\pgfpathlineto{\pgfqpoint{3.101389in}{0.699197in}}%
\pgfpathquadraticcurveto{\pgfqpoint{3.101389in}{0.718642in}}{\pgfqpoint{3.081944in}{0.718642in}}%
\pgfpathlineto{\pgfqpoint{1.976269in}{0.718642in}}%
\pgfpathquadraticcurveto{\pgfqpoint{1.956825in}{0.718642in}}{\pgfqpoint{1.956825in}{0.699197in}}%
\pgfpathlineto{\pgfqpoint{1.956825in}{0.428056in}}%
\pgfpathquadraticcurveto{\pgfqpoint{1.956825in}{0.408611in}}{\pgfqpoint{1.976269in}{0.408611in}}%
\pgfpathlineto{\pgfqpoint{1.976269in}{0.408611in}}%
\pgfpathclose%
\pgfusepath{stroke,fill}%
\end{pgfscope}%
\begin{pgfscope}%
\pgfsetrectcap%
\pgfsetroundjoin%
\pgfsetlinewidth{1.003750pt}%
\definecolor{currentstroke}{rgb}{0.000000,0.000000,1.000000}%
\pgfsetstrokecolor{currentstroke}%
\pgfsetdash{}{0pt}%
\pgfpathmoveto{\pgfqpoint{1.995714in}{0.640864in}}%
\pgfpathlineto{\pgfqpoint{2.092936in}{0.640864in}}%
\pgfpathlineto{\pgfqpoint{2.190158in}{0.640864in}}%
\pgfusepath{stroke}%
\end{pgfscope}%
\begin{pgfscope}%
\pgfsetbuttcap%
\pgfsetroundjoin%
\definecolor{currentfill}{rgb}{0.000000,0.000000,1.000000}%
\pgfsetfillcolor{currentfill}%
\pgfsetlinewidth{1.003750pt}%
\definecolor{currentstroke}{rgb}{0.000000,0.000000,1.000000}%
\pgfsetstrokecolor{currentstroke}%
\pgfsetdash{}{0pt}%
\pgfsys@defobject{currentmarker}{\pgfqpoint{-0.027778in}{-0.027778in}}{\pgfqpoint{0.027778in}{0.027778in}}{%
\pgfpathmoveto{\pgfqpoint{0.000000in}{-0.027778in}}%
\pgfpathcurveto{\pgfqpoint{0.007367in}{-0.027778in}}{\pgfqpoint{0.014433in}{-0.024851in}}{\pgfqpoint{0.019642in}{-0.019642in}}%
\pgfpathcurveto{\pgfqpoint{0.024851in}{-0.014433in}}{\pgfqpoint{0.027778in}{-0.007367in}}{\pgfqpoint{0.027778in}{0.000000in}}%
\pgfpathcurveto{\pgfqpoint{0.027778in}{0.007367in}}{\pgfqpoint{0.024851in}{0.014433in}}{\pgfqpoint{0.019642in}{0.019642in}}%
\pgfpathcurveto{\pgfqpoint{0.014433in}{0.024851in}}{\pgfqpoint{0.007367in}{0.027778in}}{\pgfqpoint{0.000000in}{0.027778in}}%
\pgfpathcurveto{\pgfqpoint{-0.007367in}{0.027778in}}{\pgfqpoint{-0.014433in}{0.024851in}}{\pgfqpoint{-0.019642in}{0.019642in}}%
\pgfpathcurveto{\pgfqpoint{-0.024851in}{0.014433in}}{\pgfqpoint{-0.027778in}{0.007367in}}{\pgfqpoint{-0.027778in}{0.000000in}}%
\pgfpathcurveto{\pgfqpoint{-0.027778in}{-0.007367in}}{\pgfqpoint{-0.024851in}{-0.014433in}}{\pgfqpoint{-0.019642in}{-0.019642in}}%
\pgfpathcurveto{\pgfqpoint{-0.014433in}{-0.024851in}}{\pgfqpoint{-0.007367in}{-0.027778in}}{\pgfqpoint{0.000000in}{-0.027778in}}%
\pgfpathlineto{\pgfqpoint{0.000000in}{-0.027778in}}%
\pgfpathclose%
\pgfusepath{stroke,fill}%
}%
\begin{pgfscope}%
\pgfsys@transformshift{2.092936in}{0.640864in}%
\pgfsys@useobject{currentmarker}{}%
\end{pgfscope}%
\end{pgfscope}%
\begin{pgfscope}%
\definecolor{textcolor}{rgb}{0.000000,0.000000,0.000000}%
\pgfsetstrokecolor{textcolor}%
\pgfsetfillcolor{textcolor}%
\pgftext[x=2.267936in,y=0.606836in,left,base]{\color{textcolor}\rmfamily\fontsize{7.000000}{8.400000}\selectfont 70\% paired data}%
\end{pgfscope}%
\begin{pgfscope}%
\pgfsetrectcap%
\pgfsetroundjoin%
\pgfsetlinewidth{1.003750pt}%
\definecolor{currentstroke}{rgb}{1.000000,0.000000,0.000000}%
\pgfsetstrokecolor{currentstroke}%
\pgfsetdash{}{0pt}%
\pgfpathmoveto{\pgfqpoint{1.995714in}{0.500432in}}%
\pgfpathlineto{\pgfqpoint{2.092936in}{0.500432in}}%
\pgfpathlineto{\pgfqpoint{2.190158in}{0.500432in}}%
\pgfusepath{stroke}%
\end{pgfscope}%
\begin{pgfscope}%
\pgfsetbuttcap%
\pgfsetmiterjoin%
\definecolor{currentfill}{rgb}{1.000000,0.000000,0.000000}%
\pgfsetfillcolor{currentfill}%
\pgfsetlinewidth{1.003750pt}%
\definecolor{currentstroke}{rgb}{1.000000,0.000000,0.000000}%
\pgfsetstrokecolor{currentstroke}%
\pgfsetdash{}{0pt}%
\pgfsys@defobject{currentmarker}{\pgfqpoint{-0.027778in}{-0.027778in}}{\pgfqpoint{0.027778in}{0.027778in}}{%
\pgfpathmoveto{\pgfqpoint{0.000000in}{0.027778in}}%
\pgfpathlineto{\pgfqpoint{-0.027778in}{-0.027778in}}%
\pgfpathlineto{\pgfqpoint{0.027778in}{-0.027778in}}%
\pgfpathlineto{\pgfqpoint{0.000000in}{0.027778in}}%
\pgfpathclose%
\pgfusepath{stroke,fill}%
}%
\begin{pgfscope}%
\pgfsys@transformshift{2.092936in}{0.500432in}%
\pgfsys@useobject{currentmarker}{}%
\end{pgfscope}%
\end{pgfscope}%
\begin{pgfscope}%
\definecolor{textcolor}{rgb}{0.000000,0.000000,0.000000}%
\pgfsetstrokecolor{textcolor}%
\pgfsetfillcolor{textcolor}%
\pgftext[x=2.267936in,y=0.466404in,left,base]{\color{textcolor}\rmfamily\fontsize{7.000000}{8.400000}\selectfont 20\% paired data}%
\end{pgfscope}%
\end{pgfpicture}%
\makeatother%
\endgroup%